\documentclass[11pt, a4paper]{article}
\pdfoutput=1 
\usepackage{jheppub}
\usepackage{amssymb,amsmath}
\usepackage{graphicx}
\usepackage{booktabs}
\usepackage{subfig}
\usepackage{enumerate}
\allowdisplaybreaks

\newcommand{\beq}{\begin{equation}}
\newcommand{\eeq}{\end{equation}}
\newcommand{\beqa}{\begin{eqnarray}}
\newcommand{\eeqa}{\end{eqnarray}}
\newcommand{\cm}{{\cal M}^0}

\newcommand{\cmb}{{\cal M}}

\newcommand{\ds}{{\rm d}\hat{\sigma}}
\newcommand{\dsigma}{{\rm d}\hat{\sigma}}
\newcommand{\dphi}{{\rm d}\Phi}
\newcommand{\wt}{\widetilde}
\newcommand{\re}{{\rm{Re}}}
\newcommand{\norm}{{\cal N}}

\newcommand{\order}[1]{{\cal O}(#1)}

\newcommand{\Q}[1]{#1_Q}
\newcommand{\Qb}[1]{#1_{\bar{Q}}}
\newcommand{\qi}[1]{\hat{#1}_q}
\newcommand{\qbi}[1]{\hat{#1}_{\bar{q}}}

\newcommand{\gl}[1]{#1_g}

\newcommand{\ph}[1]{#1_{\gamma}}

\newcommand{\ione}[2]{{\bf I}^{(1)}_{#1 #2}}
\newcommand{\Dzero}{{\cal D}_0}
\newcommand{\Done}{{\cal D}_1}
\newcommand{\cep}{C(\epsilon)}
\newcommand{\cepb}{\bar{C}(\epsilon)}
\newcommand{\asmu}{\alpha_s(\mu)}
\newcommand{\poles}{{\cal P}oles}

\newcommand{\GG}{{\bf \Gamma}}
\newcommand{\gaussf}[4]{\ensuremath{\, _2F_1 \left(#1,#2,#3;#4\right)}}
\newcommand{\Fthreetwo}[6]{\ensuremath{\, _3F_2 \left(#1,#2,#3;#4,#5;#6\right)}}
\newcommand{\LO}{\mathrm{LO}}
\newcommand{\NLO}{\mathrm{NLO}}
\newcommand{\NNLO}{\mathrm{NNLO}}
\newcommand{\RR}{\mathrm{RR}}
\newcommand{\MF}{\mathrm{MF}}
\newcommand{\RV}{\mathrm{RV}}
\newcommand{\VV}{\mathrm{VV}}
\newcommand{\VS}{\mathrm{VS}}

\newcommand{\rV}{\mathrm{V}}
\newcommand{\rS}{\mathrm{S}}
\newcommand{\rT}{\mathrm{T}}
\newcommand{\rU}{\mathrm{U}}

\newcommand{\loopint}{\int\frac{{\rm d}^dl}{(2\pi)^d}}
\def\d{\hbox{d}}

\def\e{\epsilon}

\def\ba{\begin{eqnarray}}
\def\ea{\end{eqnarray}}
\def\eps{\epsilon}
\def\bs{\boldsymbol}

\title{Top quark pair production at NNLO in the quark-antiquark channel}

\author[a,b]{Gabriel Abelof,}
\author[c,d]{Aude Gehrmann-De Ridder,}
\author[c,e]{Imre Majer}

\affiliation[a]{Department of Physics \& Astronomy, Northwestern University, Evanston, IL 60208, USA}
\affiliation[b]{High Energy Physics Division, Argonne National Laboratory, Argonne, IL 60439, USA}
\affiliation[c]{Institute for Theoretical Physics, ETH, CH-8093 Z\"urich, Switzerland}
\affiliation[d]{Physics Institute University of Z\"urich, Winterthurerstrasse 190, CH-8057, Z\"urich, Switzerland}
\affiliation[e]{Department of Physics and Astronomy, Seoul National University, 1 Gwanak-ro, Gwanak-gu, 
Seoul, 151-747, Korea}
\emailAdd{gabriel.abelof@northwestern.edu}
\emailAdd{gehra@itp.phys.ethz.ch}
\emailAdd{imre.majer@alumni.ethz.ch}


\abstract{
We present the derivation of the NNLO two-parton final state contributions to top pair production in the 
quark-antiquark channel proportionnal to the leading colour factor $N_c^2$. Together with the three and 
four-parton NNLO contributions presented in a previous publication, this enables us to complete the 
phenomenologically most important NNLO corrections to top pair hadro-production in this channel.  
We derive this two-parton contribution using the massive extension of the NNLO antenna subtraction 
formalism and implement those corrections in a parton-level event generator providing full kinematical 
information on all final state particles. In addition, we also derive the heavy quark contributions proportional  
to $N_h$. Combining the new leading-colour and heavy quark contributions together with the light quark  
contributions derived previously, we present NNLO differential distributions for LHC and Tevatron. 
We also compute the differential top quark forward-backward asymmetry at Tevatron and 
find that our results are in good agreement with the measurements by the D0 collaboration.}
   
\preprint{
\hfill
\begin{minipage}[t]{8em}\today\\    
ZU-TH 17/15\\
\end{minipage}
}

\begin{document}
\bibliographystyle{JHEP-2}
\maketitle

%
\section{Introduction}
\label{sec:intro}
The precise study of top quark properties provides a detailed probe of the Standard Model of particle physics 
and models beyond it.  Due to its small lifetime, the top quark decays before it hadronizes and it is the only 
quark whose production dynamics can be studied without having to account for hadronization effects. 
Following its discovery twenty years ago \cite{Abachi:1995iq,Abe:1995hr}, the top quark has been 
studied intensively at the Tevatron. Owing to the limited number of top quarks produced, differential studies 
there suffered from rather large uncertainties \cite{Aaltonen:2009iz,Abazov:2014vga}. The measured top 
quark forward-backward asymmetry received much attention since it was first found to deviate substantially 
from Standard Model expectations \cite{Aaltonen:2011kc}. 

With the large number of top quark pairs produced at the LHC, the study of its properties has become 
precision physics, and the measurement of very precise differential cross sections, attainable. Recently, the 
ATLAS and CMS collaborations reported measurements of normalised differential observables in 
$t\bar{t}$ production in several kinematical variables, such as the transverse momentum, rapidity and 
invariant mass of the $t\bar{t}$ system, as well as the top quark transverse momentum and rapidity 
\cite{Aad:2012hg,Aad:2014zka,Aad:2015eia,Chatrchyan:2012saa,Khachatryan:2015oqa}.  
More recently, both LHC collaborations have measured the charge asymmetry in top-pair production,
\cite{Aad:2015noh,Khachatryan:2015mna}.

These measurements will allow for a very detailed and accurate probe of the top quark production mechanism. To 
reliably interpret data, these very precise measurements must be matched onto equally accurate theoretical 
predictions, which can be obtained by including contributions up to next-to-next-to leading order (NNLO) in 
perturbative QCD. At present, NNLO corrections have been included in calculations of the inclusive total 
$t\bar{t}$ cross section in \cite{Czakon:2013goa}, for the inclusive and differential Tevatron top quark 
forward-backward asymmetry in \cite{Czakon:2014xsa}, and for differential LHC distributions in \cite{Czakon:2015owf}. 

At NNLO, perturbative calculations of collider observables are typically carried out using parton-level event 
generators. These programs generate events for all parton-level subprocesses relevant to a given final state 
configuration up to NNLO accuracy and provide full kinematical information on an event-by-event basis. 
An NNLO event generator for observables with $n$ final state particles or jets involves three main building 
blocks: the two-loop corrections to the $n$-parton final state, referred to as double-virtual contributions 
(${\rm d}\sigma^{\VV}_{\NNLO}$), the one-loop corrections to the $(n+1)$-parton final state, called real-virtual 
contributions (${\rm d}\sigma^{\RV}_{\NNLO}$), and the tree-level $(n+2)$-parton double real contribution 
(${\rm d}\sigma^{\RR}_{\NNLO}$). These three building blocks involve infrared divergences that arise from 
the exchange or emission of soft and collinear partons and cancel only in their sum. In addition, two mass
factorisation counter-terms, ${\rm{d}}\sigma^{\MF,1}_{\NNLO}$ and ${\rm {d}}\sigma^{\MF,2}_{\NNLO}$, are 
needed in the three and two-parton final states contributions respectively in order to cancel infrared 
divergences originated from initial state collinear radiation.

The combination of subprocesses of different particle multiplicity which are individually infrared divergent is a 
major challenge in the construction of NNLO parton-level event generators. Employing a 
subtraction method to regulate these infrared singularities, the NNLO partonic cross section for top pair 
production in a given partonic channel has the general structure \cite{GehrmannDeRidder:2005cm}
\beqa
\label{eq.sigNNLO}
{\rm d}\hat\sigma_{\NNLO}&=&\int_{\Phi_{4}}\left({\rm{d}}\hat\sigma_{\NNLO}^{\RR}
-{\rm{d}}\hat\sigma_{\NNLO}^{\rS}\right)+\int_{\Phi_{4}}{\rm{d}}\hat\sigma_{\NNLO}^{\rS}\nonumber\\
&+&\int_{\Phi_{3}}\left({\rm{d}}\hat\sigma_{\NNLO}^{\RV}-{\rm{d}}\hat\sigma_{\NNLO}^{\VS}\right)
+\int_{\Phi_{3}}{\rm{d}}\hat\sigma_{\NNLO}^{\VS}
+\int_{\Phi_{3}}{\rm{d}}\hat\sigma_{\NNLO}^{\MF,1}\nonumber\\
&+&\int_{\Phi_2}{\rm{d}}\hat\sigma_{\NNLO}^{\VV}+\int_{\Phi_2}{\rm{d}}\hat\sigma_{\NNLO}^{\MF,2}.
\eeqa
Two types of subtraction terms are introduced: ${\rm d} \hat\sigma^{\rS}_{\NNLO}$ for the $4$-parton final 
state, and ${\rm d} \hat\sigma^{\VS}_{\NNLO}$ for the $3$-parton final state. The former approximates the 
infrared behaviour of the double real contributions ${\rm d}\hat\sigma^{\RR}_{\NNLO}$ in their single and 
double unresolved limits, whereas the latter reproduces the single unresolved behaviour of the mixed 
real-virtual contributions ${\rm d}\hat\sigma^{\RV}_{\NNLO}$.

In the context of the present computation, we further decompose the double real subtraction term 
$\ds_{\NNLO}^{\rS}$ as in \cite{Abelof:2014fza}, which allows us to rearrange the different terms in 
eq.(\ref{eq.sigNNLO}) into the more convenient form
\beqa\label{eq.subnnlo}
\ds_{\NNLO}&=&\int_{\Phi_{4}}\left[\ds_{\NNLO}^{\RR}-\ds_{\NNLO}^{\rS}\right]\nonumber \\
&+& \int_{\Phi_{3}}\left[\ds_{\NNLO}^{\RV}-\ds_{\NNLO}^{\rT}\right] \nonumber \\
&+&\int_{\Phi_{2}}\left[\ds_{\NNLO}^{\VV}-\ds_{\NNLO}^{\rU}\right],
\eeqa
with 
\beqa
\label{eq.Tdef}  \ds_{\NNLO}^{\rT} &=& 
\phantom{ -\int_1 }\ds_{\NNLO}^{\VS}- \int_1 \ds_{\NNLO}^{\rS,1} - \ds_{\NNLO}^{\MF,1},  \\
\label{eq.Udef}  \ds_{\NNLO}^{\rU} &=& 
-\int_1 \ds_{\NNLO}^{\VS}-\int_2 \ds_{\NNLO}^{\rS,2}-\ds_{\NNLO}^{\MF,2}. 
\eeqa

The NNLO contributions to the top pair production in the quark-antiquark channel can be decomposed into 
colour factors as
\beqa
\label{eq.qqbdec}
&&\hspace{-0.3in}\ds_{q \bar{q},\NNLO}=(N_c^2-1)\bigg[N_c^2\,A +N_c \,B + C + \frac{D}{N_c} 
+\frac{E}{N_c^2} + N_l\,\left(N_c\,F_l +\frac{G_l}{N_c}\right) \nonumber \\
&&\hspace{0.3in}+ N_h \left(N_c\,F_h +\frac{G_h}{N_c} \right)+ N_l^2\,H_l \,+\,N_l \,N_h\, H_{lh} 
+N_{h}^2\,H_{h} \bigg],
\eeqa 
with $N_c$ being the number of colours, $N_l$ the number of light quark flavours, and $N_h$ the number of 
heavy quark flavours. All coefficients multiplying the different colour factors in eq.(\ref{eq.qqbdec}) are gauge
invariant and can be computed independently. 

The goals of this paper are twofold. The first aim is the derivation of the two-parton contribution given in 
eq.(\ref{eq.subnnlo}) for the colour factor $N_c^2$ within the theoretical framework of antenna subtraction 
with massive fermions. This framework has been established in the massless case in \cite{Currie:2013vh,
GehrmannDeRidder:2005cm,GehrmannDeRidder:2007jk,Glover:2010im,Currie:2013dwa,
GehrmannDeRidder:2011aa,GehrmannDeRidder:2012dg} and extended to the massive case in 
\cite{Abelof:2011jv,Abelof:2012he,Abelof:2012rv,Abelof:2014fza,GehrmannDeRidder:2009fz,Abelof:2011ap,
Abelof:2014jna}. We shall not repeat it here. Together with the three and four-parton contributions derived in 
\cite{Abelof:2014fza} the two-parton contribution yields the coefficient $A$ in eq.(\ref{eq.qqbdec}) and enables 
us to obtain theoretical predictions for the most significant NNLO contribution to top pair hadro-production in the 
quark-antiquark channel. In addition, we derive the heavy flavour contributions labeled in eq.(\ref{eq.qqbdec}) 
as $F_h$ and $G_h$, presenting a detailed discussion of the ultraviolet renormalisation of the loop amplitudes 
involved. 

Our second goal is to present phenomenological results for Tevatron and LHC energies, including the 
leading-colour and heavy quark contributions derived in this paper, as well as the light quark terms, which we completed in 
previous papers \cite{Abelof:2011ap,Abelof:2014jna}.

We employ analytic expressions for all matrix elements, subtraction terms and integrated subtraction terms 
that arise in our calculation. For the real-virtual contributions, we use {\tt OpenLoops} \cite{Cascioli:2011va} in 
combination with {\tt CutTools} \cite{Ossola:2007ax} and for the double virtual contributions we use the analytic 
two-loop matrix-elements of \cite{Bonciani:2008az,Bonciani:2009nb}.\footnote{For the gluon-gluon channel, which
is beyond the scope of the present paper, analytic expressions for the leading-colour and light quark contributions 
of the two-loop amplitudes have been presented in \cite{Bonciani:2010mn,Bonciani:2013ywa}.} As a result, at the 
real-virtual and virtual-virtual levels, the explicit infrared poles are cancelled analytically. Furthermore, the 
numerical evaluation of the finite remainders at each level included in the parton-level generator is substantially 
faster than when these terms are evaluated numerically. 

The paper is organized as follows: In section \ref{sec:integratedantennae} we present the derivation of the 
new integrated massive initial-final antennae required in the double virtual counter-term $\ds^{\rU}_{\NNLO}$ 
for the leading-colour contributions to top pair production in the quark-antiquark channel. Section 
\ref{sec:virtualvirtualNc} contains the explicit derivation of the counter-term $\ds^{\rU}_{\NNLO}$ expressed in 
terms of massive integrated dipoles. Employing those dipoles, we find that the virtual-virtual subtraction term 
has a similar structure to the one found in the massless case in the context of di-jet production at NNLO 
\cite{Currie:2013vh}. Section \ref{sec:heavy} is dedicated to the heavy quark contributions proportional to $N_h N_c$ and $N_h/N_c$. 
Section \ref{sec:Results} contains our numerical results: Various differential distributions 
are presented and the phenomenological impact of this computation is discussed. Finally section 
\ref{sec:conclusions} contains our conclusions. We enclose three appendices containing the master 
integrals required to compute the integrated antennae presented in section \ref{sec:integratedantennae}, as 
well as the phase space parametrisations needed to calculate these integrals.

%
\section{Integrated initial-final massive antennae}
\label{sec:integratedantennae}
\subsection{General features} 
One of the aims of this paper is the computation of the NNLO two-parton contribution proportional to 
$N_c^2$ for top pair production in the quark-antiquark channel. Within the antenna subtraction method, this 
involves the construction of the double virtual counter-term $\ds_{q\bar{q},\NNLO,N_c^2}^{\rU}$ which 
renders the two-parton contribution 
\beq
\label{eq.twopartonfs}
\int_{{\rm{d}}\Phi_{2\phantom{+1}}}\left[\ds_{q \bar{q},\NNLO,N_c^2}^{\VV}-\ds_{q \bar{q},\NNLO,N_c^2}^{\rU}\right]
\eeq
finite and suitable to be implemented in an NNLO parton-level generator. 

As shown in eq.(\ref{eq.Udef}), this virtual-virtual counter-term contains mass factorisation terms together 
with integrated double real and real-virtual subtraction terms. In order to compute these, the integrated forms 
of the antenna functions present in the corresponding real-virtual and double real subtraction terms are 
needed. The integrated antennae are obtained by integrating the antenna functions over the corresponding 
antenna phase space inclusively. This integration is carried out analytically, and turns the implicit soft and 
collinear singularities of the antenna functions into explicit poles in the dimensional regularisation parameter 
$\e$, which are then cancelled against the explicit poles of the double virtual matrix elements and mass 
factorisation counter-terms.  

In the context of this paper, we need the integrated forms of two massive initial-final A-type antennae 
which are presented here for the first time. These antennae involve a massless initial state quark and a 
massive quark in the final state as radiators, and one or two final state gluons which are unresolved, i.e. soft 
and/or collinear. More concretely, the calculation that we present in this paper requires the integrated forms of 
the tree-level four-parton antenna $A_4^0(1_Q,3_g,4_g,\hat{2}_{q})$ and the one-loop leading-colour 
three-parton antenna $A_3^{1,lc}(1_Q, 3_g,\hat{2}_q)$. The tree-level four-parton antenna appears 
in the double real subtraction term ($\ds_{q \bar{q},\NNLO,N_c^2}^{\rS,2}$) given in \cite{Abelof:2014fza}. It is 
needed in order to capture the infrared behaviour of the leading-colour double real contributions associated to 
the partonic process $q\bar{q} \to t\bar{t}gg$ when the two final state gluons are unresolved and 
colour-connected. The one-loop antenna appears in the real-virtual subtraction term 
$\ds_{q\bar{q},\NNLO,N_c^2}^{\VS,a}$ given in \cite{Abelof:2014fza}. It is required to capture the infrared 
behaviour of the leading-colour part of one-loop matrix element associated to the partonic process 
$q\bar{q} \to t\bar{t}g$. 

The general definition of the integrated forms of three and four-parton massive antennae in all three 
configurations (final-final, initial-final, initial-initial), as well as the corresponding phase space factorisations 
and mappings, have been presented in \cite{GehrmannDeRidder:2009fz,Abelof:2011jv,Abelof:2011ap}. Below
we will only briefly recall the definition of initial-final integrated massive antennae, as they are the only ones 
needed in the calculation presented in this paper.

Initial-final three-parton tree-level and one-loop antennae like $A_3^{l}(1_Q, 3_g,\hat{2}_q)$ ($l=0,1$) are generically 
denoted as $X^{l}_{i,jk}$, with parton $i$ in the initial state, and $j,k$ in the final state. The kinematics 
associated to these initial-final antenna functions with one massive radiator are $p_i+q\rightarrow p_j+p_k$, 
with $q^2=-Q^2<0$, $p_i^2=p_j^2=0$ and $p_k^2=m_Q^2$, and the integrated antennae are defined as 
\cite{Abelof:2011ap}
\beq
{\cal X}^{l}_{i,jk}=\frac{1}{C(\epsilon)}\frac{(Q^2+m_Q^2)}{2\pi} \int {\rm d}\Phi_2(p_j,p_k;p_i,q)X^{l}_{i,jk}.
\eeq
${\rm d}\Phi_2$ is the corresponding $2 \to 2$ phase space, and 
\beq\label{eq.ceps}
C(\epsilon)=(4\pi)^{\epsilon}\frac{e^{-\epsilon \gamma}}{8\pi^2}.
\eeq
The initial-final massive antenna phase space denoted by ${\rm d}\Phi_{X_{i,jk}}$ is given by
\beq
\label{eq:initialfinalphase}
{\rm d}\Phi_{X_{i,jk}}(p_j,p_k;p_i,q)= \frac{(Q^2+m_Q^2)}{2\pi} {\rm d}\Phi_2(p_j,p_k;p_i,q).
\eeq
In addition to the phase space integration, one-loop antennae must be also integrated over the loop 
momentum.

Initial-final four-parton antennae like $A_4^0(1_Q,3_g,4_g,\hat{2}_{q})$ are generically denoted as 
$X^{0}_{i,jkl}$, with parton $i$ in the initial state and $j,k,l$ in the final state. The associated kinematics
is $p_i+q\rightarrow p_j+p_k+p_l$, with $q^2=-Q^2$, $p_i^2=p_j^2=p_k^2=0$ and $p_l^2=m_Q^2$. The 
integrated forms of these four-parton initial-final antennae are obtained as
\beq\label{eq.aint4}
{\cal X}^0_{i,jkl}=\frac{1}{\left[C(\epsilon)\right]^2} \frac{(Q^2+m_{Q}^2)}{2\pi} \int \d\Phi_3(p_j,p_k,p_l;p_i,q) X^0_{i,jkl}\,,
\eeq  
and the corresponding antenna phase space is given by
\beq
\label{eq:initialfinalphase}
{\rm d}\Phi_{X_{i,jkl}}(p_j,p_k,p_l;p_i,q)=\frac{(Q^2+m_Q^2)}{2\pi} {\rm d}\Phi_3(p_j,p_k,p_l;p_i,q).
\eeq

After the inclusive phase space integration, the three and four-parton integrated antennae are functions of 
$Q^2$, $p_i\cdot q$ and $m_Q^2$. We change the dependence on the latter two variables for
\beq
x_0=\frac{Q^2}{Q^2+m_Q^2}
\eeq
and
\beq
x_i=\frac{Q^2+m_Q^2}{2p_i\cdot q}.
\eeq
We also change the dependence on $Q^2$ for $s_{\bar{i}(\wt{jk})}$ and $s_{\bar{i}(\wt{jkl})}$ in three and 
four-parton integrated antennae respectively, where the bars and tildes are related to the initial-final phase 
space mappings that can be found, for example, in appendix B of \cite{Abelof:2011ap}. The invariants are 
defined as $s_{ij}=2p_i\cdot p_j$ with the crossings explicitly performed on the momenta $p_i,\,p_j$, i.e. 
$p_i^0,p_j^0>0$. In integrated subtraction terms, the mapped final state momenta $\wt{p}_{jk}$ and 
$\wt{p}_{jkl}$ are relabeled in accordance to the labeling of momenta in the lower multiplicity final state, in 
such a way that the integrated antennae end up depending on invariants $s_{\bar{i}j}$. This is the notation 
that we shall follow throughout this paper. 

In all cases, the inclusive phase space integration is carried out following the standard technique of reduction 
to master integrals using integration-by-parts identities (IBP) \cite{Chetyrkin:1981qh,Tkachov:1981wb}. The 
computation of the master integrals is then carried out either directly to all-orders or iteratively, order by 
order in $\e$, using differential equations techniques \cite{Gehrmann:1999as}. 

All the master integrals found contain multiplicative factors of the form $(1-x_i)^{-n\e}$ which regulate soft 
endpoint singularities and should be kept unexpanded. All other terms can be safely expanded in $\e$. As we 
explain in detail in appendix C, in some of the master integrals found in the reduction of the one-loop antenna 
of interest here, namely $A_3^{1,lc}(1_Q, 3_g,\hat{2}_q)$, more than one $\e$-power of $(1-x_i)$ is 
encountered. Therefore, the master integrals which we denote collectively by $I_{\alpha}(x_i,x_0,\e)$ can be 
most generally expressed as: 
\beq
\label{eq.ialpha}
I_{\alpha}(x_i,x_0,\e)=\sum_{n,m} (1-x_i)^{m-n\e} R_{\alpha}^{(n)}(x_i,x_0,\epsilon).
\eeq
The integers $m$ and $n$ are specific to each master integral; the functions 
$R_{\alpha}^{(n)}(x_i,x_0,\epsilon)$ are regular as $x_i \to 1$ and can be calculated as Laurent series in $\e$.

The integrated antennae collectively denoted by ${\cal X}(x_i,x_0,\e)$ are linear combinations of master 
integrals with coefficients containing poles in $\epsilon$ as well as in $(1-x_i)$. After the masters have been 
inserted into the integrated antennae, these take the form 
\begin{equation}
{\cal X}(x_i,x_0,\e) =(1-x_i)^{-1-n\e}  {\cal R}_{{\cal X}}(x_i,x_0,\e),
\end{equation}
where ${\cal R_{\cal X}}(x_i,x_0,\e)$ is a regular function as $x_i\to 1$. The Laurent expansion of the singular 
factor $(1-x_i)^{-1-n\e}$ is done in the form of distributions: 
\beq
(1-x)^{-1-n\e}=-\frac{\delta(1-x)}{n\e}+\sum_{m=0}^{\infty}\frac{(-n\e)^m}{m!}{\cal D}_m(x)
\eeq
with
\beq
{\cal D}_m(x)=\left(\frac{\ln ^m(1-x)}{(1-x)} \right)_+\, .
\eeq

It is worth noting that in the functions $R_{\alpha}^{(n)}(x_i,x_0,\epsilon)$ and ${\cal R}_{\cal X}(x_i,x_0,\e)$, which 
are regular as $x_i \to 1$, the massless limit $x_0\to 1$ cannot in general be safely taken. This is due to the 
presence of terms of the form $\log^k(1-x_0)=\log^k(m_{Q}^2/(Q^2 +m_{Q}^2))$, which are expected and 
correspond to quasi-collinear limits of the antenna functions 
\cite{Abelof:2011jv,Catani:2002hc,GehrmannDeRidder:2009fz}.

In order to perform the Laurent expansion of the integrated antennae, we distinguish two regions: a hard 
region where $ x_i\neq 1$ and a soft region where $x_i=1$. The highest order in $\e$ needed in the expansion of 
each master integral is determined by the $\e$ and $x_i$-dependent coefficient that multiplies the integral in the 
integrated antenna. In the soft region, since the expansion in distributions generates an additional $1/\e$ 
factor, the functions $R_{\alpha}^{(n)}$ are required to one order higher in $\e$ than in the hard region. 
Further details concerning the integration of $A_4^0(1_Q,3_g,4_g,\hat{2}_{q})$ and 
$A_3^{1,lc}(1_Q, 3_g,\hat{2}_q)$ will be presented below. 

\subsection{The integrated tree-level antennae ${\cal A}_{q,Qg}^0$ and ${\cal A}^0_{q,Qgg}$}
The computation of $\ds_{q \bar{q},\NNLO,N_c^2}^{\rU}$ requires the knowledge of the integrated 
tree-level three and four-parton massive initial-final antennae ${\cal A}_{q,Qg}^0$ and ${\cal A}^0_{q,Qgg}$. 
The former has been derived in \cite{Abelof:2011jv}, and only its pole part will be recalled below. The 
integrated flavour-violating four-parton antenna ${\cal A}^0_{q,Qgg}$ is new and will be presented here for 
the first time.

\subsubsection{The integrated three-parton antenna ${\cal A}_{q,Qg}^0$} 
The pole part of the integrated antenna ${\cal A}_{q,Qg}^0$ is given by
\beq
\label{eq.poleA03}
\poles\left({\cal A}^0_{q,Qg}(\e,s_{\bar{i}j},x_i)\right)
=-2\ione{Q}{\bar{q}}(\e,s_{\bar{i}j})\delta(1-x_i) + \Gamma^{(1)}_{qq}(x_i),
\eeq
where $\ione{Q}{\bar{q}}$ is a colour-order infrared singularity operator which has been presented in 
\cite{Abelof:2011jv} and reads
\beq
\label{eq.IoneqQ}
\ione{Q}{\bar{q}}(\e,s_{Q\bar{q}})=
-\frac{e^{\e\gamma_E}}{2\Gamma(1-\e)}\left(\frac{s_{Q\bar{q}}}{\mu^2}\right)^{-\e}
\left[ \frac{1}{2\e^2}+\frac{5}{4\e}+\frac{1}{2\e}\ln\left( \frac{m_Q^2}{s_{Q\bar{q}}}\right)\right],
\eeq
and $\Gamma_{qq}^{(1)}(x)$ is a colour-ordered splitting kernel given by
\beq
\label{eq.kernelqq}
\Gamma_{qq}^{(1)}(x)=-\frac{1}{\e}\left(\frac{3}{4}\delta(1-x)+\Dzero(x)-\frac{1}{2} -\frac{1}{2}x\right).
\eeq

As can be seen from the equations above, ${\cal A}^0_{q,Qg}$ has its deepest pole at $1/{\e}^2$. In the 
virtual-virtual subtraction term that will be presented in section \ref{sec:virtualvirtualNc}, we will need products
and convolutions of two of these integrated antennae, and we will therefore require its Laurent expansion up 
to order $\e^2$. The resulting expression is too long to be presented here. It is included in an ancillary 
{\tt Mathematica} file attached to the arXiv submission of this paper.

The integrated initial-final antennae depend explicitly on one momentum fraction $x_i$ carried by the 
initial state parton. In order to combine these integrated forms with other integrated subtraction terms 
depending on both momentum fractions $x_1$ and $x_2$ carried by the incoming quark-antiquark pair, it is 
useful to make all antennae explicitly depend on both $x_1$ and $x_2$. We achieve this by introducing 
delta functions in the following way:
\beqa
\label{eq.A03x1x2a}
&&\hspace{-0.2in} {\cal A}^0_{q,Qg}(\e,s_{\bar{1}j},x_1,x_2) =
\int{\rm d}x_2{\cal A}^0_{q,Qg}(\e,s_{\bar{1}j},x_1)\delta(1-x_2) \\
\label{eq.A03x1x2b}
&&\hspace{-0.2in} {\cal A}^0_{q,Qg}(\e,s_{\bar{2}j},x_2,x_1) =
\int{\rm d}x_1{\cal A}^0_{q,Qg}(\e,s_{\bar{2}j},x_2)\delta(1-x_1) 
\eeqa
The same prescription will be applied to ${\cal A}^0_{q,Qgg}$ and ${\cal A}^{1,lc}_{q,Qg}$.

\subsubsection{The integrated four-parton antenna ${\cal A}^0_{q,Qgg}$}
To compute the integrated initial-final massive antenna ${\cal A}^0_{q,Qgg}$ we start by expressing all phase 
space integrals as cuts of two-loop four-point functions with two off-shell legs in forward scattering 
kinematics~\cite{Anastasiou:2002yz}, and reduce these four-point functions to master integrals with 
{\tt LiteRed} \cite{Lee:2012cn}. 

\begin{figure}[t]
\begin{tabular}{ccc}
\subfloat[$I_{[0]}$, $I_{[-8]}$]{\includegraphics[width=0.3\textwidth]{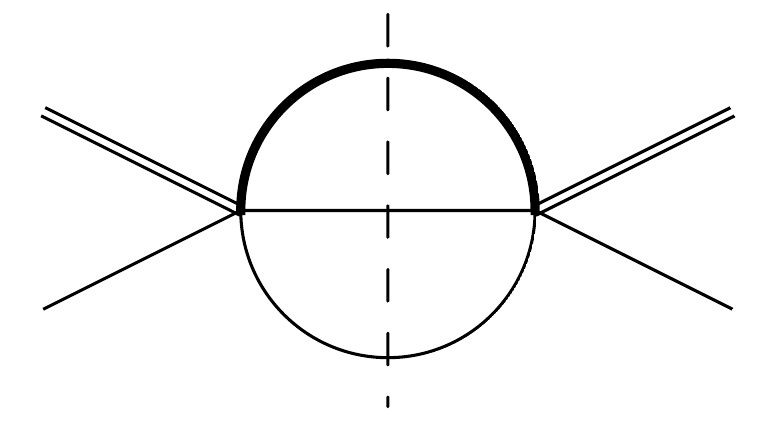}} & 
\subfloat[$I_{[4]}$]{\includegraphics[width=0.3\textwidth]{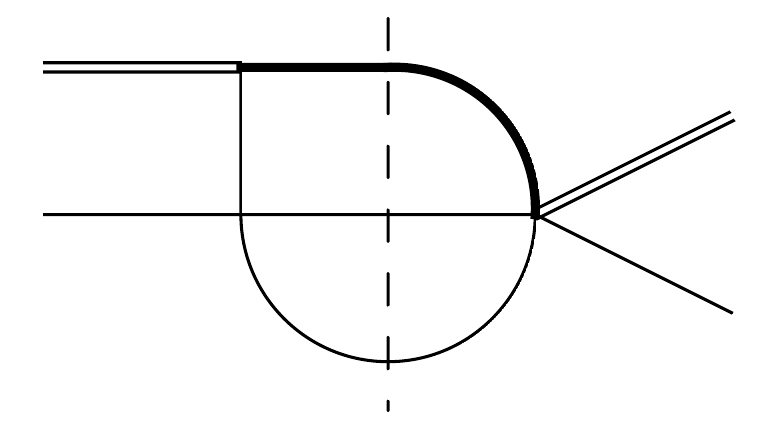}} &
\subfloat[$I_{[4,8]}$]{\includegraphics[width=0.3\textwidth]{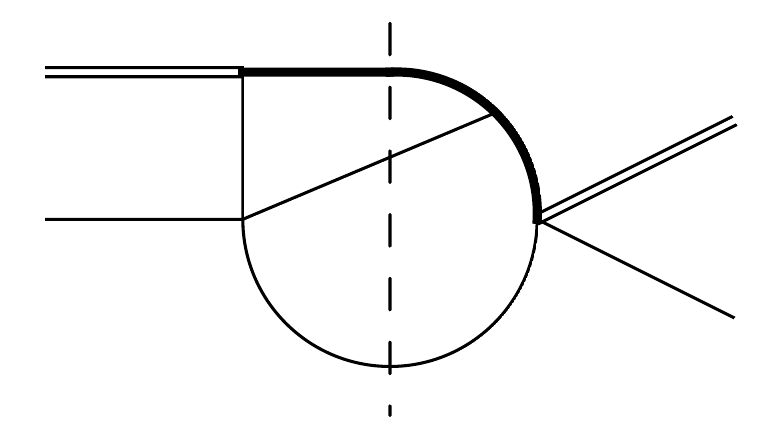}} \\
\subfloat[$I_{[4,5,8]}$]{\includegraphics[width=0.3\textwidth]{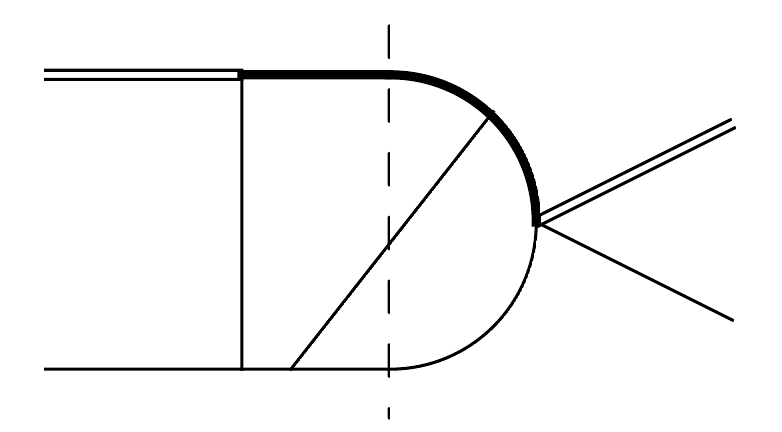}} &
\subfloat[$I_{[4,7,8]}$]{\includegraphics[width=0.3\textwidth]{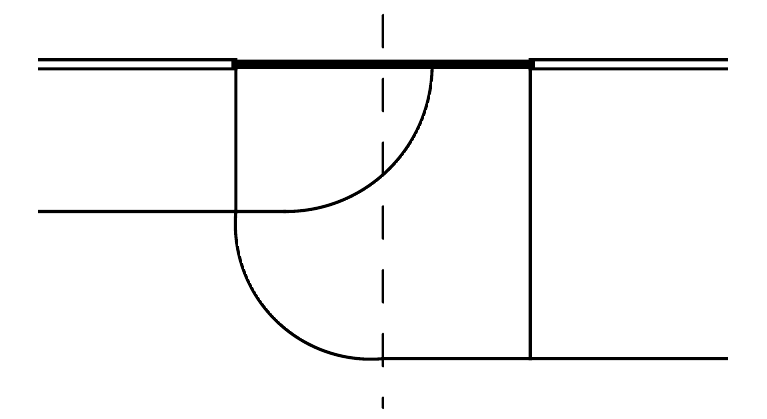}} &
\end{tabular}
\caption{Master integrals for ${\cal A}^0_{q,Qgg}$. Integrals are labelled according to the denominators 
involved (see eq.(\ref{eq.a04denos})). Bold (thin) lines are massive (massless). The double line in the external 
states represents the off-shell momentum $q$ with $q^2=-Q^2$. The cut propagators are the ones intersected 
by the dashed lines.}
\label{fig.a04masters}
\end{figure}

As mentioned above, we consider the following DIS-like kinematics 
\beq
p_2+q\to p_1+p_3+p_4
\eeq
with $p_2^2=p_3^2=p_4^2=0$, $p_1^2=m_Q^2$, $q^2=-Q^2<0$. The denominators that we find are:
\beqa
\label{eq.a04denos}
D_1&=&p_1^2-m_Q^2\nonumber\\
D_2&=&p_3^2\nonumber\\
D_3&=&p_4^2\nonumber\\
D_4&=&(p_1-q)^2\nonumber\\
D_5&=&(p_3+p_4)^2\nonumber\\
D_6&=&(p_3-p_2)^2\nonumber\\
D_7&=&(p_4-p_2)^2\nonumber\\
D_8&=&(p_1+p_3)^2-m_Q^2
\eeqa
where $D_1$, $D_2$ and $D_3$ are the cut propagators. In the reduction procedure we impose momentum 
conservation and mass-shell conditions for the external legs, and we discard all integrals that do not contain
the propagators $D_1$, $D_2$ or $D_3$ with power -1. We find the six master integrals depicted in 
fig.\ref{fig.a04masters}. Their explicit expressions, together with a 
parametrisation of the phase space measure are presented in appendices \ref{sec:misa04} and 
\ref{sec:psa04} respectively.

The integrals $I_{[0]}$, $I_{[8]}$ and $I_{[4]}$ are known \cite{Abelof:2012he,Abelof:2011jv}. We computed 
the remaining master integrals, namely, $I_{[4,8]}$, $I_{[4,5,8]}$ and $I_{[4,7,8]}$, using differential equation 
techniques, and we present them here for the first time. Although the differential equations for these three 
integrals are coupled, they can be decoupled order by order in $\e$, which enabled us to solve them 
iteratively. We fixed two of the three sets of undetermined integration constants by demanding that the 
integrals be regular as $x\to 1$ after factoring out an appropriate power of $(1-x)$. In order to fix the 
remaining set of integration constants we calculated via direct evaluation the soft limit of $I_{[4,7,8]}$, and 
employed it as a boundary condition. The details of the computation of this soft limit can be found in appendix 
\ref{sec:psa04}. 

The highest order in $\e$ with which each master integral contributes up to the finite part of 
${\cal A}^0_{q,Qgg}$ is specified in table \ref{tab.a04masters}. As mentioned before, in the soft region 
($x=1$), they must be expanded to one order higher than in the hard region ($x \neq 1$).
\begin{table}
\centering
\begin{tabular}{|c|c|c|c|}
\hline
Integral & Needed in the hard region to  & Deepest pole \\
\hline
$I_{[0]}$ & $\e^3$ & $\e^0$ \\
$I_{[-8]}$ & $\e^3$ & $\e^0$ \\
$I_{[4]}$ & $\e^2$ & $\e^0$ \\
$I_{[4,8]}$ & $\e^0$ & $\e^{0}$ \\
$I_{[4,5,8]}$ & $\e^0$ & $\e^{-3}$ \\
$I_{[4,7,8]}$ & $\e^0$ & $\e^{-3}$ \\
\hline
\end{tabular}
\caption{Deepest poles and highest order in $\e$ needed for each master integral in ${\cal A}^0_{q,Qgg}$.}
\label{tab.a04masters}
\end{table}

After expanding in distributions we find that the integrated four-parton massive antenna ${\cal A}^0_{q,Qgg}$ 
has a deepest pole at $\order{1/\e^4}$. This is expected since at the unintegrated level 
$A_4^0(\Q{1},\gl{3},\gl{4},\qi{2})$ contains double soft and triple collinear singular limits, which are employed
to reproduce the singular behaviour of the double real contributions associated to the partonic process 
$ q \bar{q} \to t \bar{t} g g$. The integrated antenna can be written in terms of harmonic polylogarithms (HPLs)
with arguments $x_i$ or $x_0$ and generalised harmonic polylogarithms (GPLs) of argument $x_0$ and 
weights involving $1/x_i$. GPLs and HPLs appear with up to trascendentality three and four respectively. 
Given its length, the full expression of this antenna up to $\order{\e^0}$ is given in the ancillary file attached to
the arXiv submission of this paper. The deepest poles read   
\beqa
&&\hspace{-0.2in}{\cal A}^0_{q,Qgg}(\e,s_{\bar{i}j},x_i)=s_{\bar{i}j}^{-2\e}\nonumber\\
&&\times\bigg\{\frac{1}{4\e^4}\delta(1-x_i)+\frac{1}{2\e^3}\bigg[1+x_i+\delta(1-x_i)\bigg( \frac{35}{24}+G(1;x_0) \bigg)
-2\Dzero(x_i)\bigg]\nonumber\\
&&\hspace{0.1in}+\frac{1}{\e^2}\bigg[\frac{(11 x_0^2x_i ^3+59 x_0^2x_i ^2-22x_0x_i ^2-118x_0x_i+2 x_i+68)}{24(1-x_0 x_i)^2}-\frac{(9+11x_i^2)}{8(1-x_i)}G(0;x_i)\nonumber\\
&&\hspace{0.2in}-2(1+x_i)G(1;x_i)+\frac{(7-x_i^2)}{4(1-x_i)}G(1;x_0)+\frac{3(1+x_i^2)}{4(1-x_i)}G\bigg(\frac{1}{x_i};x_0\bigg)\nonumber\\
&&\hspace{0.2in}+\delta(1-x_i)\bigg(\frac{331}{144}-\frac{13\pi^2}{48}+\frac{35}{24}G(1;x_0)+G(1,1;x_0) \bigg)-\Dzero(x_i)\bigg(\frac{35}{12}+2G(1;x_0)\bigg)\nonumber\\
&&\hspace{0.2in}+4\Done(x_i)\bigg]+\order{\e^{-1}}\bigg\}.\nonumber\\
\eeqa

\subsection{The integrated one-loop three-parton antenna  ${\cal A}^{1,lc}_{q,Qg}$}
The leading-colour one-loop antenna employed in \cite{Abelof:2014fza} for the construction of a real-virtual
subtraction term for partonic process $ q \bar{q} \to t \bar{t} g$ at one-loop is given by
\beq\label{eq.A13}
A_3^{1,lc}(1_Q,3_g,\hat{2}_q)=\frac{|\cmb_3^{[lc]}(1_Q,3_g,\hat{2}_q)|^2}{|\cmb_2^0((\wt{13})_Q,\hat{\bar{2}}_q)|^2}-A_3^0(1_Q,3_g,\hat{2}_q)\frac{|\cmb_2^{[lc]}((\wt{13})_Q,\hat{\bar{2}}_q)|^2}{|\cmb_2^0((\wt{13})_Q,\hat{\bar{2}}_q)|^2},
\eeq
where the subscript $[lc]$ indicates that only the leading-colour primitive amplitude (proportional to $N_c$) is kept. As mentioned above, the DIS-like kinematics associated with this antenna function is 
\beq
p_2+q\to p_1+p_3
\eeq
with $p_2^2=p_3^2=0$, $p_1^2=m_Q^2$, $q^2=-Q^2<0$. The corresponding phase space parametrization
has been derived in \cite{Abelof:2011ap} in the context of the integration of tree-level initial-final massive 
antennae. It is repeated for completeness in appendix \ref{sec:misa13}. 

The phase space integration of the second term in eq.(\ref{eq.A13}) is trivial, since the ratio
$|\cmb_2^{[lc]}((\wt{13})_Q,\hat{\bar{2}}_q)|^2/|\cmb_2^0((\wt{13})_Q,\hat{\bar{2}}_q)|^2$ only depends on 
$Q^2$, and it can be therefore pulled out of the phase space integral. Thus, only $A_3^0(1_Q,3_g,\hat{2}_q)$ 
must be integrated over the phase space, and this integral is known.

\begin{figure}[t]
\begin{tabular}{ccc}
\subfloat[$I_{[3]}$]{\includegraphics[width=0.3\textwidth]{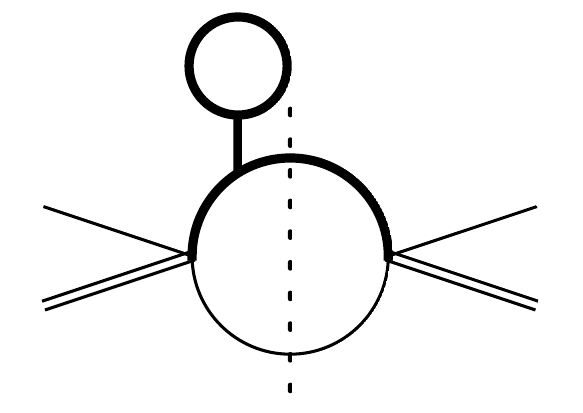}} & 
\subfloat[$I_{[3,5]}$]{\includegraphics[width=0.3\textwidth]{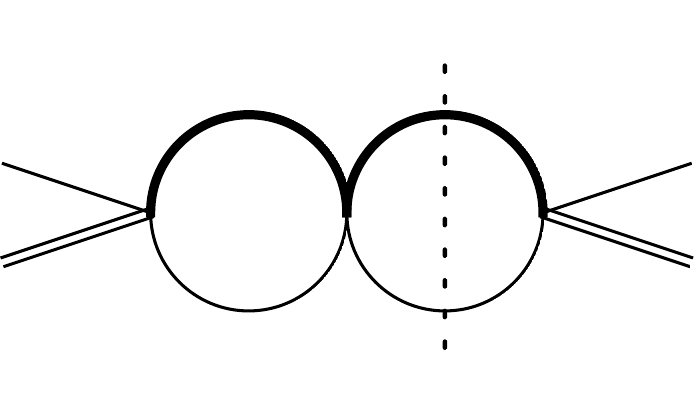}} &
\subfloat[$I_{[3,6]}$]{\includegraphics[width=0.3\textwidth]{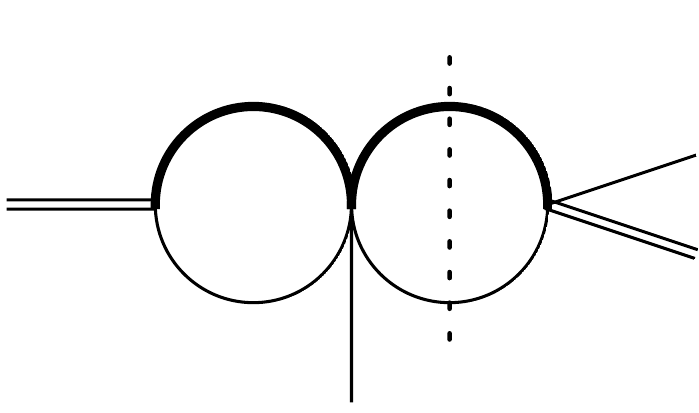}} \\
\subfloat[$I_{[4,6]}$]{\includegraphics[width=0.3\textwidth]{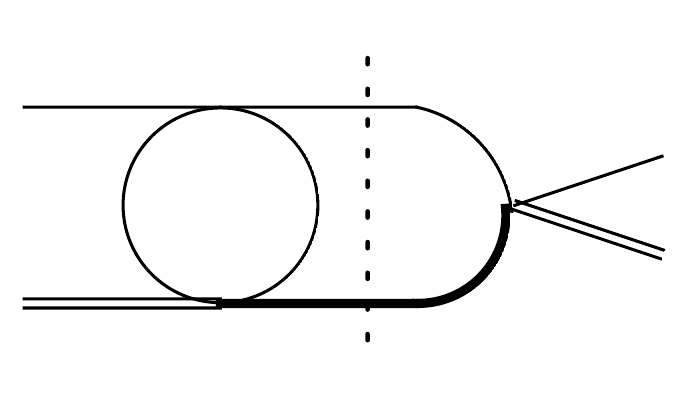}} &
\subfloat[$I_{[3,4,6]}$]{\includegraphics[width=0.3\textwidth]{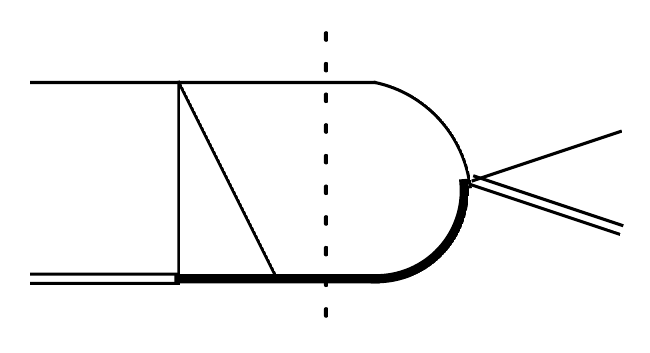}} &
\subfloat[$I_{[3,4,5,6]}$]{\includegraphics[width=0.3\textwidth]{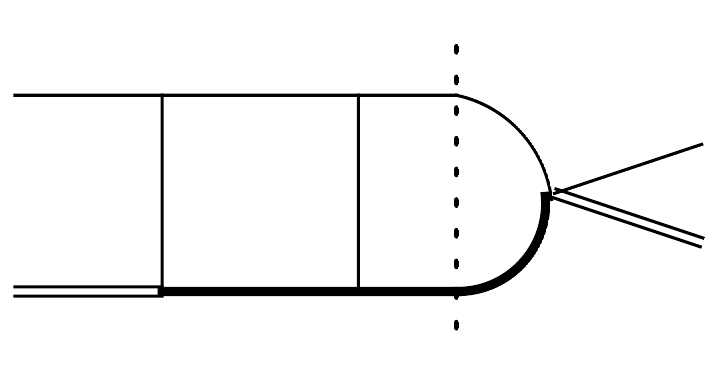}}\\
&\subfloat[$I_{[3,4,6,7]}$]{\includegraphics[width=0.3\textwidth]{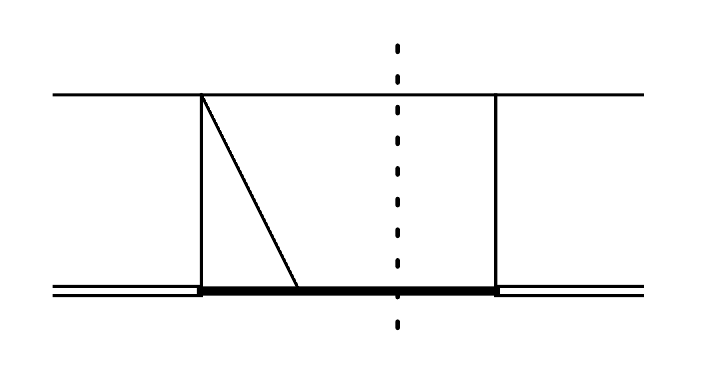}} &
\end{tabular}
\caption{Master integrals for ${\cal A}^{1,lc}_{q,Qg}$. Integrals are labelled according to the denominators 
involved (see eq.(\ref{eq.a13denos})). Bold (thin) lines are massive (massless). The double line in the external 
states represents the off-shell momentum $q$ with $q^2=-Q^2$. The cut propagators are the ones intersected 
by the dashed lines.}
\label{fig.a13masters}
\end{figure}

We must therefore focus on the first term in eq.(\ref{eq.A13}), for which we follow the same procedure as the 
one described above for ${\cal A}^0_{q,Qgg}$. The following denominators must be considered:
\beqa
\label{eq.a13denos}
&&\hspace{-0.2in}D_1 = p_1^2-m_Q^2 \nonumber\\
&&\hspace{-0.2in}D_2 = p_3^2 \nonumber\\
&&\hspace{-0.2in}D_3 =  l^2-m_Q^2\nonumber\\
&&\hspace{-0.2in}D_4 = (l-p_1)^2\nonumber\\
&&\hspace{-0.2in}D_5 = (l-p_1-p_3)^2\nonumber\\
&&\hspace{-0.2in}D_6 = (l-p_1+p_2-p_3)^2\nonumber\\
&&\hspace{-0.2in}D_7 = (p_1-q)^2,
\eeqa
where $D_1$ and $D_2$ are the cut propagators, $D_3$ to $D_6$ are the loop propagators, and 
$D_7=-s_{23}$. We find the master integrals depicted in fig.\ref{fig.a13masters}.

In addition to these new master integrals, the ultraviolet renormalised integrated antenna 
${\cal A}^{1,lc}_{q,Qg}$ also contains the known inclusive phase space integral, denoted by $I_{[0]}$ 
which was computed in \cite{Abelof:2011jv}. This integral is brought about by the phase space integration of the 
mass and strong coupling renormalisation counter-terms.

The evaluation of the loop integrals $I_{[3]}$, $I_{[3,5]}$, $I_{[3,6]}$ and $I_{[4,6]}$ is rather straightforward. It 
can be done to all orders in $\e$ by integrating the underlying loop integrals over the phase space. The 
remaining three master integrals, namely $I_{[3,4,6]}$, $I_{[3,4,5,6]}$ and $I_{[3,4,6,7]}$, cannot be computed 
in this way. Instead, we calculated them using differential equations. Unlike the case of the master integrals
needed for ${\cal A}^0_{q,Qgg}$, the integration constants cannot be in this case fixed by imposing 
regularity conditions on the integrals. This impossibility is due to the fact that each integral contains several
different powers of $(1-x)$. Therefore, in order to determine the integration constants, we employed in this 
case independently computed soft limits as boundary conditions. For $I_{[3,4,5,6]}$ we used the soft limit of 
the underlying one-loop box given in \cite{Brucherseifer:2013iv}, whereas for $I_{[3,4,6]}$ and $I_{[3,4,6,7]}$ 
we calculated the soft limit of the one-loop triangle to all orders in $\e$ using a Mellin-Barnes representation. 
More details on these derivations are given in appendix \ref{sec:misa13}. 
\begin{table}
\centering
\begin{tabular}{|c|c|c|c|}
\hline
Integral & Needed in the hard region to & Deepest pole\\
\hline
$I_{[0]}$ & $\e^3$ & $\e^{0}$ \\
$I_{[3]}$ & $\e^2$ & $\e^{-1}$ \\
$I_{[3,5]}$ & $\e^1$ & $\e^{-1}$ \\
$I_{[3,6]}$ & $\e^2$ & $\e^{-1}$ \\
$I_{[4,6]}$ & $\e^2$ & $\e^{-1}$ \\
$I_{[3,4,6]}$ & $\e^0$ & $\e^{-2}$ \\
$I_{[3,4,5,6]}$ & $\e^0$ & $\e^{-3}$ \\
$I_{[3,4,6,7]}$ & $\e^0$ & $\e^{-3}$ \\
\hline
\end{tabular}
\caption{Deepest poles and highest order in $\e$ needed for each master integral in ${\cal A}^{1,lc}_{q,Qg}$.}
\label{tab.a13masters}
\end{table}

Like the integrated four-parton antenna presented above, ${\cal A}^{1,lc}_{q,Qg}$ has its deepest pole at 
$\order{1/{\e}^4}$. It can also be written in terms of HPLs and GPLs of trascendentality four and three 
respectively. The HPLs have arguments $x_i$ or $x_0$, and the GPLs have argument $x_0$ and 
$1/x_i$ in the weights. The complete expression for ${\cal A}^{1,lc}_{q,Qg}$ is enclosed in the ancillary file
included in the arXiv submission of this paper. Its deepest poles are given by
\beqa
&&\hspace{-0.2in}{\cal A}^{1,lc}_{q,Qg}(\e,s_{\bar{i}j},x_i)=s_{\bar{i}j}^{-2\e}\nonumber\\
&&\times\bigg\{-\frac{1}{8\e^4}\delta(1-x_i)+\frac{1}{\e^3}\bigg[-\frac{(1+x_i)}{4}-\delta(1-x_i)\bigg( \frac{7}{6}+\frac{1}{4}G(1;x_0) \bigg)+\frac{1}{2}\Dzero(x_i)\bigg]\nonumber\\
&&\hspace{0.1in}+\frac{1}{\e^2}\bigg[-\frac{29+8x_0^2x_i^3+26x_0^2x_i^2-16x_0x_i^2-52x_0x_i+5x_i}{12(1-x_0 x_i)^2}+\frac{3(1+x_i^2)}{4(1-x_i)}G(0;x_i)\nonumber\\
&&\hspace{0.2in}+(1+x_i)G(1;x_i)-\frac{1}{1-x_i}G(1;x_0)+\frac{(1+x_i^2)}{2(1-x_i)}G\bigg(\frac{1}{x_i};x_0\bigg)\nonumber\\
&&\hspace{0.2in}+\delta(1-x_i)\bigg(-\frac{5}{3}+\frac{5\pi^2}{48}-\frac{17}{12}G(1;x_0)-\frac{1}{2}G(1,1;x_0) \bigg)+\Dzero(x_i)\bigg(\frac{17}{6}+G(1;x_0)\bigg)\nonumber\\
&&\hspace{0.2in}-2\Done(x_i)\bigg]+\order{\e^{-1}}\bigg\}.\nonumber\\
\eeqa

%
\section{Top quark pair production at NNLO} 
\label{sec:virtualvirtualNc}
Following the decomposition of the NNLO cross section for top pair production in two, three and four-parton final state contributions as shown in eqs.(\ref{eq.subnnlo}) and (\ref{eq.Udef}), in this section we 
present the two-parton final state for the coefficient $A$ in eq.(\ref{eq.qqbdec}). In particular, 
we shall construct the virtual-virtual subtraction term $\ds^{\rU}_{q\bar{q},\NNLO,N_c^2}$ as a 
combination of integrated real-real and real-virtual subtraction terms, and mass factorisation counter-terms. 
We will show that these ingredients can be arranged in such a way that $\ds^{\rU}_{q\bar{q},\NNLO,N_c^2}$ 
can be expressed in terms of so-called integrated dipoles, which facilitate the cancellation of explicit infrared
poles in the two-parton final state contribution.

For massless jet observables computed at NNLO with antenna subtraction, the general formalism allowing to 
write the real-virtual and virtual-virtual counter-terms using integrated dipoles was presented in 
\cite{Currie:2013vh}. There it was shown that massless integrated dipoles, denoted as $\bs{J}_{2}^{(\ell)}$ 
with $\ell=1,2$, emerge naturally when rearranging the mass-factorisation counter-terms and integrated 
double real and real-virtual subtraction terms into three different pieces that are free of collinear initial state 
singularities. It was furthermore observed that the colour-ordered $\bs{J}_{2}^{(\ell)}$'s are related to Catani's 
one and two-loop infrared singularity operators $\bs{I}_{ij}^{(\ell)}(\eps)$ \cite{Catani:1998bh}.

As we will see below, the virtual-virtual subtraction term $\ds^{\rU}_{q\bar{q},\NNLO,N_c^2}$ for heavy quark
pair production is similar in structure to the massless case. For the purpose of making this similarity manifest, 
we define a new type of massive integrated dipoles $\bs{J}_{2}^{(\ell)}$. In particular, the dipole 
$\bs{J}_{2}^{(1)}$ that we find in the present calculation of this leading colour contribution is related to the colour-ordered infrared singularity 
operator $\ione{Q}{{\bar{q}}}$, defined in \cite{Abelof:2011jv,Abelof:2014fza} and recalled in eq.(\ref{eq.IoneqQ}).

\subsection{NNLO contributions to top quark pair production in the $q\bar{q}$ channel} 
In order to set up our conventions for normalisation factors and matrix elements, we start by recalling the 
leading order partonic cross section for heavy quark pair production. It can be written as
\beq\label{eq.qqblo}
\ds_{q\bar{q},\LO}=\norm_{\LO}^{\:q\bar{q}}\int\dphi_2(p_3,p_4;p_1,p_2)\:|\cm_4(\Q{3},\Qb{4},\qbi{2},\qi{1})|^2 J^{(2)}_2(p_3,p_4),
\eeq
where, $\dphi_2(p_3,p_4;p_1,p_2)$ is the $2 \to 2$ partonic phase space, and $J^{(2)}_2(p_3,p_4)$ is a 
so-called measurement function, which ensures that a pair of final state massive quarks of momenta $p_3$ 
and $p_4$ are observed. $\cm_4(...)$ is a colour-ordered and coupling-stripped tree-level amplitude. It is 
related to the full amplitude through the (trivial) colour decomposition
\beq\label{eq.coldecqqblo}
{\cal M}^0_{q_1\bar{q}_2\rightarrow Q_3 \bar{Q}_4}=g_s^2\left( \delta_{i_3i_1}\delta_{i_2i_4}-\frac{1}{N_c}\delta_{i_3i_4}\delta_{i_2i_1}\right)\cm_4(\Q{3},\Qb{4},\qbi{2},\qi{1}).
\eeq
The normalisation factor is
\beq\label{eq.normlo}
\norm_{\LO}^{\:q\bar{q}}=\frac{1}{2s}\:\left( \frac{\asmu}{2\pi}\right)^2\:\frac{\cepb^2}{\cep^2}\:\frac{(N_c^2-1)}{4N_c^2},
\eeq
where $s$ is the energy squared in the hadronic center-of-mass frame. Included in $\norm_{\LO}^{\:q\bar{q}}$ 
are the flux factor, as well as the sums and averages over colour and spin. The constant $\cep$ has been 
defined in eq.(\ref{eq.ceps}), while $\cepb$ is  given by
\beq
\cepb=(4\pi)^{\e} e^{-\e \gamma_E},
\eeq
providing the useful relation
\beq
g_s^2=4\pi\alpha_s=\left( \frac{\alpha_s}{2\pi}\right)\frac{\cepb}{\cep}.
\eeq

The leading-colour double real corrections to heavy quark pair production are due to the tree-level partonic 
process $q\bar{q} \to Q \bar{Q} gg$. They read
\beqa
\label{eq.RR}
\lefteqn{\ds_{q\bar{q},NNLO,N_c^2}^{RR}=}\nonumber\\
&&\norm_{NNLO}^{\RR,q\bar{q}}\,N_c^2\,\dphi_4(p_3,p_4,p_5,p_6;p_1,p_2)
|{\cal M}^0_{q_1\bar{q}_2\rightarrow Q_3\bar{Q}_4g_5 g_6}|^2\Big|_{N_c^2} J_2^{(4)}(p_3,p_4,p_5,p_6),
\eeqa
where $\dphi_4$ is the $2 \to 4$ phase space, and 
$|{\cal M}^0_{q_1\bar{q}_2\rightarrow Q_3\bar{Q}_4g_5g_6}|^2$ is the square of the full coupling-stripped 
tree-level amplitude normalised to $(N_c^2-1)$. This factor is included in the overall normalisation $\norm_{\NNLO}^{\RR,q\bar{q}}$, which is given by
\beq\label{eq.Nrrnnlo}
\norm_{\NNLO}^{\RR,q\bar{q}}=\norm_{LO}^{q\bar{q}}\left(\frac{\asmu}{2\pi}\right)^2\frac{\cepb^2}{\cep^2}.
\eeq
The subscript $N_c^2$ on the matrix element squared indicates that only the terms proportional to this colour 
factor are to be kept.

Given the fact that the measurement function $J_2^{(4)}$ allows the final state gluons with momentum $p_5$ 
and $p_6$ to become unresolved, i.e. soft or collinear, eq.(\ref{eq.RR}) contains infrared divergences. These 
divergences are implicit, in the sense that they only become explicit as poles in $\e$ after the integration over 
the phase space is performed. The antenna subtraction term that regulates these infrared divergences has
been derived in \cite{Abelof:2014fza}.

The mixed real-virtual contributions are given by the phase space integral of the interfered one-loop and 
tree-level amplitudes for the $2\to 3$ process $ q \bar{q} \to Q \bar{Q} g$. They read
\beqa\label{eq.RV}
&&\hspace{-0.3in}\ds_{q \bar{q},\NNLO,N_c^2}^{\RV}=\norm_{\NNLO}^{\RV,q\bar{q}}\,N_c^2\, \int\frac{{\rm d}x_1}{x_1}\frac{{\rm d}x_2}{x_2}\dphi_3(p_3,p_4,p_5;x_1p_1,x_2p_2)\delta(1-x_1)\delta(1-x_2)\nonumber\\
&&\hspace{1in}\times\,2\re \left({\cal M}^1_{q_1 \bar{q}_2\rightarrow Q_3 \bar{Q}_4 g_5} {\cal M}^{0\,\dagger}_{q_1\bar{q}_2\rightarrow Q_3 \bar{Q}_4 g_5}\right)\Bigg|_{N_c^2}J_2^{(3)}(p_3,p_4,p_5),\nonumber\\
\eeqa
with the normalisation factor $\norm_{\NNLO}^{\RV,q\bar{q}}$ given by
\beq
\label{eq.NnnloRV}
\norm_{\NNLO}^{\RV,q\bar{q}}=\norm_{LO}^{q\bar{q}}\left(\frac{\asmu}{2\pi}\right)^2\cepb=\norm_{\NNLO}^{\RR,q\bar{q}}\:\cep.
\eeq

The real-virtual contributions in eq.(\ref{eq.RV}) contain explicit ultraviolet and infrared divergences as well as implicit infrared ones. The explicit poles in $\e$ originate from the loop integration in 
${\cal M}^1_{q_1\bar{q}_2\rightarrow t_3 \bar{t}_4 g_5}$, whereas the implicit singularities are due to the 
phase space integration over regions where the matrix elements diverge: the soft limit $p_5\rightarrow 0$ and 
the collinear limits $p_1||p_5$, $p_2||p_5$. While the ultraviolet poles are cancelled upon renormalisation, we 
employ a subtraction term $\ds_{q \bar{q},\NNLO,N_c^2}^{\rT}$ to deal with the infrared ones. This 
subtraction term has the twofold purpose of canceling the explicit poles, whose structure is well known 
\cite{Catani:2000ef}, while simultaneously regularising the phase space integrand in the soft and collinear 
limits. Its explicit construction was derived in \cite{Abelof:2014fza}.

Finally, the double virtual contributions have not been derived before and are the object of study in the next 
section. They involve the interference of a two-loop $2\rightarrow 2$ matrix element with its tree-level 
counterpart, and a one-loop amplitude squared. They can be written as, 
\beqa
\label{eq.setup.sigmannloVV}
&&\hspace{-0.5in}\ds_{q \bar{q},\NNLO,N_c^2}^{\VV}=\norm_{\NNLO}^{\VV,q\bar{q}}\,N_c^2\int\frac{{\rm d}x_1}{x_1}\frac{{\rm d}x_2}{x_2}\dphi_2(p_3,p_4;x_1p_1,x_2p_2)\delta(1-x_1)\delta(1-x_2)\nonumber\\
&&\hspace{0.5in}\times\bigg[ 2\re \left({\cal M}^2_{q_1\bar{q}_2\rightarrow Q_3 \bar{Q}_4} {\cal M}^{0\,\dagger}_{q_1\bar{q}_2\rightarrow Q_3 \bar{Q}_4}\right)+|{\cal M}^1_{q_1\bar{q}_2\rightarrow Q_3 \bar{Q}_4}|^2\bigg]\Bigg|_{N_c^2}\hspace{-0.075in}J_2^{(2)}(p_3,p_4)
\eeqa
where the normalisation factor $\norm_{\NNLO}^{\VV,q\bar{q}}$ is given by
\beq
\label{eq:NnnloVV}
\norm_{\NNLO}^{\VV,q\bar{q}}=\norm_{\LO}^{q\bar{q}}\left(\frac{\asmu}{2\pi}\right)^2\cepb^2=\norm_{\NNLO}^{\RR,q\bar{q}}\:\cep^2=\norm_{\NNLO}^{\RV,q\bar{q}}\:\cep.
\eeq
 
After ultraviolet renormalisation, these double virtual contributions contain explicit infrared poles up to order 
four, which originate from the loop integration. There are no implicit poles, as the measurement function 
$J_2^{(2)}$ does not allow any final state particle to be unresolved. As we shall see below, all explicit infrared 
poles in eq.(\ref{eq.setup.sigmannloVV}) are captured and cancelled by those in the subtraction term 
$\ds^{\rU}_{q \bar{q},\NNLO,N_c^2}$.

For the two-loop matrix element in eq.(\ref{eq.setup.sigmannloVV}) we employ the analytic results of 
\cite{Bonciani:2009nb}. The square of the one-loop amplitude was calculated in \cite{Korner:2008bn}. 
However, we independently computed our own expression, written in terms of GPLs. For the numerical 
evaluation of the HPLs in the double virtual amplitudes we use {\tt Chaplin} \cite{Buehler:2011ev}, and for the 
genuine GPLs, the {\tt GiNaC} implementation of \cite{Vollinga:2004sn}. We derived an expansion of both
terms in $\ds_{q \bar{q},\NNLO,N_c^2}^{\VV}$ at the production threshold ($\hat{s}\sim 4m_Q^2$) in order
to circumvent the numerical instabilities that occur when evaluating the GPLs in this limit.

Before proceeding to the construction of the double virtual counter-term, it is worth commenting on the 
purpose of the $x_1$ and $x_2$ dependence in eqs.(\ref{eq.RV}) and (\ref{eq.setup.sigmannloVV}). This 
seemingly trivial dependance introduced in $\ds^{\RV}_{q\bar{q},\NNLO,N_c^2}$ and 
$\ds^{\VV}_{q\bar{q},\NNLO,N_c^2}$ through delta functions, facilitates the combination with the counter-
terms $\ds_{q \bar{q},\NNLO,N_c^2}^{\rT}$ and $\ds_{q\bar{q},\NNLO,N_c^2}^{\rU}$, which contain
integrated antennae and splitting kernels that are non-trivial functions of the momentum fractions 
$x_1$ and $x_2$. In general, three regions can be distinguished: the soft ($x_1=x_2=1$), the collinear 
($x_1=1,\,x_2\neq 1$ and $x_1\neq 1\,x_2=1$) and the hard ($x_1\neq 1\,x_2\neq 1$). The two delta 
functions in the above equations imply that the real-virtual and the virtual-virtual corrections only contribute in 
the soft region. Their respective counter-terms, on the other hand, contribute in all regions.

\subsection{Structure of the virtual-virtual subtraction term $\ds_{q \bar{q},\NNLO,N_c^2}^{\rU}$}
In general, the virtual-virtual subtraction term denoted as $\ds_{\NNLO}^{\rU}$ is built from mass factorisation
counter-terms, as well as integrated double real and real-virtual subtraction terms. In the present case, it is 
given by
\beqa\label{eq.dsuqqb}
\ds^{\rU}_{q\bar{q},\NNLO,N_c^2}&=&
-\int_2\ds^{\rS,b\:4}_{q\bar{q},\NNLO,N_c^2}
-\int_2\ds^{\rS,d}_{q\bar{q},\NNLO,N_c^2}
-\ds^{\MF,2}_{q\bar{q},\NNLO,N_c^2} \nonumber \\
&& -\int_1 \ds^{\VS,a}_{q\bar{q},\NNLO,N_c^2}
-\int_1 \ds^{\VS,b}_{q\bar{q},\NNLO,N_c^2}
-\int_1 \ds^{\VS,d}_{q\bar{q},\NNLO,N_c^2}.
\eeqa

The unintegrated forms of all these terms have been given in \cite{Abelof:2014fza}. 
$\int_2\ds^{\rS,b\:4}_{q\bar{q},\NNLO,N_c^2}$ and $\int_1 \ds^{\VS,a}_{q\bar{q},\NNLO,N_c^2}$ contain the 
new massive initial-final integrated antennae presented in section \ref{sec:integratedantennae}, namely the 
tree-level four-parton antenna $A_{q,Qgg}^0$ and the one-loop antenna $A^{1,lc}_{q,Qg}$. In general,
an additional integrated real-virtual subtraction term, related to a double real subtraction term capturing 
almost-colour-unconnected unresolved limits, and denoted as $\int_1 \ds^{\VS,c}_{\NNLO}$, should be 
present in eq.(\ref{eq.dsuqqb}). Its absence in the present case is due to the fact that at the real-real level
the partonic process contains only two gluons.

Following \cite{Currie:2013vh} we decompose our virtual-virtual counter-term as 
\beq
\label{eq.splitu}
\ds_{q\bar{q},\NNLO,N_c^2}^{\rU}=
\ds_{q\bar{q},\NNLO,N_c^2}^{\rU,a} 
+\ds_{q\bar{q},\NNLO,N_c^2}^{\rU,b}
+\ds_{q\bar{q},\NNLO,N_c^2}^{\rU,c},
\eeq
where, by construction, each term will be free of explicit initial state collinear poles. In order to achieve this,
some rearrangements in the integrated real-virtual subtraction terms 
$\int_1\ds^{\VS,a}_{q\bar{q},\NNLO,N_c^2}$ are necessary.
 
We decompose the real-virtual subtraction term $\ds^{\VS,a}_{q\bar{q},\NNLO,N_c^2}$ in two parts:
\beq
\int_1 \ds_{q\bar{q},\NNLO,N_c^2}^{\VS,a}=
\int_1 \ds_{q\bar{q},\NNLO,N_c^2}^{\VS,a,1} 
+\int_1 \ds_{q\bar{q},\NNLO,N_c^2}^{\VS,a,2}.
\eeq
In the first term, which will be included in $\ds_{q\bar{q},\NNLO,N_c^2}^{\rU,a}$, we group the contributions
of the form ${\cal X}^0_3 |{\cal M}^{1}_{4}|^2$, whereas in the second term, which will be a part of 
$\ds_{q\bar{q},\NNLO,N_c^2}^{\rU,c}$, we put the subtraction terms of the form  
${\cal X}^1_3 |{\cal M}^{0}_{4}|^2$.

The integrated form of the ultraviolet-scale-compensating subtraction term given in eq.(8.22) of 
\cite{Abelof:2014fza} reads
\beqa
\label{eq.dsvsd}
&&\hspace{-0.15in}\int_{1} \ds_{q\bar{q},\NNLO,N_c^2}^{\VS,d}=\norm_{\NNLO}^{\VV,q\bar{q}}\,N_c^2\int \frac{{\rm d}x_1}{x_1} \frac{{\rm d}x_2}{x_2} {\rm d}\Phi_2(p_3,p_4; x_1p_1,x_2p_2)  \nonumber\\
&& \hspace{0.1in}\times \frac{b_0}{\e} \bigg \{ \left [ \left (\frac{|s_{\bar{1} 3}|}{\mu^2}\right)^{-\e} -1 \right ] 
{\cal A}^{0}_{q,Q g}(\e,s_{\bar{1}3},x_1,x_2)+ \left [\left (\frac{|s_{\bar{2} 4}|}{\mu^2}\right)^{-\e} -1 \right ]
{\cal A}^{0}_{q,Q g}(\e,s_{\bar{2}4},x_2,x_1)\bigg \}\nonumber\\
&&\hspace{0.7in}\phantom{\bigg[}
\times|\cm_4(\Q{3},\Qb{4},\hat{\bar{2}}_{\bar{q}},\hat{\bar{1}}_{q})|^2J_2^{(2)}(p_3,p_4), \nonumber\\
\eeqa
where $b_0$ is the $N_c$ coefficient of the QCD beta function at one loop, i.e. $b_0=11/6$. The square
brackets in the second line may be expanded to produce two terms: The terms with the pre-factor 
$(s_{\bar{i}j}/\mu^2)^{-\e}$ will be part of $\ds_{q\bar{q},\NNLO,N_c^2}^{\rU,c}$, while those terms originated
from the factor $-1$ in the square brackets with be part of $\ds_{q\bar{q},\NNLO,N_c^2}^{\rU,a}$.

In order to render each of the three pieces of the virtual-virtual subtraction term in eq.(\ref{eq.splitu}) free of 
initial state collinear singularities, we must split the mass factorisation counter-term 
$\ds_{q\bar{q},\NNLO,N_c^2}^{\MF,2}$ into three parts. This construction is detailed below.

\subsection{The mass factorisation counter-term $\ds_{q\bar{q},\NNLO,N_c^2}^{\MF,2}$}
In general, for a given partonic process initiated by two partons $i$ and $j$ with momenta $p_1$ and $p_2$, 
the mass factorisation counter-term $\ds_{ij,\NNLO}^{\MF,2}$ to be included at NNLO at the virtual-virtual 
level reads \cite{Currie:2013vh,GehrmannDeRidder:2011aa,GehrmannDeRidder:2012dg},
\beqa
\label{eq.mfnnlo2}
&&\hspace{-0.1in}\ds_{ij,\NNLO}^{\MF,2}=
-\left(\frac{\alpha_s}{2 \pi}\right)^2\cepb^2\int\frac{{\rm d}x_1}{x_1}\frac{{\rm d}x_2}{x_2} 
\sum_{k,l} {\bf \Gamma}^{(2)}_{ij,kl}(x_1,x_2) \ds_{kl,\LO}(x_1p_1,x_2p_2)\nonumber\\
&&\hspace{-0.04in}-\left(\frac{\alpha_s}{2 \pi}\right)\cepb\int\frac{{\rm d}x_1}{x_1}\frac{{\rm d}x_2}{x_2}
\sum_{k,l} {\bf \Gamma}^{(1)}_{ij,kl}(x_1,x_2) \Big[ \ds_{kl,\NLO}^{\rV} 
+\ds_{kl,\NLO}^{\MF}+ \int_1\ds_{kl,\NLO}^{\rS} \Big](x_1 p_1,x_2p_2), \nonumber\\
\eeqa
where $\ds_{kl,\NLO}^{\rV}$, $\ds_{kl,\NLO}^{\MF}$ and $\int_1\ds_{kl,\NLO}^{\rS}$ are the NLO virtual
contributions, mass factorisation counter-terms and integrated subtraction terms respectively, and the 
splitting kernels ${\bf \Gamma}^{(1)}_{ij,kl}$ and ${\bf \Gamma}^{(2)}_{ij,kl}$ are given by
\beqa
\label{eq.kernels11}
&&\hspace{-0.2in}{\bf \Gamma}^{(1)}_{ij,kl}(x_1,x_2)=
\delta_{ki}\delta(1-x_1){\bf \Gamma}^{(1)}_{lj}(x_2)
+\delta_{lj}\delta(1-x_2){\bf \Gamma}^{(1)}_{ki}(x_1)\\
&&\hspace{-0.2in}{\bf \Gamma}^{(2)}_{ij,kl}(x_1,x_2)=
\delta_{ki}\delta(1-x_1){\bf \Gamma}^{(2)}_{lj}(x_2)+
\delta_{lj}\delta(1-x_2){\bf \Gamma}^{(2)}_{ki}(x_1)
+{\bf \Gamma}^{(1)}_{ki}(x_1){\bf \Gamma}^{(1)}_{lj}(x_2).
\eeqa
To simplify the construction of the double virtual subtraction term $\ds_{ij,\NNLO}^{\rU}$, 
${\bf \Gamma}^{(2)}_{ij,kl}$ can be written as suggested in \cite{Currie:2013vh} as 
\beq
\label{eq:bargam2}
{\bf \Gamma}_{ij;kl}^{(2)}(x_1,x_2)=
\overline{{\bf \Gamma}}_{ij;kl}^{(2)}(x_1,x_2)
-\frac{\beta_{0}}{\e}{\bf \Gamma}_{ij;kl}^{(1)}(x_1,x_2)
+\frac{1}{2}\big[{\bf \Gamma}_{ij;ab}^{(1)}\otimes{\bf \Gamma}_{ab;kl}^{(1)}\big](x_1,x_2),
\eeq
such that,
\beq
\label{eq.bargamma2}
\overline{{\bf \Gamma}}_{ij;kl}^{(2)}(x_1,x_2)=\overline{{\bf \Gamma}}_{ik}^{(2)}(x_1)\delta_{jl}\delta(1-x_2)+\overline{{\bf \Gamma}}_{jl}^{(2)}(x_2)\delta_{ik}\delta(1-x_1).
\eeq
The reduced kernel $\overline{{\bf \Gamma}}_{ij}^{(2)}(z)$ is related to the usual Altarelli-Parisi splitting functions \cite{Altarelli:1977zs} as
\beqa\label{eq.overlinegamma2}
\overline{{\bf \Gamma}}_{ij}^{(2)}(x)&=&-\frac{1}{2\e}\bigg({\bf P}_{ij}^{1}(x)+\frac{b_{0}}{\e}{\bf P}_{ij}^{0}(x)\bigg).
\eeqa

Following \cite{Currie:2013vh} the double virtual mass factorisation can be split into three terms,
\beq
\dsigma_{ij,\NNLO}^{\MF,2}=\dsigma_{ij,\NNLO}^{\MF,2,a}+\dsigma_{ij,\NNLO}^{\MF,2,b}+\dsigma_{ij,\NNLO}^{\MF,2,c},
\eeq
where the individual contributions are given by:
\beqa
&&\hspace{-0.4in}\dsigma_{ij,\NNLO}^{\MF,2,a}=-\left(\frac{\alpha_s}{2\pi}\right)\cepb\int\frac{{\rm{d}}x_{1}}{x_{1}}\frac{{\rm{d}}x_{2}}{x_{2}}\ {\bf \Gamma}_{ij;kl}^{(1)}\bigg(\dsigma_{kl,\NLO}^{\rV}-\left(\frac{\alpha_s}{2 \pi}\right)\cepb\frac{b_{0}}{\eps}\dsigma_{kl,\LO}\bigg),\label{eq:dsmf2A}\\
&&\hspace{-0.4in}\dsigma_{ij,\NNLO}^{\MF,2,b}=+\left(\frac{\alpha_s}{2 \pi}\right)\cepb\int\frac{{\rm{d}}x_{1}}{x_{1}}\frac{{\rm{d}}x_{2}}{x_{2}}\bigg\{-{\bf \Gamma}_{ij;kl}^{(1)}\
\left(\ds_{kl,\NLO}^{\MF}+ \int_1\ds_{kl,\NLO}^{\rS} \right)
\nonumber \\
&&\hspace{-0.4in}\phantom{\dsigma_{ij,\NNLO}^{\MF,2,b}=} -\left(\frac{\alpha_s}{2 \pi}\right)\cepb\frac{1}{2}\big[{\bf \Gamma}_{ij;ab}^{(1)}\otimes {\bf \Gamma}_{ab;kl}^{(1)}\big]\dsigma_{kl,\LO}\bigg\},\label{eq:dsmf2B}\\
&&\hspace{-0.4in}\dsigma_{ij,\NNLO}^{\MF,2,c}=-\left(\frac{\alpha_s}{2 \pi}\right)^2\cepb^2\int\frac{{\rm{d}}x_{1}}{x_{1}}\frac{{\rm{d}}x_{2}}{x_{2}}\ \overline{{\bf \Gamma}}_{ij;kl}^{(2)}\ \dsigma_{kl,\LO}.\label{eq:dsmf2C}
\eeqa

We apply the generic definitions given above to obtain the mass factorisation counter-terms  
$\dsigma_{q\bar{q},\NNLO,N_c^2}^{\MF,2,x}$ ($x=a,b,c$) for the leading-colour NNLO corrections to 
top pair hadro-production in the quark-antiquark channel. For $\dsigma_{q\bar{q},\NNLO,N_c^2}^{\MF,2,a}$
we find
\beqa
\label{eq.sigmf2a}
&&\hspace{-0.5in}\ds_{q\bar{q},\NNLO,N_c^2}^{\MF,2,a}=- N_c^2\left(\frac{\alpha_s}{2\pi}\right)\cepb\int\frac{{\rm{d}}x_{1}}{x_{1}}\frac{{\rm{d}}x_{2}}{x_{2}}\ \Gamma_{qq;qq}^{(1)}(x_1,x_2)\nonumber \\
&&\hspace{0.7in}\times\bigg(\ds_{q\bar{q},\NLO,N_c}^{\rV}-\left(\frac{\alpha_s}{2 \pi}\right)\cepb\frac{b_{0}}{\e}\dsigma_{q\bar{q},\LO}\bigg)(x_1p_1,x_2 p_2),
\eeqa 
where $\ds_{q\bar{q},\NLO,N_c}^{\rV}$ is the leading-colour NLO virtual cross section, and 
\beq
\label{eq.oneloopkernelgeneral}
\Gamma^{(1)}_{qq,qq}(x_1,x_2)= \delta(1-x_2) \,\Gamma^{(1)}_{qq}(x_1)   +\delta(1-x_1)\,\Gamma^{(1)}_{qq}(x_2), 
\eeq
with $\Gamma^{(1)}_{qq}(x)$ given in eq.(\ref{eq.kernelqq}).

The second part of the mass factorisation counter-term, $\ds_{q\bar{q},\NNLO,N_c^2}^{\MF,2,b}$, reads
\beqa
\label{eq:dsmf2B} 
\dsigma_{q\bar{q},\NNLO,N_c^2}^{\MF,2,b}&=&+\left(\frac{\alpha_s}{2 \pi}\right)\cepb\int\frac{{\rm{d}}x_{1}}{x_{1}}\frac{{\rm{d}}x_{2}}{x_{2}}\bigg\{-\Gamma_{qq;qq}^{(1)}(x_1,x_2)
\left(\ds_{q\bar{q},\NLO,N_c}^{\MF}+ \int_1\ds_{q\bar{q},\NLO,N_c}^{\rS} \right)
\nonumber \\
&& -\left(\frac{\alpha_s}{2 \pi}\right)\cepb \frac{1}{2}\big[\Gamma_{qq;qq}^{(1)}\otimes\Gamma_{qq;qq}^{(1)}\big](x_1,x_2) \dsigma_{q\bar{q},\LO}\bigg\}.\nonumber\\
\eeqa
Using the expressions for $\ds_{q\bar{q},\NLO,N_c}^{\MF}$ and  $\int_1\ds_{q\bar{q},\NLO,N_c}^{\rS}$ given 
in \cite{Abelof:2014fza}, we get 
\beqa
\label{eq:dsmf2B2} 
&&\hspace{-0.15in}\dsigma_{q\bar{q},\NNLO}^{\MF,2,b}=\cepb^2\left(\frac{\alpha_s}{2 \pi}\right)^2\int \frac{{\rm{d}}x_{1}}{x_{1}}\frac{{\rm{d}}x_{2}}{x_{2}}  
 \times \bigg\{\big[\Gamma_{qq;qq}^{(1)}\otimes\Gamma_{qq;qq}^{(1)}\big] (x_1,x_2)
 \nonumber\\ 
 &&-\big[\Gamma_{qq;qq}^{(1)}\otimes {\cal A}^{0}_{q,Qg}(\e,s_{\bar{1}3})\big](x_1,x_2) 
-\big[\Gamma_{qq;qq}^{(1)}\otimes {\cal A}^{0}_{q,Qg}(\e,s_{\bar{2}4})\big](x_2,x_1) 
\bigg\}\dsigma_{q\bar{q},\LO}(x_1p_1,x_2p_2).\nonumber\\
\eeqa
The convolution of two functions $f(x_1,x_2)$ and $g(y_1,y_2)$ is defined as
\beq
\label{eq.conv2}
[f \otimes g](z_1,z_2) \equiv \int {\rm d}x_1 {\rm d}x_2{\rm d}y_1{\rm d}y_2 f(x_1,x_2) g(y_1,y_2) 
\delta(z_1-x_1y_1)\delta(z_2-x_2y_2),
\eeq
in such a way that the convolutions appearing in eq.(\ref{eq:dsmf2B2}) take the following forms
\beqa
&&\hspace{-0.3in}\big[\Gamma_{qq;qq}^{(1)}\otimes\Gamma_{qq;qq}^{(1)}\big](x_1,x_2)=
\delta(1-x_2)\big[\Gamma_{qq}^{(1)}\otimes\Gamma_{qq}^{(1)}\big](x_1)+ \delta(1-x_1)\big[\Gamma_{qq}^{(1)}\otimes\Gamma_{qq}^{(1)}\big](x_2)\nonumber\\
&&\hspace{-0.3in}\phantom{\big[\Gamma_{qq;qq}^{(1)}\otimes\Gamma_{qq;qq}^{(1)}\big](x_1,x_2)}+2 \Gamma_{qq}^{(1)}(x_1) \Gamma_{qq}^{(1)}(x_2) \\ \nonumber\\
&&\hspace{-0.3in}\big[\Gamma_{qq;qq}^{(1)}\otimes {\cal A}^{0}_{q,Qg}(\e,s_{\bar{1}j})\big](x_1,x_2)=
\delta(1-x_2)\big[\Gamma_{qq}^{(1)}\otimes {\cal A}^{0}_{q,Qg}(\e,s_{\bar{1},j})\big](x_1)\nonumber\\
&&\hspace{-0.3in}\phantom{\big[\Gamma_{qq;qq}^{(1)}\otimes {\cal A}^{0}_{q,Qg}(\e,s_{\bar{1}j})\big](x_1,x_2)}+ \Gamma_{qq}^{(1)}(x_2) {\cal A}^{0}_{q,Qg}(\e,s_{\bar{1},j},x_1) \\ \nonumber\\
&&\hspace{-0.3in}\big[\Gamma_{qq;qq}^{(1)}\otimes {\cal A}^{0}_{q,Qg}(\e,s_{\bar{2}j})\big](x_2,x_1)=
\delta(1-x_1)\big[\Gamma_{qq}^{(1)}\otimes {\cal A}^{0}_{q,Qg}(\e,s_{\bar{2},j})\big](x_2) \nonumber \\
&&\hspace{-0.3in}\phantom{\big[\Gamma_{qq;qq}^{(1)}\otimes {\cal A}^{0}_{q,Qg}(\e,s_{\bar{2}j})\big](x_2,x_1)} + \Gamma_{qq}^{(1)}(x_1) {\cal A}^{0}_{q,Qg}(\e,s_{\bar{2},j},x_2)
\eeqa
with
\beq
[f \otimes g] (x)=\int {\rm {d}}z_1{\rm {d}}z_2f(z_1)g(z_2)\delta(x - z_1z_2).
\eeq

The convolution of the form $\big[\Gamma_{qq}^{(1)}   \otimes {\cal A}_{q,Qg}^0(\epsilon,s)\big]$ has not been computed before. It is needed up to finite order and reads 
\beqa
&&\hspace{-0.2in}\big[\Gamma_{qq}^{(1)}   \otimes {\cal A}_{q,Qg}^0(\epsilon,s)\big](x) = s^{-\e}\times\bigg\{ \frac{1}{\e^3}\bigg[\frac{3}{8}\delta (1-x)+\frac{1}{2}\mathcal{D}_0(x)-\frac{1+x}{4} \bigg]\nonumber\\
&&+\frac{1}{\e^2} \bigg[  \left(\frac{x+1}{4} -\frac{3}{8}\delta(1-x)+\frac{1}{2}\Dzero(x)\right)G(1;x_0)-\left( \frac{3(1+x)}{4} -\frac{1}{1-x}\right)G(0;x)\nonumber\\
&&\hspace{0.5in}+(1+x)G(1;x)+\frac{-3x+5}{8} + \left(\frac{3}{8}+\zeta_2\right)\delta (1-x)\nonumber\\
&&\hspace{0.5in}-\frac{1}{4}\Dzero(x)-2\Done(x) \bigg]+\order{\e^{-1}}\bigg\}.
\eeqa
The complete expression can be found in the ancillary {\tt Mathematica} file attached to the arXiv submission
of this paper.

Finally, the mass factorisation counter-term denoted as  $\ds_{q\bar{q},\NNLO,N_c^2}^{\MF,2,c}$ reads
\beq
\label{eq:dsmf2C}
\dsigma_{q\bar{q},\NNLO,N_c^2}^{\MF,2,c}= 
-N_c^2\left(\frac{\alpha_s}{2\pi}\right)^2\cepb^2
\int\frac{{\rm{d}}x_{1}}{x_{1}}\frac{{\rm{d}}x_{2}}{x_{2}}\ \overline{\Gamma}_{qq;qq}^{(2)}(x_1,x_2) \dsigma_{q\bar{q},\LO}(x_1p_1,x_2 p_2),
\eeq
with the leading-colour two-loop kernel $\overline{\Gamma}_{qq;qq}^{(2)}(x_1,x_2)$ defined as in 
eq.(\ref{eq.bargamma2}) and
\beqa
&&\hspace{-0.4in}\overline{\Gamma}^{(2)}_{qq}(x)=\frac{1}{2\e}\Gamma_{qq}^{(1)}(x)\nonumber\\
&&\hspace{-0.4in}\phantom{\overline{\Gamma}^{(2)}_{qq}(x)}-\frac{1}{\e}\bigg[
\bigg( \frac{67}{72} -\frac{1}{24} \pi^2 +\frac{13}{48} \ln(x)-\frac{1}{4}\ln(x)\,\ln(1-x)+\frac{1}{8}\,\ln^2(x) \bigg) p_{qq}(x) \nonumber\\
&&\hspace{-0.4in}\phantom{\overline{\Gamma}^{(2)}_{qq}(x)}+ \frac{25}{24}\, (1-x) +\frac{1}{16} \Big (1-3 x -(1+x)\ln(x)\Big) \ln(x)+\Big(\frac{43}{192} +\frac{13 \pi^2}{144}\Big) \delta(1-x)\bigg].\nonumber\\
\eeqa
As usual, $p_{qq}$ is given by
\beq
p_{qq}(x)=\frac{2}{1-x} -1 -x.
\eeq

\subsection{The subtraction term $\ds_{q\bar{q},\NNLO,N_c^2}^{\rU,a}$}
\label{sec:virtualvirtualNc2}
The virtual-virtual subtraction term $\ds_{q\bar{q},\NNLO,N_c^2}^{\rU,a}$ is given by
\beqa
\label{eq.dsua}
&&\ds^{\rU,a}_{q\bar{q},\NNLO,N_c^2}= -\int_1 \ds^{\VS,a,1}_{q\bar{q},\NNLO,N_c^2}
-\int_1 \ds^{\VS,d,1}_{q\bar{q},\NNLO,N_c^2} -\ds^{\MF,2,a}_{q\bar{q},\NNLO,N_c^2}
\nonumber \\
&&\phantom{\ds^{\rU,a}_{q\bar{q},\NNLO,N_c^2}}= -{\cal N}_{\NNLO}^{\VV,q\bar{q}}\,N_{c}^2
\int \frac{{\rm d}x_1}{x_1} \frac{{\rm d}x_2}{x_2} {\rm d}\Phi_2(p_3,p_4; x_1p_1,x_2p_2) 
 \nonumber \\
&&\hspace{1.2in} \times \bigg[\bigg ({\cal A}^{0}_{q,Q g}(\e,s_{\bar{1}3},x_1,x_2) -\GG^{(1)}_{qq}(x_1)
\delta(1-x_2)\bigg) \nonumber\\
&&\hspace{1.275in} +  \bigg( {\cal A}^{0}_{q,Q g}(\e,s_{\bar{2}4},x_2,x_1) -\GG^{(1)}_{qq}(x_2)\delta(1-x_1)\bigg)\bigg] \nonumber\\
&&\hspace{1.2in} \times \bigg\{
|\cmb_4^{1,[lc]}(\Q{3},\Qb{4},\hat{\bar{2}}_{\bar{q}},\hat{\bar{1}}_{q})|^2
-\frac{b_0}{\e}\,|\cm_4(\Q{3},\Qb{4},\hat{\bar{2}}_{\bar{q}},\hat{\bar{1}}_{q})|^2  \bigg \} 
J_2^{(2)}(p_3,p_4) \nonumber\\
\eeqa
where $\cmb_4^{1,[lc]}(\Q{3},\Qb{4},\hat{\bar{2}}_{\bar{q}},\hat{\bar{1}}_{q})$ is a shorthand notation for the interference of 
the one-loop leading-colour primitive amplitude 
$\cmb_4^{1,[lc]}(\Q{3},\Qb{4},\hat{\bar{2}}_{\bar{q}},\hat{\bar{1}}_{q})$ and its tree-level counterpart. The content 
of the square bracket is manifestly free of initial state collinear divergences as is the whole virtual-virtual 
subtraction term $\ds^{\rU,a}_{q\bar{q},\NNLO,N_c^2}$. 

In analogy to the massless case \cite{Currie:2013vh}, and using a similar notation, we can reformulate 
eq.(\ref{eq.dsua}) in terms of integrated dipoles. Indeed, we can define the integrated massive NLO dipoles 
as:
\beqa
\label{eq.J21a}
&&\hspace{-0.2in}{\bf J}_{2}^{1}(\bar{1},J_Q)= {\cal A}^{0}_{q,Q g}(\e,s_{\bar{1}J_Q},x_1,x_2)
-\Gamma^{(1)}_{qq}(x_1)\delta(1-x_2)\\
\label{eq.J21b}
&&\hspace{-0.2in}{\bf J}_{2}^{1}(\bar{2},J_{\bar{Q}})= {\cal A}^{0}_{q,Q g}(\e,s_{\bar{2} J_{\bar{Q}}},x_2,x_1)
-\Gamma^{(1)}_{qq}(x_2)\delta(1-x_1).
\eeqa
Using the expression in eq.(\ref{eq.poleA03}) for the pole part of ${\cal A}^{0}_{q,Q g}$ one can immediately 
see that these massive integrated dipoles of the form ${\bf J}_{2}^{1}$ are directly related to the massive 
colour-ordered infrared singularity operator $\ione{Q}{\bar{q}}$. 
 
Furthermore, the integrated subtraction term $\ds^{\rU,a}_{q\bar{q},\NNLO,N_c^2}$ may be rewritten as 
\beqa
&&\hspace{-0.15in}\ds^{\rU,a}_{q\bar{q},\NNLO,N_c^2}=-{\cal N}_{\NNLO}^{\VV,q\bar{q}}\,N_{c}^2
\int \frac{{\rm d}x_1}{x_1} \frac{{\rm d}x_2}{x_2} {\rm d}\Phi_2(p_3,p_4; x_1p_1,x_2p_2) \Big ({\bf J}_{2}^{1} (\bar{1},3_Q)+{\bf J}_{2}^{1} (\bar{2},4_{\bar{Q}}) \Big) \nonumber \\
&&\hspace{1.3in} \times
\left \{ |\cmb_4^{1,[lc]}(\Q{3},\Qb{4},\hat{\bar{2}}_{\bar{q}},\hat{\bar{1}}_{q})|^2
-\frac{b_0}{\e}\,|\cm_4(\Q{3},\Qb{4},\hat{\bar{2}}_{\bar{q}},\hat{\bar{1}}_{q})|^2  \right \} 
J_2^{(2)}(p_3,p_4). \nonumber\\
\eeqa
In general, for an $n$-jet hadron collider observable computed at NNLO, $\ds^{\rU,a}_{\NNLO}$ contains 
integrated dipoles of three types: final-final, initial-final and initial initial. In our case, for the leading colour contributions, 
the initial-final dipoles defined in eqs.(\ref{eq.J21a}) and (\ref{eq.J21b}) are sufficient.

\subsection{The subtraction term $\ds_{q\bar{q},\NNLO,N_c^2}^{\rU,b}$}
For the virtual-virtual subtraction term $\ds^{\rU,b}_{q\bar{q},\NNLO,N_c^2}$ we obtain:
 \beqa 
\label{eq.dsub}
&&\hspace{-0.1in} \ds^{\rU,b}_{q\bar{q},\NNLO,N_c^2}= -\int_2 \ds^{\rS,d}_{q\bar{q},\NNLO,N_c^2}
-\int_1 \ds^{\VS,b,1}_{q\bar{q},\NNLO,N_c^2} -\ds^{\MF,2,b}_{q\bar{q},\NNLO,N_c^2}
\nonumber \\
&&\hspace{-0.1in}\phantom{\ds^{\rU,b}_{q\bar{q},\NNLO,N_c^2}}= -{\cal N}_{\NNLO}^{\VV,q\bar{q}}\,N_{c}^2
\int \frac{{\rm d}x_1}{x_1} \frac{{\rm d}x_2}{x_2} {\rm d}\Phi_2(p_3,p_4; x_1p_1,x_2p_2)\, \bigg\{ \big[\Gamma_{qq;qq}^{(1)}\otimes\Gamma_{qq;qq}^{(1)}\big] (x_1,x_2)
 \nonumber \\
&&\hspace{0.15in}\phantom{\ds^{\rU,b}_{q\bar{q},\NNLO,N_c^2}} 
-\big[\Gamma_{qq;qq}^{(1)}\otimes {\cal A}^{0}_{q,Qg}(\e,s_{\bar{2}4})\big](x_2,x_1)-\big[\Gamma_{qq;qq}^{(1)}\otimes {\cal A}^{0}_{q,Qg}(\e,s_{\bar{1}3})\big](x_1,x_2)
\nonumber \\
&&\hspace{0.15in}\phantom{\ds^{\rU,b}_{q\bar{q},\NNLO,N_c^2}}   
 + \big[ {\cal A}^{0}_{q,Qg}(\e,s_{\bar{1}3})\otimes
{\cal A}^{0}_{q,Qg}(\e,s_{\bar{2}4}) \big](x_1,x_2)  \bigg\} 
|\cm_4(\Q{3},\Qb{4},\hat{\bar{2}}_{\bar{q}},\hat{\bar{1}}_{q})|^2 J_2^{(2)}(p_3,p_4). \nonumber\\
\eeqa
Using eqs.(\ref{eq.A03x1x2a}) and (\ref{eq.A03x1x2b}) it can be noticed that the convolution of two antennae
\linebreak
$\big[ {\cal A}^{0}_{q,Qg}(\e,s_{\bar{1}3})\otimes{\cal A}^{0}_{q,Qg}(\e,s_{\bar{2}4}) \big](x_1,x_2)$ is in fact 
only a product. Furthermore, it can be seen that $\ds^{\rU,b}_{q\bar{q},\NNLO,N_c^2}$ is not free of 
explicit initial state collinear poles. In order to remedy this, we define
\beq
 \ds^{\rU,\tilde{b}}_{q\bar{q},\NNLO,N_c^2}=\ds^{\rU,b}_{q\bar{q},\NNLO,N_c^2}-\delta\ds^{\rU}_{q\bar{q},\NNLO,N_c^2}
\eeq
with
\beqa
&&\hspace{-0.2in}\delta\ds^{\rU}_{q\bar{q},\NNLO,N_c^2}={\cal N}_{\NNLO}^{\VV,q\bar{q}}\,N_{c}^2
\int \frac{{\rm d}x_1}{x_1} \frac{{\rm d}x_2}{x_2} {\rm d}\Phi_2(p_3,p_4; x_1p_1,x_2p_2)\nonumber \\
&& \times \frac{1}{2} \left \{ \Big[{\cal A}^{0}_{q,Qg}(\e,s_{\bar{1}3})\otimes
{\cal A}^{0}_{q,Qg}(\e,s_{\bar{1}3}) \Big](x_1,x_2)  
+ \Big[{\cal A}^{0}_{q,Qg}(\e,s_{\bar{2}4})\otimes
{\cal A}^{0}_{q,Qg}(\e,s_{\bar{2}4}) \Big](x_2,x_1)
\right \}\nonumber\\
&&\hspace{0.4in}\phantom{\bigg[}\times |\cm_4(\Q{3},\Qb{4},\hat{\bar{2}}_{\bar{q}},\hat{\bar{1}}_{q})|^2 J_2^{(2)}(p_3,p_4). \nonumber\\
\eeqa
After this modification, $\ds^{\rU,\tilde{b}}_{q\bar{q},\NNLO,N_c^2}$ can be written as a convolution of 
integrated massive dipoles, with no infrared collinear singularities left: 
\beqa 
\lefteqn{\ds^{\rU,\tilde{b}}_{q\bar{q},\NNLO,N_c^2}=
- {\cal N}_{\NNLO}^{\VV,q\bar{q}}\,N_{c}^2
\int \frac{{\rm d}x_1}{x_1} \frac{{\rm d}x_2}{x_2} {\rm d}\Phi_2(p_3,p_4; x_1p_1,x_2p_2)} \nonumber \\
&&\times\frac{1}{2} \left \{ 
\Big [{\bf J}_{2}^{1} (\bar{1},3_Q) + {\bf J}_{2}^{1} (\bar{2},4_{\bar{Q}})\Big]
\otimes \Big[{\bf J}_{2}^{1} (\bar{1},3_Q) +{\bf J}_{2}^{1} (\bar{2},4_{\bar{Q}}) \Big] \right\}|\cm_4(\Q{3},\Qb{4},\hat{\bar{2}}_{\bar{q}},\hat{\bar{1}}_{q})|^2  J_2^{(2)}(p_3,p_4).
\nonumber\\
\eeqa 
It is worth noting that, in general, the terms that we added to $\ds^{\rU,b}_{q\bar{q},\NNLO,N_c^2}$ 
in order to construct $\ds^{\rU,\tilde{b}}_{q\bar{q},\NNLO,N_c^2}$ are provided by the integrated subtraction 
term of the form $\int_{1} \ds^{\VS,c}_{\NNLO}$. When those terms are present, which is not the case for our 
computation as discussed previously, $\ds^{\rU,b}_{\NNLO}$ is naturally written as a convolution of a sum of 
integrated dipoles without any further additions required. 

Finally, we remark that for the sum of $\ds^{\rU,b}_{q\bar{q},\NNLO,N_c^2}$ and $\ds^{\rU,c}_{q\bar{q},\NNLO,N_c^2}$ to 
remain unchanged, $\ds^{\rU,c}_{q\bar{q},\NNLO,N_c^2}$ will be redefined as we will see below.

 \subsection{The subtraction term $\ds^{\rU,c}_{q\bar{q},\NNLO,N_c^2}$}
For the integrated virtual-virtual subtraction term $\ds^{\rU,c}_{q\bar{q},\NNLO,N_c^2}$ we have:
\beqa
 \ds^{\rU,c}_{q\bar{q},\NNLO,N_c^2}&=&
-\int_2 \ds^{\rS,b,4}_{q\bar{q},\NNLO,N_c^2}
-\int_1 \ds^{\VS,a,2}_{q\bar{q},\NNLO,N_c^2}
-\int_1 \ds^{\VS,d,2}_{q\bar{q},\NNLO,N_c^2} 
-\ds^{\MF,2,c}_{q\bar{q},\NNLO,N_c^2}
\nonumber \\
&=& -{\cal N}_{\NNLO}^{\VV,q\bar{q}}\,N_c^2
\int \frac{{\rm d}x_1}{x_1} \frac{{\rm d}x_2}{x_2} {\rm d}\Phi_2(p_3,p_4; x_1p_1,x_2p_2) 
 \nonumber \\
 && \times 
 \bigg\{ {\cal A}^{0}_{q,Qgg}(\e,s_{\bar{1}3},x_1,x_2) + {\cal A}^{0}_{q,Qgg}(\e,s_{\bar{2}4},x_2,x_1)
\nonumber \\
 &&\hspace{0.12in}+ {\cal A}^{1,lc}_{q,Qg}(\e,s_{\bar{1}3},x_1,x_2) + {\cal A}^{1,lc}_{q,Qg}(\e,s_{\bar{2}4},x_2,x_1)
 \nonumber \\
 &&\hspace{0.12in} + \frac{b_0}{\e}\, \Big[
 \left( \frac{s_{\bar{1}3}}{\mu^2} \right)^{-\e}  {\cal A}^{0}_{q,Qg}(\e,s_{\bar{1}3},x_1,x_2)
 +\left( \frac{s_{\bar{2}3}}{\mu^2} \right)^{-\e}  {\cal A}^{0}_{q,Qg}(\e,s_{\bar{2}4},x_2,x_1)\Big]
 \nonumber \\
 &&\hspace{0.12in}-\overline{\Gamma}_{qq,qq}^{(2)}(x_1,x_2) \bigg \}|\cm_4(\Q{3},\Qb{4},\hat{\bar{2}}_{\bar{q}},\hat{\bar{1}}_{q})|^2 \, J_2^{(2)}(p_3,p_4).\nonumber\\
\eeqa
 
The expression in the above equation is not free of initial state collinear poles. It becomes so, however, once 
we add back the contributions subtracted from $\ds^{\rU,b}_{q\bar{q},\NNLO,N_c^2}$ in order to render it free from
initial state collinear singularities. Indeed we find that
\beq
 \ds^{\rU,\tilde{c}}_{q\bar{q},\NNLO,N_c^2}=\ds^{\rU,c}_{q\bar{q},\NNLO,N_c^2}+\delta\ds^{\rU}_{q\bar{q},\NNLO,N_c^2}
\eeq
has a similar structure than the one found for $\ds^{\rU,c}_{\NNLO}$ in the massless case, and allows for the 
definition of new initial-final massive integrated dipoles:
 \beqa
&&\hspace{-0.2in}{\bf J}_{2}^{2} (\bar{1},3_Q)= {\cal A}^{0}_{q,Qgg}(\e,s_{\bar{1}3},x_1,x_2)
  +{\cal A}^{1,lc}_{q,Qg}(\e,s_{\bar{1}3},x_1,x_2) 
 +\frac{b_0}{\e} \left( \frac{s_{\bar{1}3}}{\mu^2} \right)^{-\e}
 \nonumber\\
&&\hspace{-0.2in}\phantom{{\bf J}_{2}^{2} (\bar{1},3_Q)} - \overline{\Gamma}_{qq}^{(2)}(x_1)\delta(1-x_2) 
-\frac{1}{2}\Big[{\cal A}^{0}_{q,Qg}(\e,s_{\bar{1}3})\otimes
{\cal A}^{0}_{q,Qg}(\e,s_{\bar{1}3}) \Big](x_1,x_2) \\ \nonumber\\
&&\hspace{-0.2in}{\bf J}_{2}^{2} (\bar{2},4_{\bar{Q}})= {\cal A}^{0}_{q,Qgg}(\e,s_{\bar{2}4},x_2,x_1)
 +{\cal A}^{1,lc}_{q,Qg}(\e,s_{\bar{2}4},x_2,x_1) 
 +\frac{b_0}{\e} \left( \frac{s_{\bar{2}4}}{\mu^2}\right )^{-\e}
  \nonumber\\
 &&\hspace{-0.2in}\phantom{{\bf J}_{2}^{2} (\bar{2},4_{\bar{Q}})}- \overline{\Gamma}_{qq}^{(2)}(x_2)\delta(1-x_1)
  -\frac{1}{2} \Big[{\cal A}^{0}_{q,Qg}(\e,s_{\bar{2}4})\otimes
{\cal A}^{0}_{q,Qg}(\e,s_{\bar{2}4}) \Big](x_2,x_1). 
\eeqa 
In terms of these massive integrated dipoles we have
\beqa
&&\hspace{-0.3in}\ds^{\rU,\tilde{c}}_{q\bar{q},\NNLO,N_c^2}=
-{\cal N}_{\NNLO}^{\VV,q\bar{q}}\,N_{c}^2
\int \frac{{\rm d}x_1}{x_1} \frac{{\rm d}x_2}{x_2} {\rm d}\Phi_2(p_3,p_4; x_1p_1,x_2p_2) 
\nonumber \\
 &&\hspace{0.4in} \times \Big ({\bf J}_{2}^{2} (\bar{1},3_Q)+ {\bf J}_{2}^{2} (\bar{2},4_{\bar{Q}})\Big)|\cm_4(\Q{3},\Qb{4},\hat{\bar{2}}_{\bar{q}},\hat{\bar{1}}_{q})|^2 \, J_2^{(2)}(p_3,p_4).
\eeqa

As in the massless case, one can see that the integrated dipoles present here are related to integrated 
subtraction terms which involve genuine NNLO objects. This feature can be regarded as an additional check 
on the correctness of the construction of the subtraction terms at real-real, real-virtual and virtual-virtual levels 
in this NNLO computation.   

\subsection{Explicit pole cancellation}
With the explicit expressions included above one can show analytically that each of the virtual-virtual 
subtraction terms  $\ds_{q\bar{q},\NNLO,N_c^2}^{\rU,a}$, $\ds_{q\bar{q},\NNLO,N_c^2}^{\rU,\tilde{b}}$ and 
$\ds_{q\bar{q},\NNLO,N_c^2}^{\rU,\tilde{c}}$ is free of explicit initial state collinear poles and that the sum of those building blocks is such that
\beq\label{eq.poles2}
\poles\left(\ds_{q\bar{q},\NNLO,N_c^2}^{\VV}-\ds_{q\bar{q},\NNLO,N_c^2}^{\rU}\  \right)=0.
\eeq
This is an ultimate check that demonstrates that we have correctly implemented the subtractions terms at 
real-real, real-virtual and virtual-virtual levels for the leading-colour contributions to 
$q \bar{q} \to Q \bar{Q}+X$ at NNLO.

%
 \section{Heavy quark contributions}
\label{sec:heavy}
In this section we present the heavy quark contributions to the top pair production cross section in the 
quark-antiquark channel computed at ${\cal O}(\alpha_s^4)$. They are proportional to the 
colour factors $N_h N_c$ (leading-colour) and $N_h/N_c$ (subleading-colour), and correspond to the
coefficients $F_h$ and $G_h$ in eq.(\ref{eq.qqbdec}). 

Both $F_h$ and $G_h$ receive contributions in the two and three-parton final states, and require, 
in both final states, subtraction terms to cancel explicit and implicit infrared divergences. For all loop
matrix elements involved we perform the ultraviolet renormalisation in the so-called decoupling scheme,
with the gluon wave functions and the heavy quark mass and wave functions renormalised on-shell, and
the strong coupling $\alpha_s$ renormalised in the $\overline{\rm MS}$ scheme with $N_l$ active flavours.
The two-loop amplitudes employed in our calculation \cite{Bonciani:2009nb}, were originally computed with
the strong coupling renormalized in the $\overline{\rm MS}$ scheme with $N_F=N_h +N_l$ active flavours,
and therefore required a conversion to the decoupling scheme.

In this section we will explicitly construct the two and three-parton contributions to the $N_h N_c$ and 
$N_h/N_c$ colour factors, and we will show how the amplitudes computed in \cite{Bonciani:2009nb} in the
full-flavour scheme can be converted to the decoupling scheme via a finite renormalization of $\alpha_s$.

 \subsection{Real-virtual contributions}
In the antenna subtraction framework, the three-parton heavy-quark contribution to top pair production at
NNLO in the $q\bar{q}$ channel is given by
\beq\label{eq.3pcNh}
\int_{\dphi_3}\Big[\ds_{q \bar{q},\NNLO,N_h}^{\RV}-\ds_{q \bar{q},\NNLO,N_h}^{\rT}\Big]. 
\eeq
Following the colour decomposition of the real-virtual matrix-element as presented in \cite{Abelof:2014jna}
$\ds_{q \bar{q},\NNLO,N_h}^{\RV}$ can be written as 
\beqa
\label{eq.qqbNhRV}
&&\hspace{-0.3in}\ds_{q\bar{q},\NNLO,N_h}^{\RV}=\norm_{\NNLO}^{q\bar{q},\RV}\:N_h \int\frac{{\rm d}x_1}{x_1}\frac{{\rm d}x_2}{x_2}\dphi_3(p_3,p_4,p_5;x_1p_1,x_2p_2)\delta(1-x_1)\delta(1-x_2)\nonumber\\
&&\times\bigg\{ N_c\bigg[|\cmb^{1,[h]}_{5}(\Q{3},\gl{5},\qi{1};;\qbi{2},\Qb{4})|^2+|\cmb^{1,[h]}_{5}(\Q{3},\qi{1};;\qbi{2},\gl{5},\Qb{4})|^2\bigg]\nonumber \\
&&\hspace{0.075in}+\frac{1}{N_c}\bigg[|\cmb^{1,[h]}_{5}(\Q{3},\gl{5},\Qb{4};;\qbi{2},\qi{1})|^2+|\cmb^{1,[h]}_{5}(\Q{3},\Qb{4};;\qbi{2},\gl{5},\qi{1})|^2\nonumber\\
&&\hspace{0.5in}-2|\cmb^{1,[h]}_{5}(\Q{3},\Qb{4},\qbi{2},\qi{1},\ph{5})|^2\bigg]\bigg\}J_2^{(3)}(p_3,p_4,p_5),
\eeqa
where we have used the following shorthand notation: 
\beq
|\cmb^{1,[h]}_{5}(\ldots)|^2=2\re(\cmb^{1,[h]}_{5}(\ldots)\cm_5(\ldots)^{*})
\eeq
with $\cm_5$ and $\cmb^{1,[h]}_{5}$ being respectively the tree-level and the $N_h$ coefficient of the one-loop 
amplitude associated to the process $q \bar{q} \to t \bar{t} g$. In addition, in eq.(\ref{eq.qqbNhRV}) we have 
defined the one-loop amplitude $\cmb_{5}^{1,[h]}(\Q{3},\Qb{4},\qbi{2},\qi{1},\ph{5})$ as
\beqa
&&\hspace{-0.4in}\cmb_{5}^{1,[h]}(\Q{3},\Qb{4},\qbi{2},\qi{1},\ph{5})=\cmb_{5}^{1,[h]}(\Q{3},\gl{5},\qi{1};;\qbi{2},\Qb{4})+\cmb_{5}^{1,[h]}(\Q{3},\qi{1};;\qbi{2},\gl{5},\Qb{4})\nonumber\\
&&\hspace{-0.4in}\phantom{\cmb_{5}^{1,[h]}(\Q{3},\Qb{4},\qbi{2},\qi{1},\ph{5})}=\cmb_{5}^{1,[h]}(\Q{3},\gl{5},\Qb{4};;\qbi{2},\qi{1})+\cmb_{5}^{1,[h]}(\Q{3},\Qb{4};;\qbi{2},\gl{5},\qi{1}),
\eeqa
where the gluon is photon-like. 

After UV renormalisation in the decoupling scheme $\ds^{\RV}_{q\bar{q},\NNLO,N_h}$ is free of explicit 
$\e$-poles. It contains, however, infrared implicit poles: The phase space integral is divergent due to 
integration over the soft limit $p_5\to 0$ and the collinear limits $p_1||p_5$ and $p_2||p_5$, where the matrix 
elements are singular. The subtraction term $\ds_{q \bar{q},\NNLO,N_h}^{\rT}$ in eq.(\ref{eq.3pcNh}) 
regularises these implicit divergences of the phase space integral. We shall present it below.

\subsection{Real virtual subtraction term $\ds_{q \bar{q},\NNLO,N_h}^{\rT}$}
Both the leading-colour ($N_hN_c$) and subleading-colour ($N_h/N_c$) real-virtual subtraction terms only
contain terms commonly known in the antenna subtraction literature as $\ds^{\VS,a (1)}$. They are built as 
products of tree-level antennae and reduced one-loop matrix elements squared. Unlike in the most general
case, there are no integrated double real subtraction terms, no mass factorisation counter terms $\ds^{\MF,1}$,
nor subtraction terms involving one-loop antennae. The absence of one-loop antennae in 
$\ds_{q \bar{q},\NNLO,N_h}^{\rT}$ is due to the fact that the $N_h$ coefficient of the antennae $A_3^1$,
which would be in principle be required, vanishes in the decoupling scheme. It is also worth mentioning
that in the decoupling scheme the reduced one-loop matrix element needed in 
$\ds_{q \bar{q},\NNLO,N_h}^{\rT}$, i.e. the $N_h$ part of the one-loop amplitude for the process 
$ q \bar{q} \to t \bar{t}$, is finite.
 
Explicitly, we find, 
\beqa
\label{eq.dsqqbvsanh}
&&\hspace{-0.4in}\ds^{\rT}_{q\bar{q},\NNLO,N_h} = \norm_{\NNLO}^{q\bar{q},\RV}\,N_h\int\frac{{\rm d}x_1}{x_1}\frac{{\rm d}x_2}{x_2}\,\dphi_3(p_3,p_4,p_5; x_1p_1,x_2p_2)\delta(1-x_1)\delta(1-x_2)\nonumber\\
&&\hspace{0.2in}\times\bigg\{ N_c\bigg[ A_3^0(\Q{3},\gl{5},\qi{\bar{1}})|\cmb_{4}^{1,[h]}(\Q{(\wt{35})},\Qb{4},\qbi{\bar{2}},\qi{\bar{\bar{1}}})|^2 J_2^{(2)}(p_{\wt{35}},p_4)\nonumber\\
&&\hspace{0.56in}+A_3^0(\Qb{4},\gl{5},\qbi{\bar{2}})|\cmb_{4}^{1,[h]}(\Q{3},\Qb{(\wt{45})},\qbi{\bar{\bar{2}}},\qi{\bar{1}})|^2 J_2^{(2)}(p_3,p_{\wt{45}})\phantom{\bigg[}\nonumber\\
&&\hspace{0.25in}+\frac{1}{N_c}\bigg[ 2A_3^0(\Q{3},\gl{5},\qbi{\bar{2}})|\cmb_{4}^{1,[h]}(\Q{(\wt{35})},\Qb{4},\qbi{\bar{\bar{2}}},\qi{\bar{1}})|^2 J_2^{(2)}(p_{\wt{35}},p_4)\phantom{\bigg[}\nonumber\\
&&\hspace{0.56in}+2A_3^0(\Qb{4},\gl{5},\qi{\bar{1}})|\cmb_{4}^{[1,h]}(\Q{3},\Qb{(\wt{45})},\qbi{\bar{2}},\qi{\bar{\bar{1}}})|^2 J_2^{(2)}(p_3,p_{\wt{45}})\phantom{\bigg[}\nonumber\\
&&\hspace{0.56in}-2A_3^0(\Q{3},\gl{5},\qi{\bar{1}})|\cmb_{4}^{1,[h]}(\Q{(\wt{35})},\Qb{4},\qbi{\bar{2}},\qi{\bar{\bar{1}}})|^2 J_2^{(2)}(p_{\wt{35}},p_4)\phantom{\bigg[}\nonumber\\
&&\hspace{0.56in}-2A_3^0(\Qb{4},\gl{5},\qbi{\bar{2}})|\cmb_{4}^{1,[h]}(\Q{3},\Qb{(\wt{45})},\qbi{\bar{\bar{2}}},\qi{\bar{1}})|^2 J_2^{(2)}(p_3,p_{\wt{45}})\phantom{\bigg[}\nonumber\\
&&\hspace{0.56in}-A_3^0(\Q{3},\gl{5},\Qb{4})|\cmb_{4}^{1,[h]}(\Q{(\wt{35})},\Qb{(\wt{45})},\qbi{\bar{2}},\qi{\bar{1}})|^2 J_2^{(2)}(p_{\wt{35}},p_{\wt{45}})\phantom{\bigg[}\nonumber\\
&&\hspace{0.56in}-A_3^0(\qbi{\bar{2}},\gl{5},\qi{\bar{1}})|\cmb_{4}^{1,[h]}(\Q{\tilde{3}},\Qb{\tilde{4}},\qbi{\bar{\bar{2}}},\qi{\bar{\bar{1}}})|^2 J_2^{(2)}(\wt{p}_3,\wt{p}_4)\phantom{\bigg[}
\bigg]\bigg\}. 
\eeqa
Since, as we mentioned above, $|\cmb_{4}^{1,[h]}(\Q{3},\Qb{4},\qbi{\bar{2}},\qi{\bar{1}})|^2$ has no $\e$-poles, the 
subtraction term in eq.(\ref{eq.dsqqbvsanh}) does not contain any explicit infrared singularities.

Being a pure subtraction term, $\ds^{\rT}_{q\bar{q},\NNLO,N_h}$ must be added back in integrated form to the 
cross section at the two-parton level. The required A-type integrated massive and massless antennae are all known
\cite{GehrmannDeRidder:2009fz,Abelof:2011jv}.

%
\subsection{Double-virtual contributions $\ds^{\VV}_{q \bar{q},\NNLO,N_h}$}
In this section we present the heavy quark NNLO two-parton contributions to $q\bar{q} \to t\bar{t}$ in the 
decoupling scheme, which are given by
\beq
\int_{{\rm{d}}\Phi_{2\phantom{+1}}}\left[\ds_{q\bar{q},\NNLO,N_h}^{\VV}-\ds_{q \bar{q},\NNLO,N_h}^{\rU}\right].
\eeq
We shall focus in particular on the construction of the double-virtual counterterm \linebreak
$\ds_{q \bar{q},\NNLO,N_h}^{\rU}$, split into a leading-colour ($N_hN_c$) and a subleading-colour 
($N_h/Nc$) part.

The contributions from the double-virtual matrix elements can be written as 
\beqa\label{eq:nnlovv}
\lefteqn{\ds^{\VV}_{q \bar{q},\NNLO,N_h}= \norm_{\NNLO}^{\VV,q\bar{q}}N_h \int\frac{{\rm d}x_1}{x_1}\frac{{\rm d}x_2}{x_2}\dphi_{2}(p_3,\ldots,p_{4};x_1p_1,x_2p_2)}\nonumber\\
&&\hspace{0.2in}\times\delta(1-x_1)\delta(1-x_2)|{\cal M}^2_{4}(\Q{3},\Qb{4},\qbi{2},\qi{1})|^2\;J_{2}^{(2)}(p_3,p_{4})\;,
\eeqa
with $\norm_{\NNLO}^{\VV,q\bar{q}}$ given in eq.(\ref{eq:NnnloVV}), and with the abbreviation
\beqa\label{eq:m2def}
|{\cal M}^2_{4}(\Q{3},\Qb{4},\qbi{2},\qi{1})|^2=
\bigg[ 2\re \left({\cal M}^2_{q_1\bar{q}_2\rightarrow t_3 \bar{t}_4} 
{\cal M}^{0\,\dagger}_{q_1\bar{q}_2\rightarrow t_3 \bar{t}_4}\right)
+|{\cal M}^1_{q_1\bar{q}_2\rightarrow t_3 \bar{t}_4}|^2\bigg]\Bigg|_{N_h}.
\eeqa
For the $N_h$ part of the two-loop matrix element in eq.(\ref{eq:m2def}) we employ the analytic 
results of \cite{Bonciani:2008az} converted to the decoupling scheme via a finite renormalisation 
of $\alpha_s$ that will be described below. The ``one-loop squared'' term has been computed analytically 
in \cite{Korner:2008bn}. We re-derived it ourselves, also analytically, and use our own result in our event 
generator. We further compared these results with those provided by Roberto Bonciani in both 
renormalisation schemes and found full agreement.

After UV-renormalisation, $\ds^{\VV}_{q \bar{q},\NNLO,N_h}$ contains explicit infrared poles, that are 
cancelled by the virtual-virtual subtraction term $\ds^{\rU}_{q \bar{q},\NNLO,N_h}$ derived below.

%
\subsection{Virtual-virtual subtraction term $\ds^{\rU}_{q \bar{q},\NNLO,N_h}$}
Following the decomposition of the mass factorisation counter term $\ds^{\MF,2}$ and $\ds^{\rU}$ explained 
in section \ref{sec:virtualvirtualNc}, we find that $\ds^{\rU}_{q \bar{q},\NNLO,N_h}$ only contains terms 
denoted there as $\ds^{\rU,a}$, involving only the mass factorisation counterterm 
$\ds^{\MF,2,a}_{q\bar{q},NNLO,N_h}$ and the integrated form of the real-virtual counter term derived in 
eq.(\ref{eq.dsqqbvsanh}).

The mass factorisation counterterm is given by
\beq
\ds^{\MF,2,a}_{q\bar{q},\NNLO,N_h}\hspace{-0.03in}=- \cepb \left(\frac{\alpha_s}{2\pi}\right)N_h\left( \frac{N_c^2-1}{N_c}\right)\int \frac{{\rm d}x_1}{x_1} \frac{{\rm d}x_2}{x_2}\Gamma_{qq;qq}^{(1)}(x_1,x_2)\ds_{q\bar{q},\NLO,N_h}^{\rV}(x_1p_1,x_2 p_2) 
\eeq
with $\Gamma_{qq;qq}^{(1)}$ given in eq.(\ref{eq.oneloopkernelgeneral}), and where 
$\ds_{q\bar{q},\NLO,N_h}^{\rV}$ the $N_h$ coefficient of thes the NLO virtual cross section.  

We decompose the double virtual counter term into a leading and a subleading-colour part:
\beq
\ds^{\rU}_{q\bar{q},\NNLO,N_h}=\ds^{\rU}_{q\bar{q},\NNLO,N_hN_c} + \ds^{\rU}_{q\bar{q},\NNLO,N_h/N_c}.
\eeq

The leading-colour part is given by
\beqa\label{eq.hqlc}
&&\hspace{-0.3in}\ds^{\rU}_{q\bar{q},\NNLO,N_hN_c}= -{\cal N}_{\NNLO}^{\VV,q\bar{q}}\,N_hN_c
\int \frac{{\rm d}x_1}{x_1} \frac{{\rm d}x_2}{x_2} {\rm d}\Phi_2(p_3,p_4; x_1p_1,x_2p_2) 
\nonumber \\
&& \times \bigg [ \bigg ({\cal A}^{0}_{q,Q g}(\e,s_{\bar{1}3},x_1,x_2) -\Gamma^{(1)}_{qq}(x_1)
\delta(1-x_2)\bigg) \nonumber\\
&&\hspace{0.075in} + \bigg( {\cal A}^{0}_{q,Q g}(\e,s_{\bar{2}4},x_2,x_1) -\Gamma^{(1)}_{qq}(x_2)\delta(1-x_1)\bigg)\bigg]\,|\cmb_4^{1,[h]}(\Q{3},\Qb{4},\hat{\bar{2}}_{\bar{q}},\hat{\bar{1}}_{q})|^2  J_2^{(2)}(p_3,p_4) \nonumber\\
\eeqa
The terms present in the square bracket are manifestly free of initial-state collinear divergences 
as is the whole double virtual subtraction term $\ds^{\rU}_{q\bar{q},\NNLO,N_hN_c}$. 
Using the same integrated massive NLO dipoles as in section \ref{sec:virtualvirtualNc2} for the 
leading-colour double virtual counter-term contributions, we can rewrite eq.(\ref{eq.hqlc}) as:
\beqa
\label{eq.eq.dsunhlcdip}
&&\hspace{-0.3in}\ds^{\rU}_{q\bar{q},\NNLO,N_hN_c}=-{\cal N}_{\NNLO}^{\VV,q\bar{q}}\,N_hN_c\int \frac{{\rm d}x_1}{x_1} \frac{{\rm d}x_2}{x_2} {\rm d}\Phi_2(p_3,p_4; x_1p_1,x_2p_2) \nonumber \\
&&  \times \Big ({\bf J}_{2}^{1} (\bar{1},3_Q)+{\bf J}_{2}^{1} (\bar{2},4_{\bar{Q}}) \Big)|\cmb_4^{1,[h]}(\Q{3},\Qb{4},\hat{\bar{2}}_{\bar{q}},\hat{\bar{1}}_{q})|^2  J_2^{(2)}(p_3,p_4).
\eeqa
 
The subleading-colour heavy quark counter term $\ds^{\rU}_{q\bar{q},\NNLO,N_h/N_c}$ reads, 
\beqa
\label{eq.dsunhslc}
&&\hspace{-0.1in}\ds^{\rU}_{q\bar{q},\NNLO,N_h/N_c}= -{\cal N}_{\NNLO}^{\VV,q\bar{q}}\frac{N_h}{N_c}
\int \frac{{\rm d}x_1}{x_1} \frac{{\rm d}x_2}{x_2} {\rm d}\Phi_2(p_3,p_4; x_1p_1,x_2p_2) 
 \nonumber \\
&&\hspace{0.38in} \times\bigg ( 2{\cal A}^{0}_{q,Q g}(\e,s_{\bar{2}3},x_2,x_1) +  2{\cal A}^{0}_{q,Q g}(\e,s_{\bar{1}4},x_1,x_2) -2{\cal A}^{0}_{q,Q g}(\e,s_{\bar{1}3},x_1,x_2) \nonumber \\
&&\hspace{0.5in} -  2{\cal A}^{0}_{q,Q g}(\e,s_{\bar{2}4},x_2,x_1) -{\cal A}^{0}_{Q g \bar{Q}}(\e,s_{34},x_1,x_2) -{\cal A}^{0}_{q\bar{q}, g}(\e,s_{\bar{1}\bar{2}},x_1,x_2)\phantom{\bigg(} \nonumber \\
&&\hspace{0.5in}+\Gamma^{(1)}_{qq}(x_1)
\delta(1-x_2) +\Gamma^{(1)}_{qq}(x_2) \delta(1-x_1) \bigg ) |\cmb_4^{1,[h]}(\Q{3},\Qb{4},\hat{\bar{2}}_{\bar{q}},\hat{\bar{1}}_{q})|^2  J_2^{(2)}(p_3,p_4). \nonumber\\
\eeqa
This subtraction term can also be written in terms of integrated dipoles. In addition to the initial-final 
massive dipoles defined in eqs.(\ref{eq.J21a}-\ref{eq.J21b}), this requires a new final-final massive dipole
defined as
\beq
{\bf J}_{2}^{1} (I_Q, K_{\bar{Q}})={\cal A}^{0}_{Q g \bar{Q}}(\e,s_{IK}),
\eeq
as well as a massless initial-initial dipole defined in \cite{Currie:2013vh} and given by
\beq
{\bf J}_{2}^{1} ( \hat{\bar{1}}_{q} , \hat{\bar{2}}_{\bar{q}})=
{\cal A}^{0}_{q \bar{q}, g}(x_1,x_2) -\Gamma^{(1)}_{qq}(x_1)
\delta(1-x_2) -\Gamma^{(1)}_{qq}(x_2) \delta(1-x_1).
\eeq
In terms of these three type of integrated dipoles, the virtual-virtual subtraction term $\ds^{\rU}_{q\bar{q},\NNLO,N_h/N_c}$
takes the following form: 
\beqa
\label{eq.eq.dsunhslcdip}
&&\hspace{-0.2in}\ds^{\rU}_{q\bar{q},\NNLO,N_h/N_c}=-{\cal N}_{\NNLO}^{\VV,q\bar{q}}\,\frac{N_{h}}{N_{c}}
\int \frac{{\rm d}x_1}{x_1} \frac{{\rm d}x_2}{x_2} {\rm d}\Phi_2(p_3,p_4; x_1p_1,x_2p_2) \nonumber \\
&& \times 
\Big ( 2 {\bf J}_{2}^{1} (\bar{2},3_Q)+2 {\bf J}_{2}^{1} (\bar{1},4_{\bar{Q}}) 
-2 {\bf J}_{2}^{1} (\bar{1},3_Q)-2 {\bf J}_{2}^{1} (\bar{2},4_{\bar{Q}}) 
-{\bf J}_{2}^{1} (3_Q ,4_{\bar{Q}})-{\bf J}_{2}^{1} (\bar{1}, \bar{2}) 
\Big) \nonumber \\
& & \hspace{0.5in} \times  |\cmb_4^{1,[h]}(\Q{3},\Qb{4},\hat{\bar{2}}_{\bar{q}},\hat{\bar{1}}_{q})|^2 J_2^{(2)}(p_3,p_4). 
\eeqa

The pole parts of each of the integrated dipoles in eqs.(\ref{eq.eq.dsunhlcdip}) and (\ref{eq.eq.dsunhslcdip}) 
is given by a specific colour-ordered infrared singularity operator $\ione{i}{j}$ given in \cite{Abelof:2014fza}.
In terms of those operators we find that the pole part of the entire double-virtual heavy-quark
counter term reads:
\beqa
&&\hspace{-0.5in}\poles\bigg(\ds^{\rU}_{q\bar{q},\NNLO,N_hN_c}+\ds^{\rU}_{q\bar{q},\NNLO,N_h/N_c}\bigg)\nonumber\\
&&\hspace{-0.1in}=-N_h{\cal N}_{\NNLO}^{\VV,q\bar{q}}\int\frac{{\rm d}x_1}{x_1}\frac{{\rm d}x_2}{x_2} \dphi_2(p_3,p_4;x_1 p_1,x_2 p_2)\delta(1-x_1)\delta(1-x_2)\nonumber\\
&&\hspace{0.3in}\bigg[8N_c\ione{Q}{\bar{q}}(\e,s_{13})-\frac{4}{N_c}\Big( 4\ione{Q}{\bar{q}}(\e,s_{13})-4\ione{Q}{\bar{q}}(\e,s_{23})+\ione{q}{\bar{q}}(\e,s_{12})\nonumber\\
&&\hspace{0.6in}+\ione{Q}{\bar{Q}}(\e,s_{34})\Big)\bigg]\re(R_h)|\cmb_4^{1,[h]}(\Q{3},\Qb{4},\hat{\bar{2}}_{\bar{q}},\hat{\bar{1}}_{q})|^2\,J_2^{(2)}(p_3,p_4).
\eeqa

As mentioned above, the renormalized one-loop amplitude $\cmb_4^{1,[h]}$ is finite. It is furthermore
proportional to its tree-level counterpart:
\beq
\cmb_4^{1,[h]}(\Q{3},\Qb{4},\qbi{2},\qi{1})=R_h\,\cm_4(\Q{3},\Qb{4},\qbi{2},\qi{1}),
\eeq
with the factor $R_h$ in the equation above given by
\beq
R_h=-\frac{1}{9}\Big(8-3\beta^2\Big)-\frac{\beta}{6}\Big(3-\beta^2\Big)H(0;x)-i\pi\frac{\beta}{6}\Big(3-\beta^2\Big)+\order{\e}
\eeq
and with $\beta = \sqrt{1-4m_Q^2/s_{12}}$. Using this expression for the one-loop amplitude, we can also express the 
pole structure of the ``one-loop squared'' contributions in $\ds^{\VV}_{q\bar{q},\NNLO,N_h}$ given in eq.(\ref{eq:nnlovv})
in terms of infrared singularity operators. For the leading-colour part we find 
\beqa
&&\hspace{-0.3in}\poles\bigg(2\re(\cmb_4^{1,[lc]}(\Q{3},\Qb{4},\qbi{2},\qi{1})\cmb_4^{1,[h]}(\Q{3},\Qb{4},\qbi{2},\qi{1})^{\dagger})\bigg)=\nonumber\\
&&\hspace{0.3in}4\ione{Q}{\bar{q}}(\e,s_{13})\re(R_h)|\cm_4(\Q{3},\Qb{4},\qbi{2},\qi{1})|^2
\eeqa
and for the subleading colour part, we get, 
\beqa
&&\hspace{-0.1in}\poles\bigg(2\re(\cmb_4^{1,[slc]}(\Q{3},\Qb{4},\qbi{2},\qi{1})\cmb_4^{1,[h]}(\Q{3},\Qb{4},\qbi{2},\qi{1})^{\dagger})\bigg)=\nonumber\\
&&\hspace{0.3in}2\Big(4\ione{Q}{\bar{q}}(\e,s_{13})-4\ione{Q}{\bar{q}}(\e,s_{23})+\ione{q}{\bar{q}}(\e,s_{12})+\ione{Q}{\bar{Q}}(\e,s_{34})\Big)\re(R_h)|\cm_4(\Q{3},\Qb{4},\qbi{2},\qi{1})|^2.\nonumber\\
\eeqa
Taking the pole part of the integrated subtraction terms given above (where we have used momentum conservation 
to set $s_{24}=s_{13}$ and $s_{14}=s_{23}$) enabled us to verify the pole structure of the two-loop amplitudes in the 
decoupling scheme. 

We conclude this section by showing how the one and two-loop amplitudes computed in \cite{Bonciani:2009nb} 
in the full-flavour scheme can be converted to the decoupling scheme with $N_l$ active flavours. This conversion 
can be achieved with the well known finite renormalisation of $\alpha_s$ (see e.g. \cite{Mitov:2006xs,Chetyrkin:1997un})
\beq
\label{eq:2alphas}
\alpha_s^{(N_F)}=\alpha_s^{(N_l)}\Bigg[1+\Bigg(\frac{\alpha_s^{(N_l)}}{2\pi}\Bigg)\xi_1+\Bigg(\frac{\alpha_s^{(N_l)}}{2\pi}\Bigg)^2\xi_2+{\cal O}(\alpha_s^3)\Bigg],
\eeq
with $\alpha_s^{(N_F)}$ denoting the strong coupling in the full theory, and $\alpha_s^{(N_l)}$ the strong coupling in 
the decoupling regime. The renormalisation constants $\xi_i$ are given by
\beqa
&&\hspace{-0.6in}\xi_1=\frac{N_h}{3}L_{\mu,\epsilon}\\
&&\hspace{-0.6in}\xi_2=\bigg[\frac{N_h^2}{9}+\epsilon\,\frac{N_h}{12}(5C_A+3C_F)\bigg]L_{\mu,\epsilon}^2+\frac{N_h}{6}(5C_A+3C_F)L_{\mu,\epsilon}-N_h\bigg(\frac{4C_A}{9}-\frac{15C_F}{8}\bigg),
\eeqa
with
\beq
L_{\mu,\epsilon}=\frac{(4\pi)^\epsilon}{\epsilon}\Bigg[\Gamma(1+\epsilon)\bigg(\frac{\mu^2}{m_Q^2}\bigg)^\epsilon-e^{-\epsilon\gamma_E}\Bigg].
\eeq

We can use eq.(\ref{eq:2alphas}) to relate an amplitude ${\cal M}$ computed in the full theory to 
its counterpart in the decoupling scheme, denoted as $\overline{\cal M}$. We have, 
\begin{eqnarray}
&&\hspace{-0.4in}{\cal M}=\left(4\pi\alpha_s^{(N_F)}\right)\Bigg\{{\cal M}^{(0)}+\Bigg(\frac{\alpha_s^{(N_F)}}{2\pi}\Bigg){\cal M}^{(1)}+\Bigg(\frac{\alpha_s^{(N_F)}}{2\pi}\Bigg)^2{\cal M}^{(2)}+{\cal O}(\alpha_s^3)\Bigg\}\nonumber\\ \nonumber\\ 
&&\hspace{-0.4in}\phantom{{\cal M}}=\left(4\pi\alpha_s^{(N_l)}\right)\Bigg\{{\cal M}^{(0)}+\Bigg(\frac{\alpha_s^{(N_l)}}{2\pi}\Bigg)\bigg[{\cal M}^{(1)}+\xi_1{\cal M}^{(0)}\bigg]\nonumber\\
&&\hspace{0.6in}+\Bigg(\frac{\alpha_s^{(N_l)}}{2\pi}\Bigg)^2\bigg[{\cal M}^{(2)}+2\xi_1{\cal M}^{(1)}+\xi_2{\cal M}^{(0)}\bigg]+{\cal O}(\alpha_s^3)\Bigg\},
\end{eqnarray}
such that for the one and two-loop amplitudes we find respectively, 
\beqa
&&\hspace{-0.3in}\overline{\cal M}^{(1)}={\cal M}^{(1)}+\xi_1{\cal M}^{(0)}\\
&&\hspace{-0.3in}\overline{\cal M}^{(2)}={\cal M}^{(2)}+2\xi_1{\cal M}^{(1)}+\xi_2{\cal M}^{(0)}={\cal M}^{(2)}+2\xi_1\overline{\cal M}^{(1)}+\left(\xi_2-2\left(\xi_1\right)^2\right){\cal M}^{(0)}.
\eeqa

%
\section{Numerical Results}
\label{sec:Results}
Together with the double real and real-virtual leading-colour contributions derived in \cite{Abelof:2014fza}, 
the two-parton channel presented in section \ref{sec:virtualvirtualNc} completes the calculation of the 
coefficient $A$ in eq.(\ref{eq.qqbdec}). We implemented this coefficient along with  $F_l$, $F_h$, $G_l$, 
$G_h$, $H_l$, $H_h$ and $H_{lh}$, in a Monte Carlo parton-level event generator based on the set up of 
eq.(\ref{eq.subnnlo}). While the inclusion of the coefficients $H_X$ is rather straightforward, as they are all 
infrared finite and only receive contributions from double virtual matrix elements, the coefficients $F_X$ and 
$G_X$ are non-trivial. $F_l$ and $G_l$ were presented in \cite{Abelof:2014jna}, and $F_h$ and $G_h$ were
derived in this publication in section \ref{sec:heavy}.

Using an adaptation of the phase space generator employed in the NNLO di-jet calculation of 
\cite{GehrmannDeRidder:2013mf}, our program, written in {\tt Fortran}, provides full kinematical information on 
an event-by-event basis, thus allowing for the evaluation of differential distributions for top pair production 
including exact NNLO results for the quark-antiquark channel. Naturally, the calculation of total cross sections 
is also possible. 

In this section, we present numerical results for LHC and Tevatron. For LHC, with $\sqrt{s}=8$ TeV, we show 
differential distributions in the transverse momentum of the top quark $p_T^t$, the top quark rapidity $y^{t}$, 
the invariant mass of the $t\bar{t}$ system $m_{t\bar{t}}$ and its rapidity $y^{t\bar{t}}$. For Tevatron we show 
differential cross sections in $p_T^t$, $y^{t}$ and $m_{t\bar{t}}$, as well as in the absolute value of the top 
quark rapidity $|y^{t}|$. In a separate sub-section we present our results for the forward-backward asymmetry 
$A_{FB}$, differential in the rapidity difference $|\Delta y^{t\bar{t}}|=|y^t-y^{\bar{t}}|$ as well 
as in $m_{t\bar{t}}$ and $p_T^{t\bar{t}}$. 

All the results presented below include all partonic channels at LO and NLO. At NNLO,  
the contributions to the quark-antiquark channel from the colour factors $N_c^2$, $N_lN_c$, $N_l/N_c$, 
$N_hN_c$, $N_h/N_c$, $N_l^2$, $N_h^2$,$N_lN_h$ and $N_hN_c$ and $N_h/N_c$ 
are included. We use the pole mass of the top quark $m_t = 173.3$ GeV, and the PDF sets 
MSTW2008 68cl. The factorisation and renormalisation scales are set equal to the top quark mass 
$\mu_R = \mu_F = m_t$ throughout. Where provided, scale variations correspond to the range 
$m_t/2\leq m_t\leq 2m_t$.

\begin{table}
\begin{centering}
\begin{tabular}{|ccccc|}
\hline
& LO & NLO  &  NNLO (LC)  & NNLO (Ref \cite{Baernreuther:2012ws}) \\
\hline 
Tevatron & $\hspace{0.25in}6.069\hspace{0.25in}$ & $\hspace{0.25in}5.845\hspace{0.25in}$ & $\hspace{0.25in}6.425\hspace{0.25in}$ & $\hspace{0.25in}6.118\hspace{0.25in}$ \\ 
LHC 7     & $\hspace{0.25in}29.76\hspace{0.25in}$ & $\hspace{0.25in}28.21\hspace{0.25in}$ & $\hspace{0.25in}29.51\hspace{0.25in}$ & $\hspace{0.25in}28.59\hspace{0.25in}$ \\
LHC 8     & $\hspace{0.25in}37.60\hspace{0.25in}$ & $\hspace{0.25in}35.67\hspace{0.25in}$ & $\hspace{0.25in}37.31\hspace{0.25in}$ & $\hspace{0.25in}36.22\hspace{0.25in}$ \\
LHC 14   & $\hspace{0.25in}91.22\hspace{0.25in}$ & $\hspace{0.25in}85.91\hspace{0.25in}$ & $\hspace{0.25in}89.96\hspace{0.25in}$ & $\hspace{0.25in}87.99\hspace{0.25in}$ \\
\hline
\end{tabular}
\caption{Contributions to the total hadronic cross section from the quark-antiquark channel at LO, NLO and
NNLO at different collider energies. Cross sections are in [pb], with $\mu_R = \mu_F = m_t$. 
Our NNLO predictions are contained in the third column, labelled as NNLO (LC), as they are dominated 
with the leading colour (LC) part.}
\label{tab.totxsecs}
\end{centering}
\end{table}

The total hadronic cross section in the quark-antiquark channel is presented in table \ref{tab.totxsecs} for 
Tevatron, and LHC with center-of-mass energies $7$, $8$ and $14$ TeV. From the comparison with the 
full NNLO calculation of \cite{Baernreuther:2012ws} it can be seen that, due to the omission of 
subleading-colour pieces at $\order{\alpha_s^4}$, our results consistently overestimate the 
NNLO correction to the total cross section, indicating that the combined contribution of the coefficients 
$C$ and $E$ in eq.(\ref{eq.qqbdec}) is negative and non-negligible. It should be noted that the size of these
subleading-colour terms decreases with the hadronic center-of-mass energy, as can be seen from the fact
that the discrepancies between or results and those of \cite{Baernreuther:2012ws} at NNLO are $5\%$ for 
Tevatron, $3.25\%$ for LHC 7, $3\%$ for LHC 8 and $2.25\%$ for LHC 14.

\subsection{Differential distributions for LHC}
\begin{figure}[t]
\begin{tabular}{cc}
\subfloat[]{\includegraphics[width=0.5\textwidth]{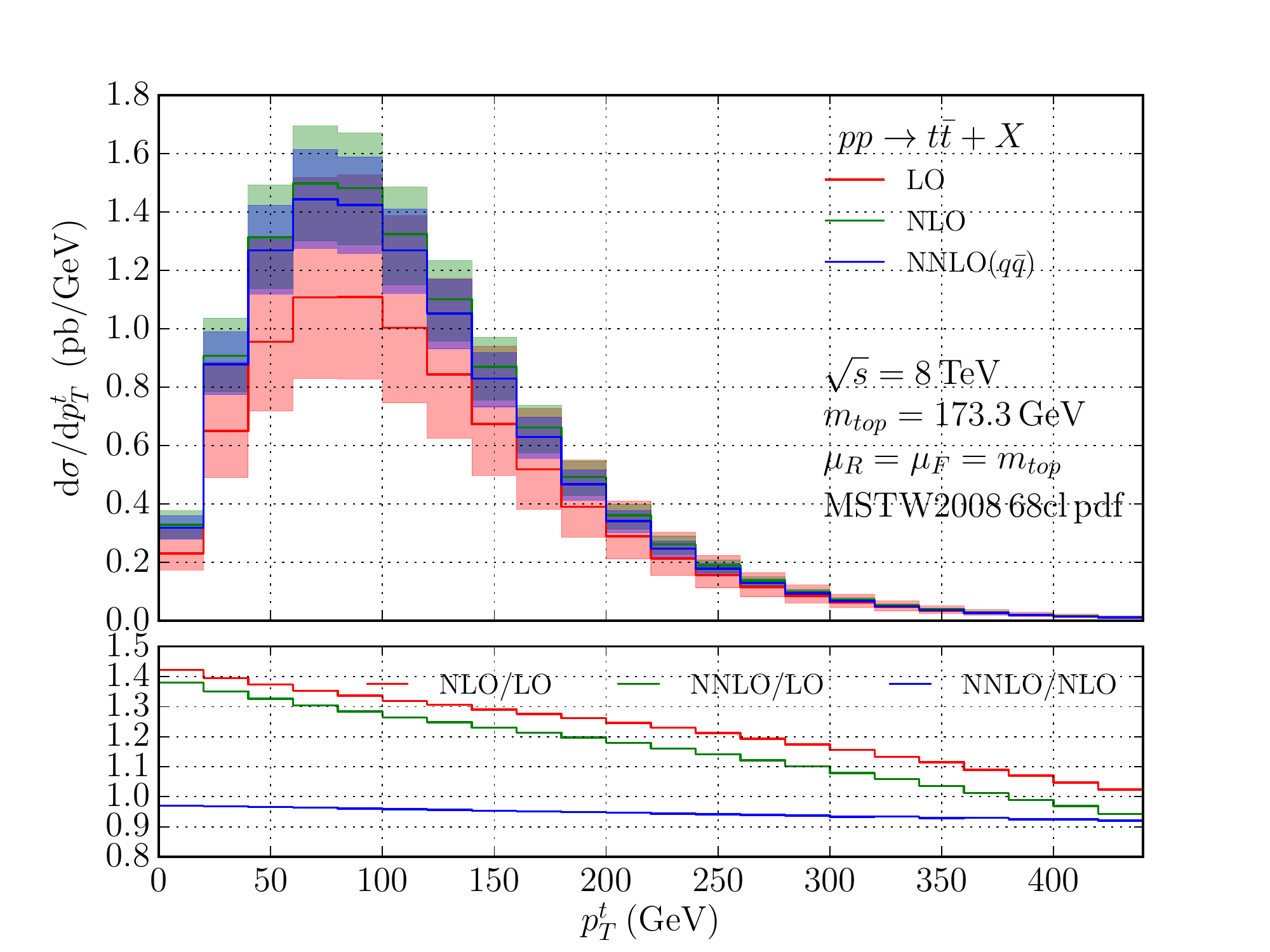}} & 
\subfloat[]{\includegraphics[width=0.5\textwidth]{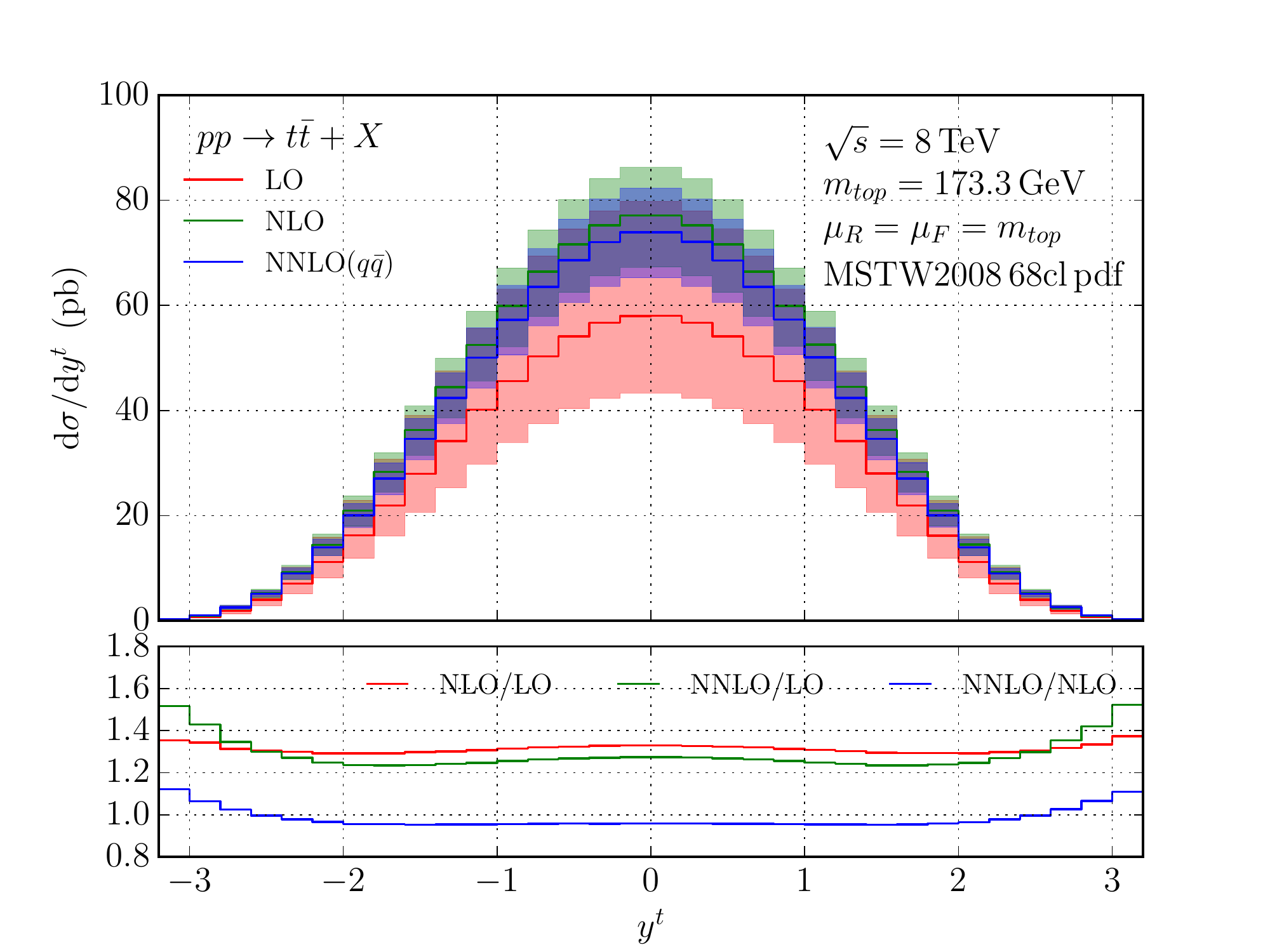}} \\ 
\subfloat[]{\includegraphics[width=0.5\textwidth]{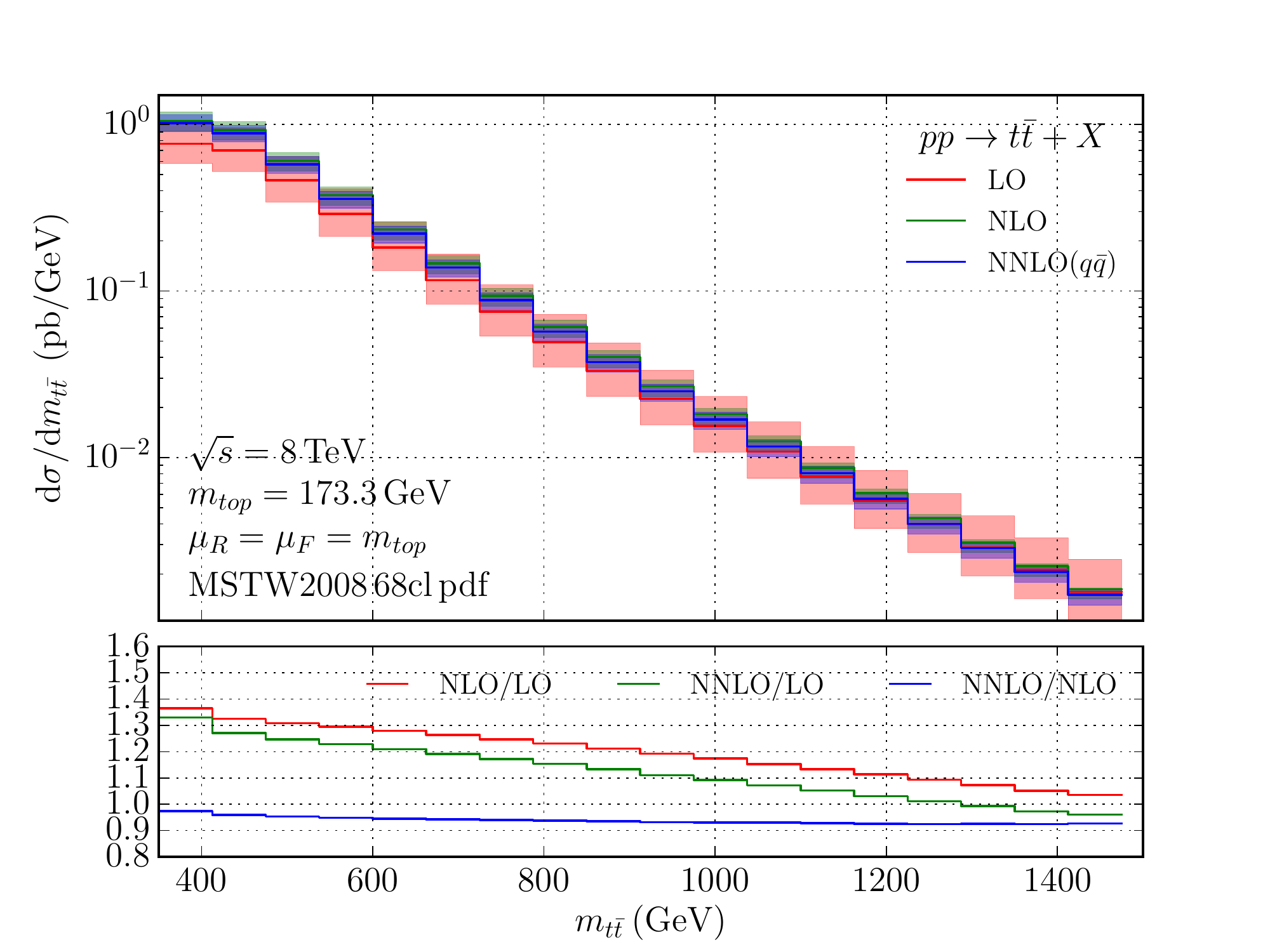}} &
\subfloat[]{\includegraphics[width=0.5\textwidth]{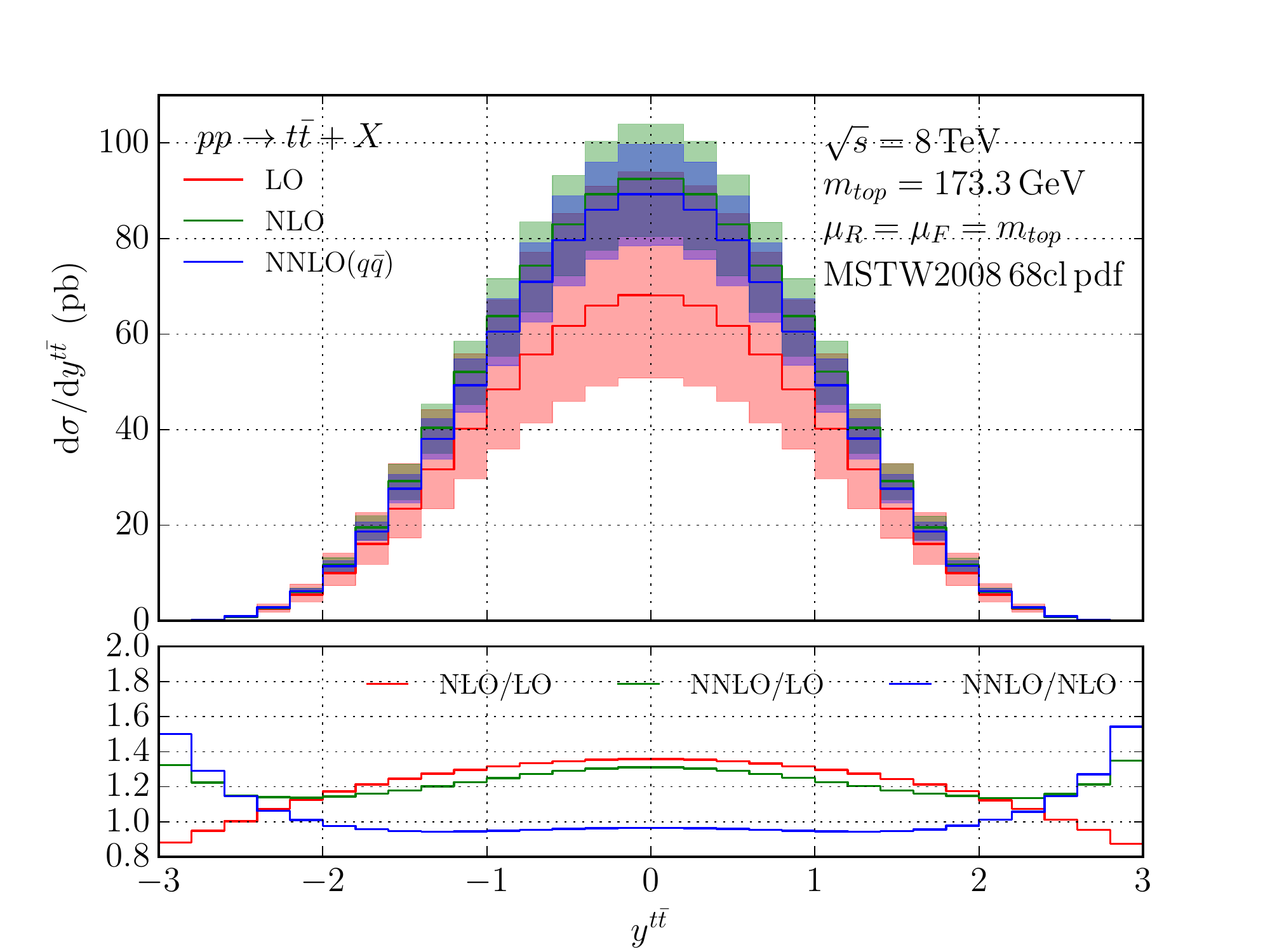}} 
\end{tabular}
\caption{Differential distributions for LHC ($\sqrt{s}=8$ TeV) in: (a) top quark transverse 
momentum $p_T^t$, (b) top quark rapidity $y^{t}$, (c) invariant mass of the $t\bar{t}$ system $m_{t\bar{t}}$, 
(d) rapidity of the $t\bar{t}$ system $y^{t\bar{t}}$. The NNLO contributions included are in the quark-antiquark 
channel only. Renormalisation and factorisation scales are set equal $\mu_R=\mu_F=\mu$ and varied as 
$m_t/2\leq\mu\leq2m_t$.}
\label{fig.LHC8dist}
\end{figure}

In fig.\ref{fig.LHC8dist} we present differential distributions for top pair production in $pp$ collisions
with $\sqrt{s}=8$ TeV at LO, NLO and NNLO, together with the corresponding $k$-factors. As expected, the 
impact of the NNLO corrections in the $q\bar{q}$ channel on LHC cross sections is in general mild, given 
the dominance of the gluon-gluon initiated process in this scenario. 

From the ratios NNLO/NLO in the lower panels of figs. (a) and (c), it can be seen that the NNLO corrections 
decrease the $p_T^t$ and $m_{t\bar{t}}$ distributions over the entire spectra considered. The decreases 
range between $3\%$ and $7\%$, being in both cases more pronounced in the tails of the distributions.

In the distributions in $y^t$ and $y^{t\bar{t}}$, given in figs. (b) and (d), it can be seen that in comparison with the 
full NLO result, NNLO corrections in the quark-antiquark channel shift the cross sections downwards in the 
central region by $5\%$. In the very forward and backwards ends of the spectra the impact of these 
corrections is more substantial, causing an upwards shift of $12\%$ in the $y^t$ distribution, and over 
$50\%$ in the $y^{t\bar{t}}$ distribution.

All plots in fig.\ref{fig.LHC8dist} show a slight reduction in the scale uncertainty when the NNLO corrections 
to the $q\bar{q}$ channel are included.

\begin{figure}[t]
\begin{tabular}{cc}
\subfloat[]{\includegraphics[width=0.5\textwidth]{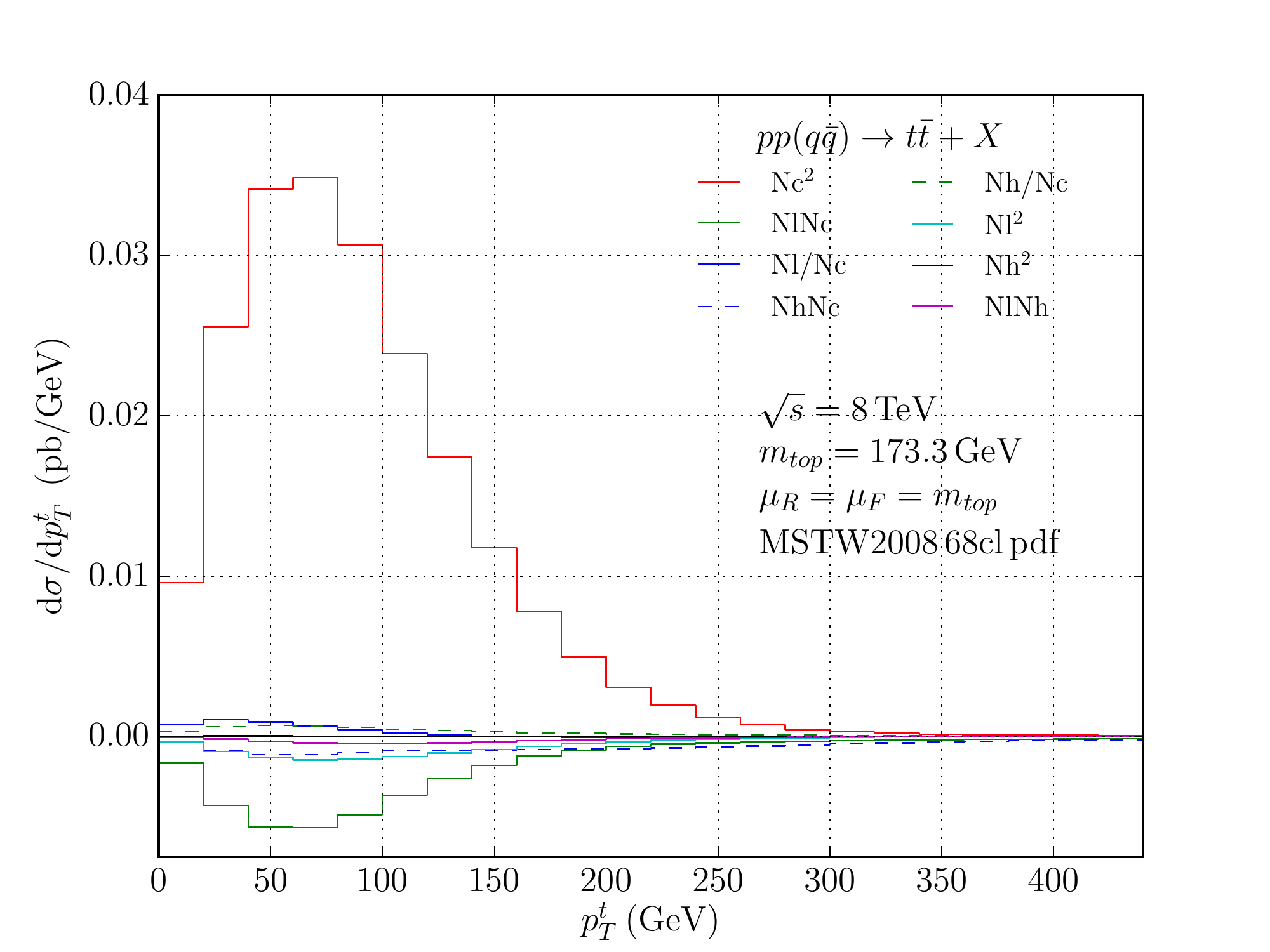}} & 
\subfloat[]{\includegraphics[width=0.5\textwidth]{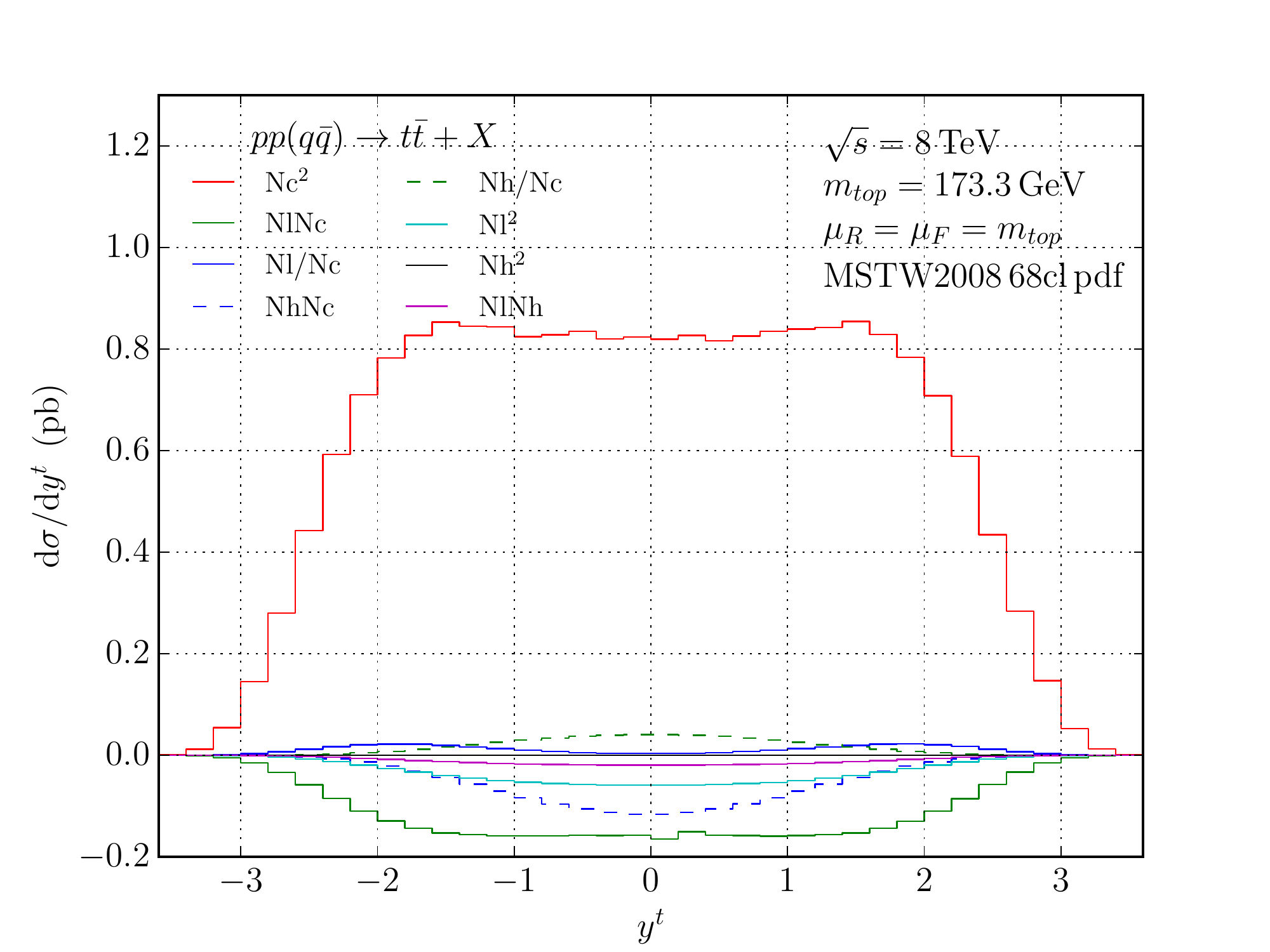}}
\end{tabular}
\caption{Contributions to the NNLO QCD corrections to $pp(q\bar{q})\to t\bar{t}+X$ from the different colour factors included in our computation.}
\label{fig.LHC8colour}
\end{figure}

In order to assess the relative size of the contributions from the different colour factors included at NNLO in 
our calculation, in fig.\ref{fig.LHC8colour} we show the breakdown of the NNLO corrections to the $q\bar{q}$
channel as functions of $p_T^t$ and $y^t$. We find that the leading-colour piece, proportional to $N_c^2$, 
contributes most significantly, followed by $N_lN_c$, which was calculated in \cite{Abelof:2014fza}. The 
contributions from all other colour factors are very small.

\begin{figure}[t]
\begin{tabular}{cc}
\subfloat[]{\includegraphics[width=0.5\textwidth]{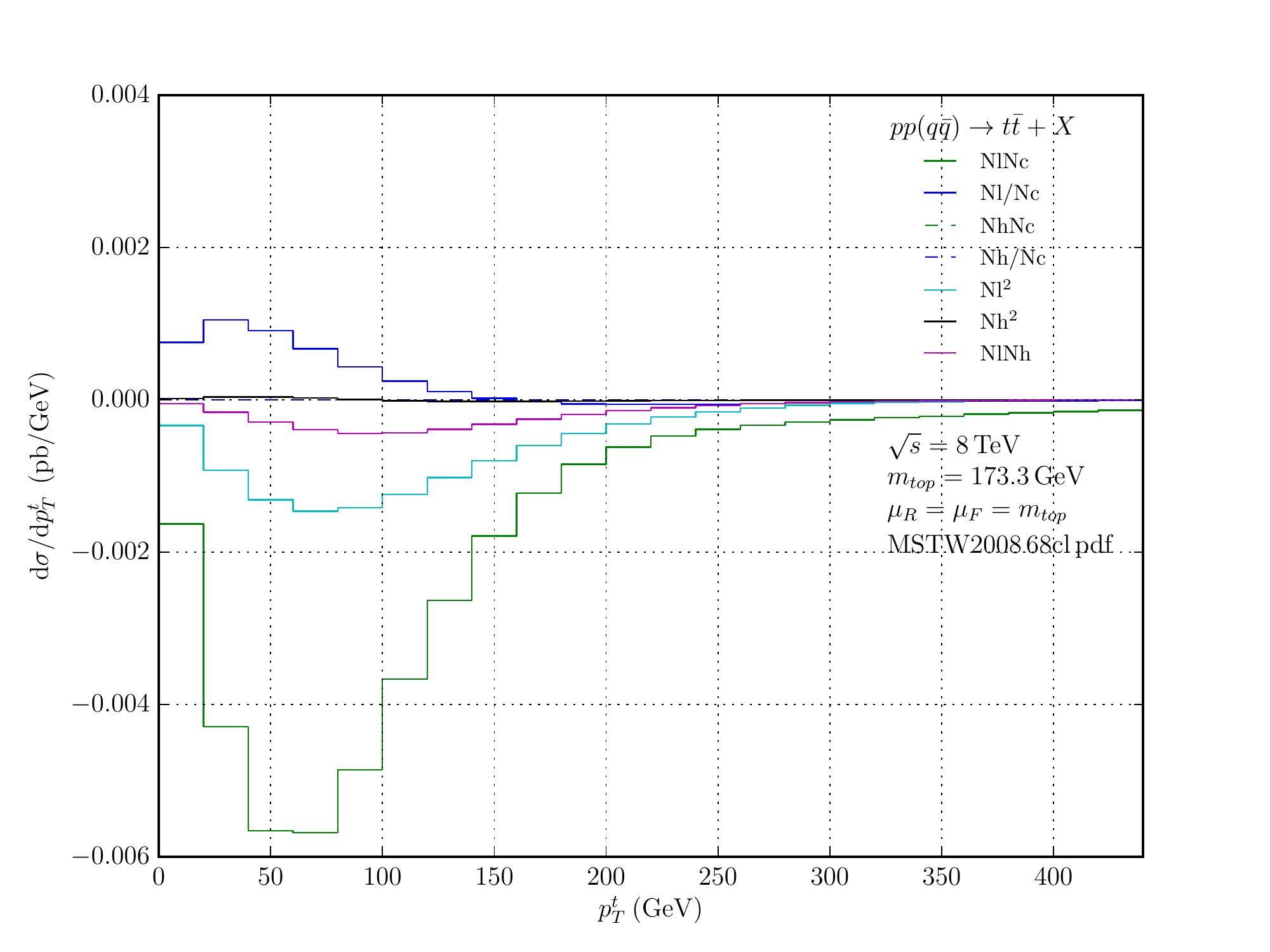}} & 
\subfloat[]{\includegraphics[width=0.5\textwidth]{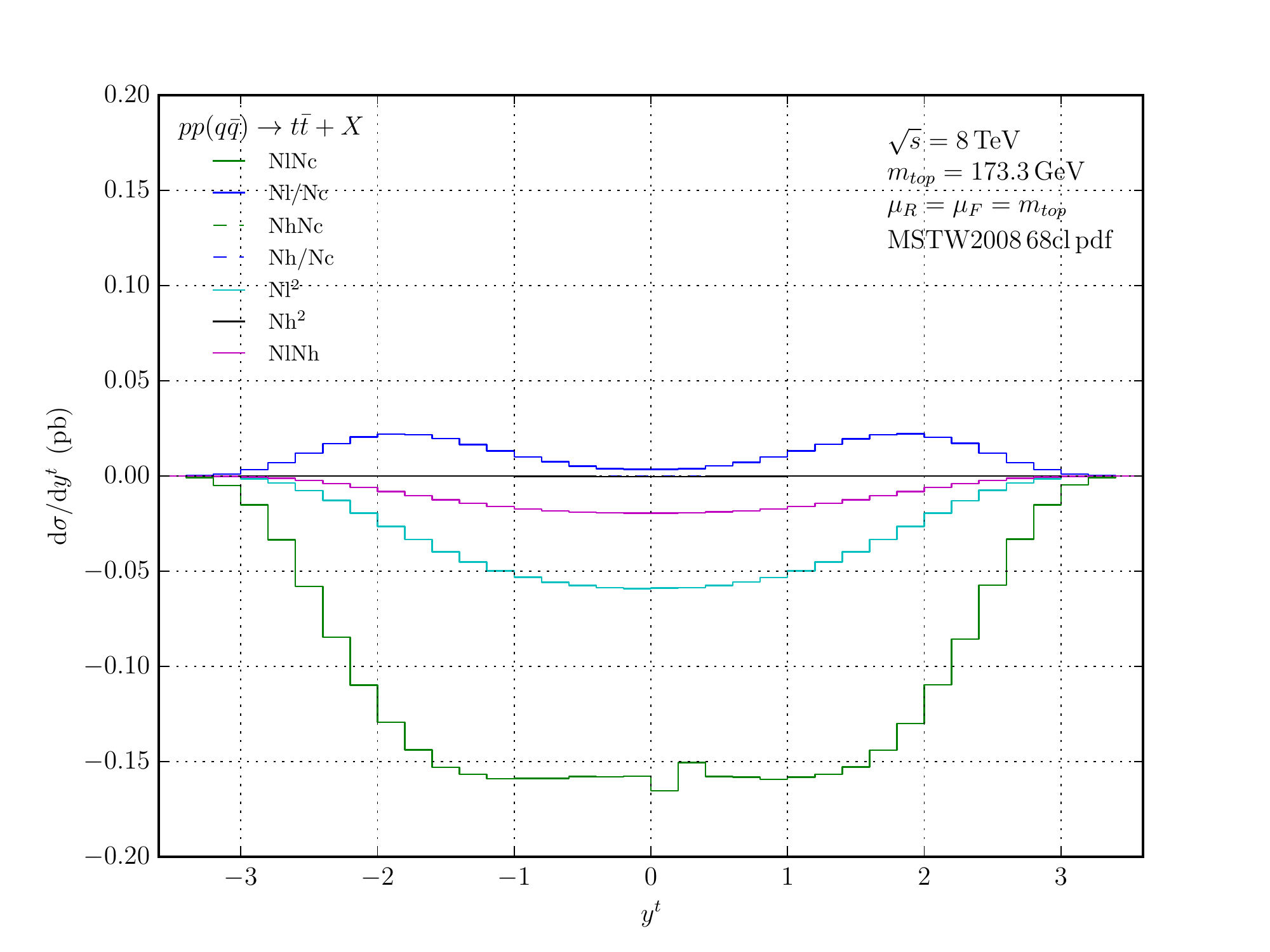}}
\end{tabular}
\caption{Contributions to the NNLO QCD corrections to $pp(q\bar{q})\to t\bar{t}+X$ from fermionic colour factors.}
\label{fig.LHC8colourheavy}
\end{figure}

To further disentangle the light and the heavy quark contributions, in fig \ref{fig.LHC8colourheavy}
we show only the fermionic contributions, omitting the $N_c^2$ colour factor.

\subsection{Differential distributions for Tevatron}
\begin{figure}[t]
\begin{tabular}{cc}
\subfloat[]{\includegraphics[width=0.5\textwidth]{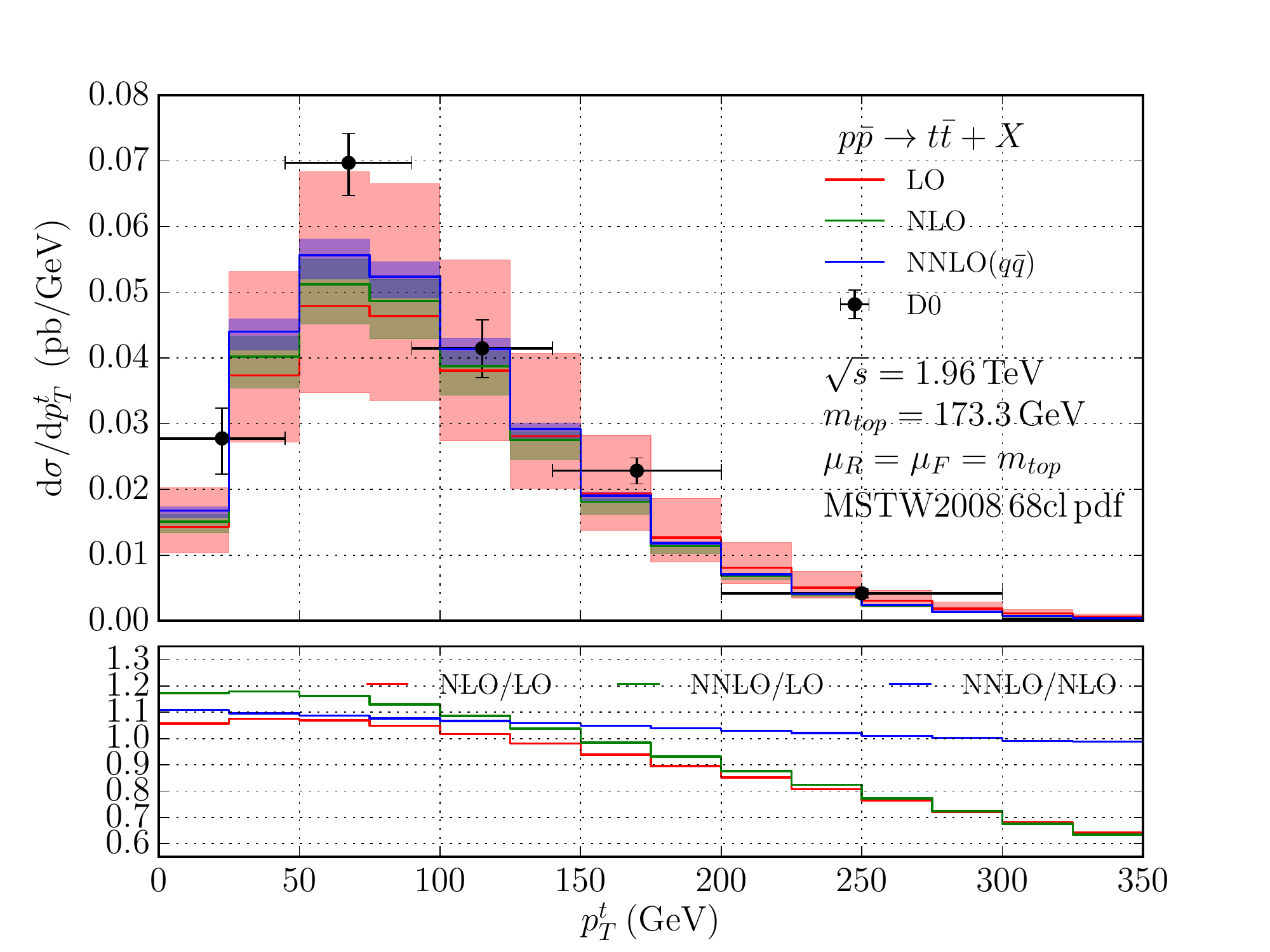}} & 
\subfloat[]{\includegraphics[width=0.5\textwidth]{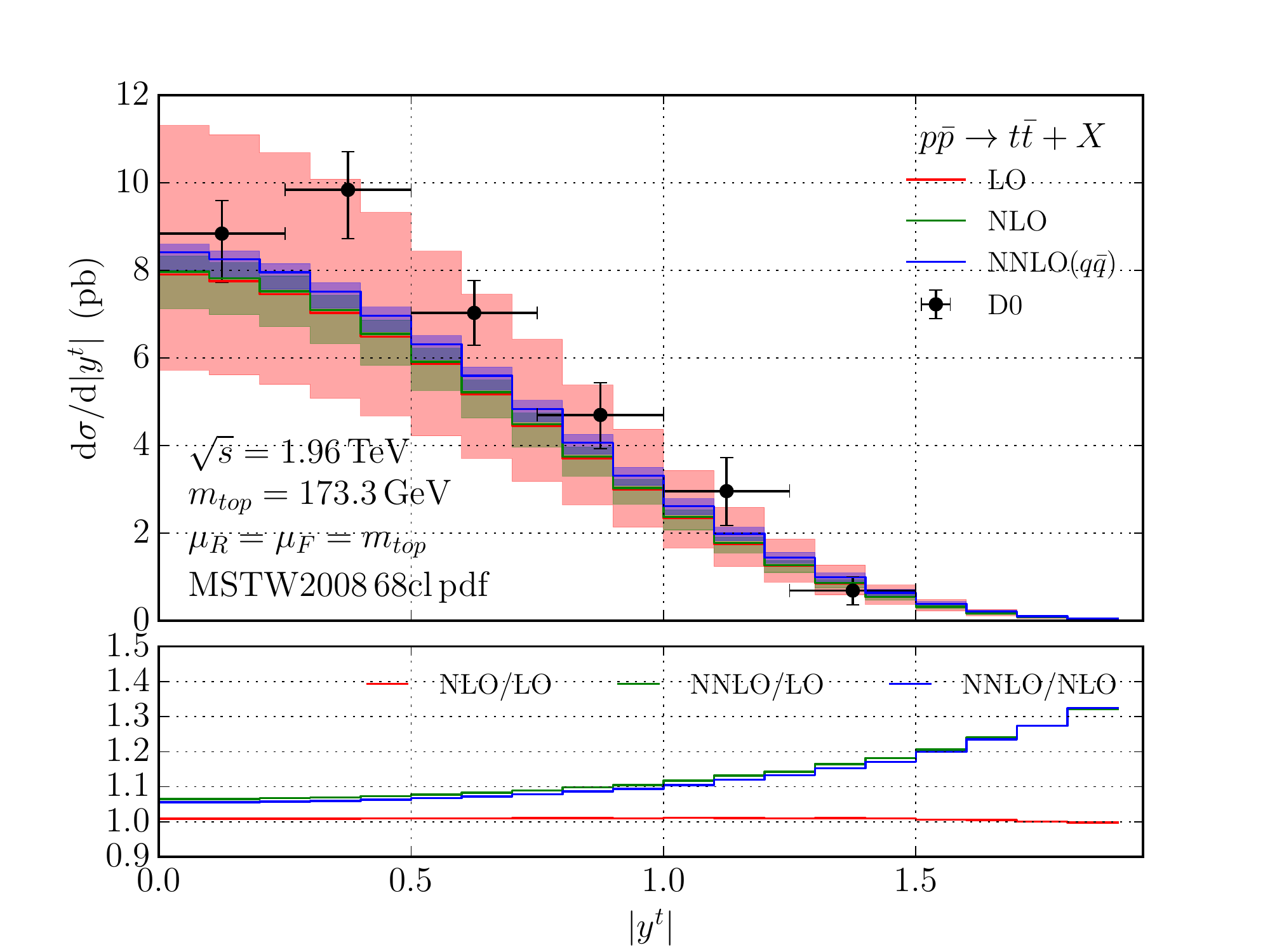}} \\ 
\subfloat[]{\includegraphics[width=0.5\textwidth]{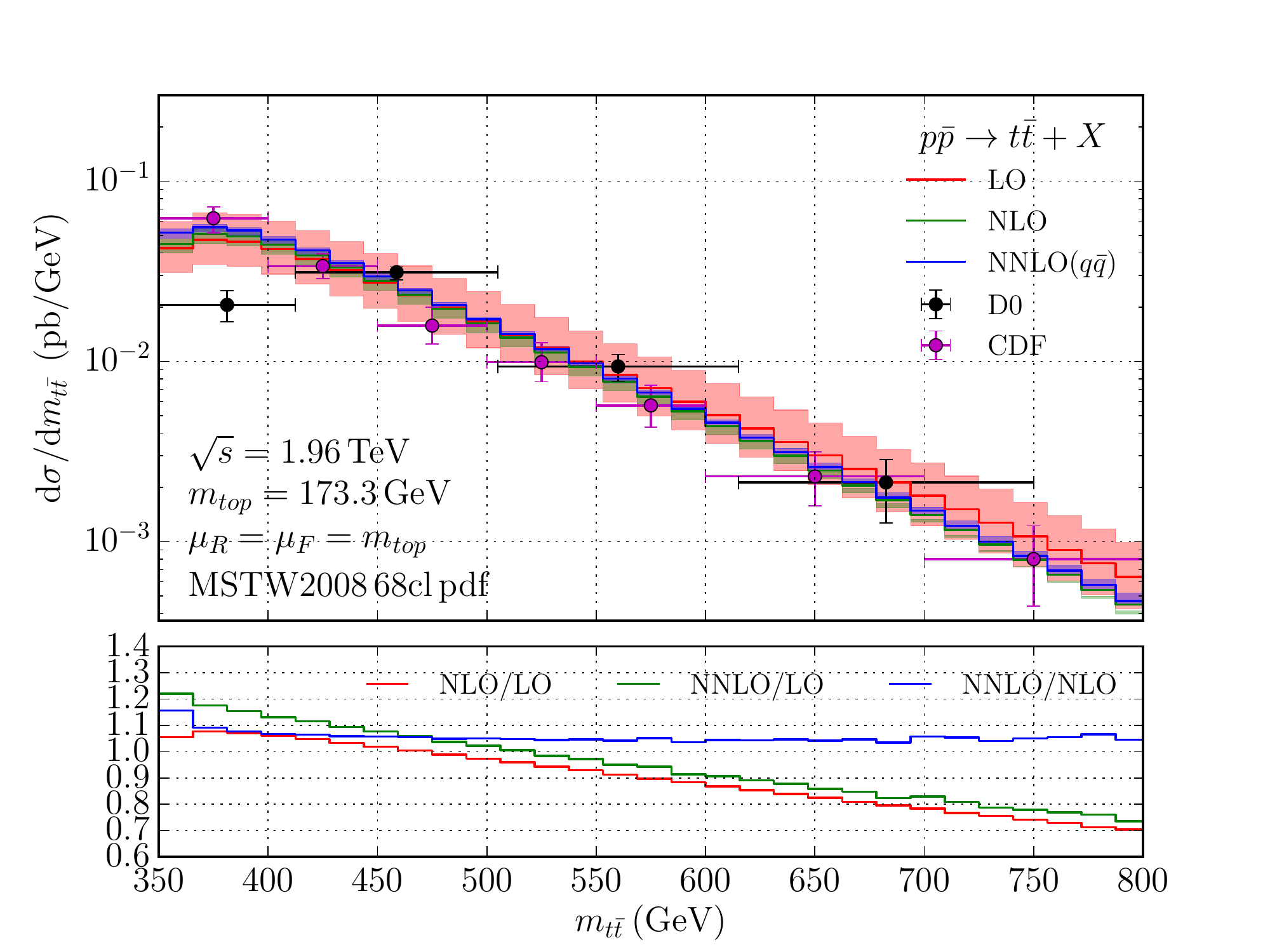}} &
\subfloat[]{\includegraphics[width=0.5\textwidth]{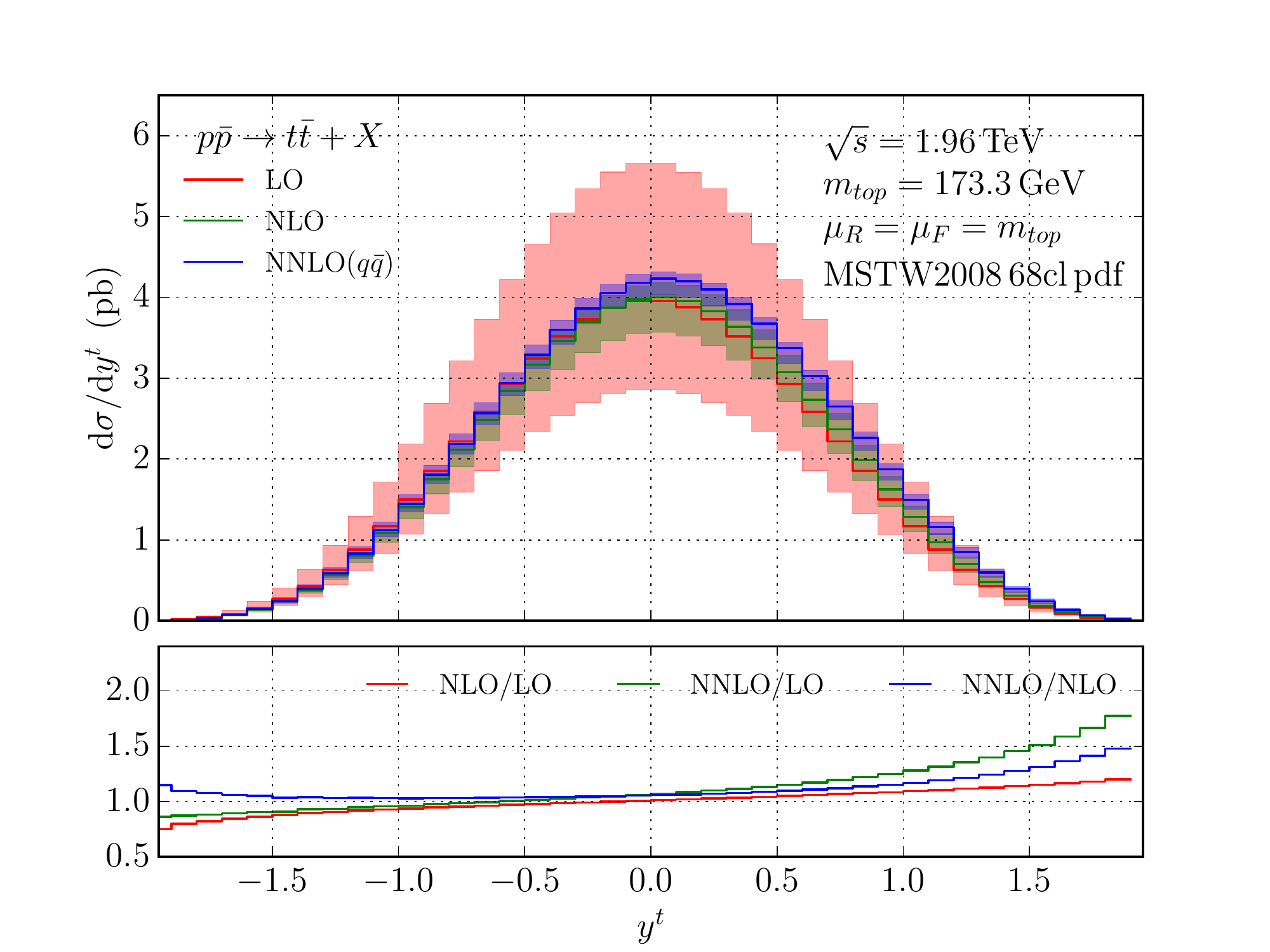}} 
\end{tabular}
\caption{Differential distributions for Tevatron in: (a) top quark transverse momentum $p_T^t$, 
(b) absolute value of the top quark rapidity $|y^{t}|$, (c) top quark rapidity $y^{t}$, (d) invariant mass of the 
$t\bar{t}$ system $m_{t\bar{t}}$. The NNLO contributions included are in the quark-antiquark channel only. 
Renormalisation and factorisation scales are set equal $\mu_R=\mu_F=\mu$ and varied as 
$m_{top}/2\leq\mu\leq2m_{top}$. Experimental data points from CDF and D0 are taken from the results in the 
$\ell$ + jets channel presented in \cite{Aaltonen:2009iz} and \cite{Abazov:2014vga} respectively.}
\label{fig.tevdist}
\end{figure}

In fig.\ref{fig.tevdist} we present differential distributions for top pair production in $p\bar{p}$ 
collisions with $\sqrt{s}=1.96$ TeV at LO, NLO and NNLO, along with the corresponding ratios. Given the 
dominance of the quark-antiquark channel in this scenario, the impact of the NNLO corrections included in our 
calculation is more important here than in the LHC distributions presented above. In all cases we find good 
agreement with experimental data, and a significant reduction in the scale uncertainty at NNLO.

As can be seen from the distribution in $p_T^t$ given in fig.(a), in the lower part of the spectrum, NNLO 
corrections introduce a $10\%$ shift with respect to the NLO prediction. This shift decreases as 
$p_T^t$ increases, becoming negligible in the tail of the distribution. In the distribution in $|y^t|$ given in 
fig.(b), NNLO corrections to the $q\bar{q}$ channel introduce an increasing shift of up to $30\%$ with respect to 
the NLO result. A significant reduction in the scale uncertainty is also observed. In fig.(c) we show the inclusive 
cross section for top pair hadro-production as a function of the invariant mass of the top-antitop system. NNLO 
corrections, in this case, cause a positive shift over the entire spectrum ranging from $15\%$ near the production 
threshold to approximately $5\%$ in the tail of the distribution. 

It is well known that QCD corrections to the $y^t$ distribution in $p\bar{p}$ collisions cause a shift towards the 
forward region. This fact can be observed in our result shown in fig.(d), where it can also be seen that NNLO 
corrections modify this shift in comparison to the NLO result. As we will see in the next section, this will have 
an effect in the NNLO result for the forward backward asymmetry.

\begin{figure}[t]
\begin{tabular}{cc}
\subfloat[]{\includegraphics[width=0.5\textwidth]{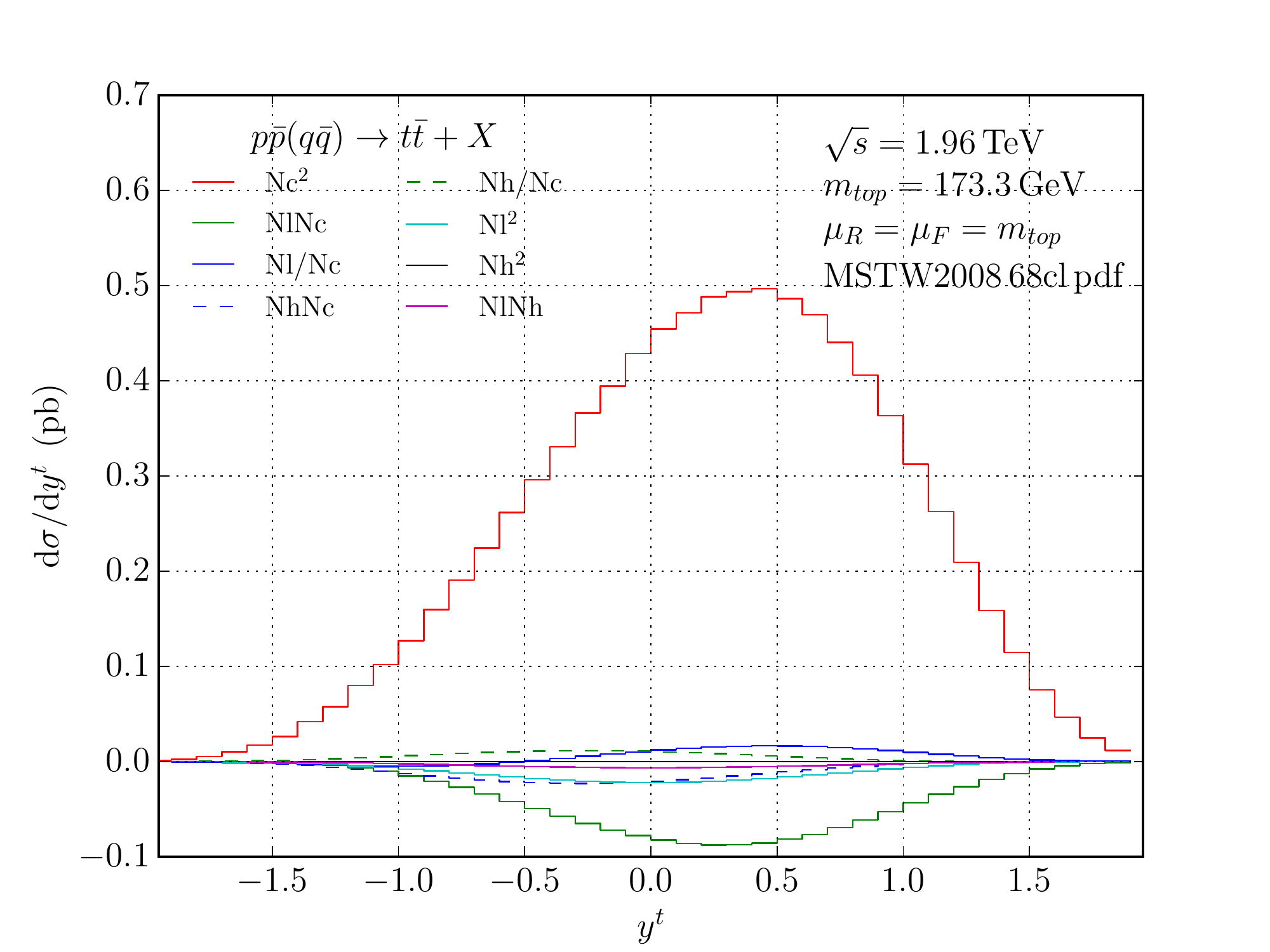}} & 
\subfloat[]{\includegraphics[width=0.5\textwidth]{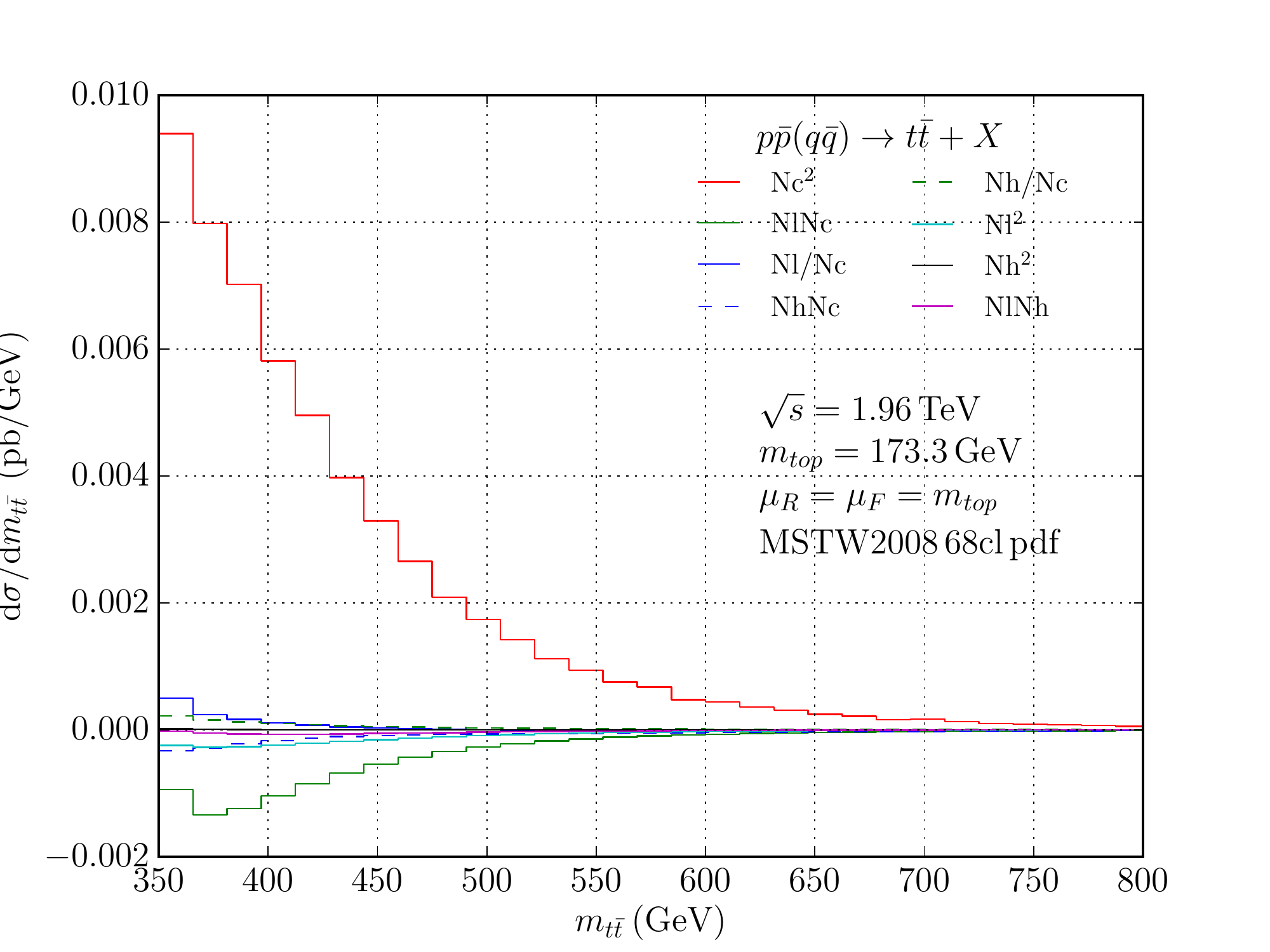}}
\end{tabular}
\caption{Contributions to the NNLO QCD corrections to $p\bar{p}(q\bar{q})\to t\bar{t}+X$ from the different colour factors included in our computation.}
\label{fig.tevcolour}
\end{figure}

\begin{figure}[t]
\begin{tabular}{cc}
\subfloat[]{\includegraphics[width=0.5\textwidth]{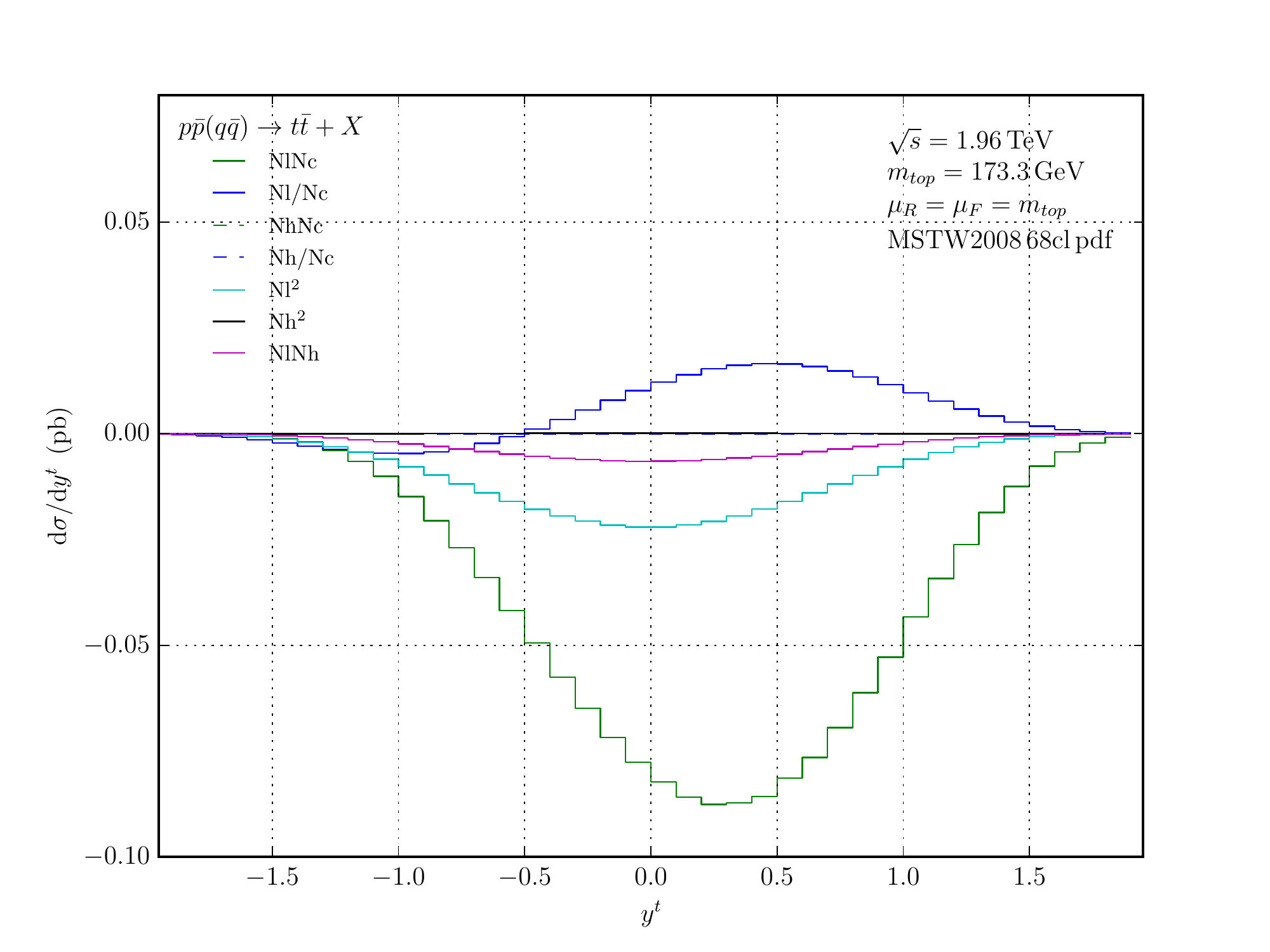}} & 
\subfloat[]{\includegraphics[width=0.5\textwidth]{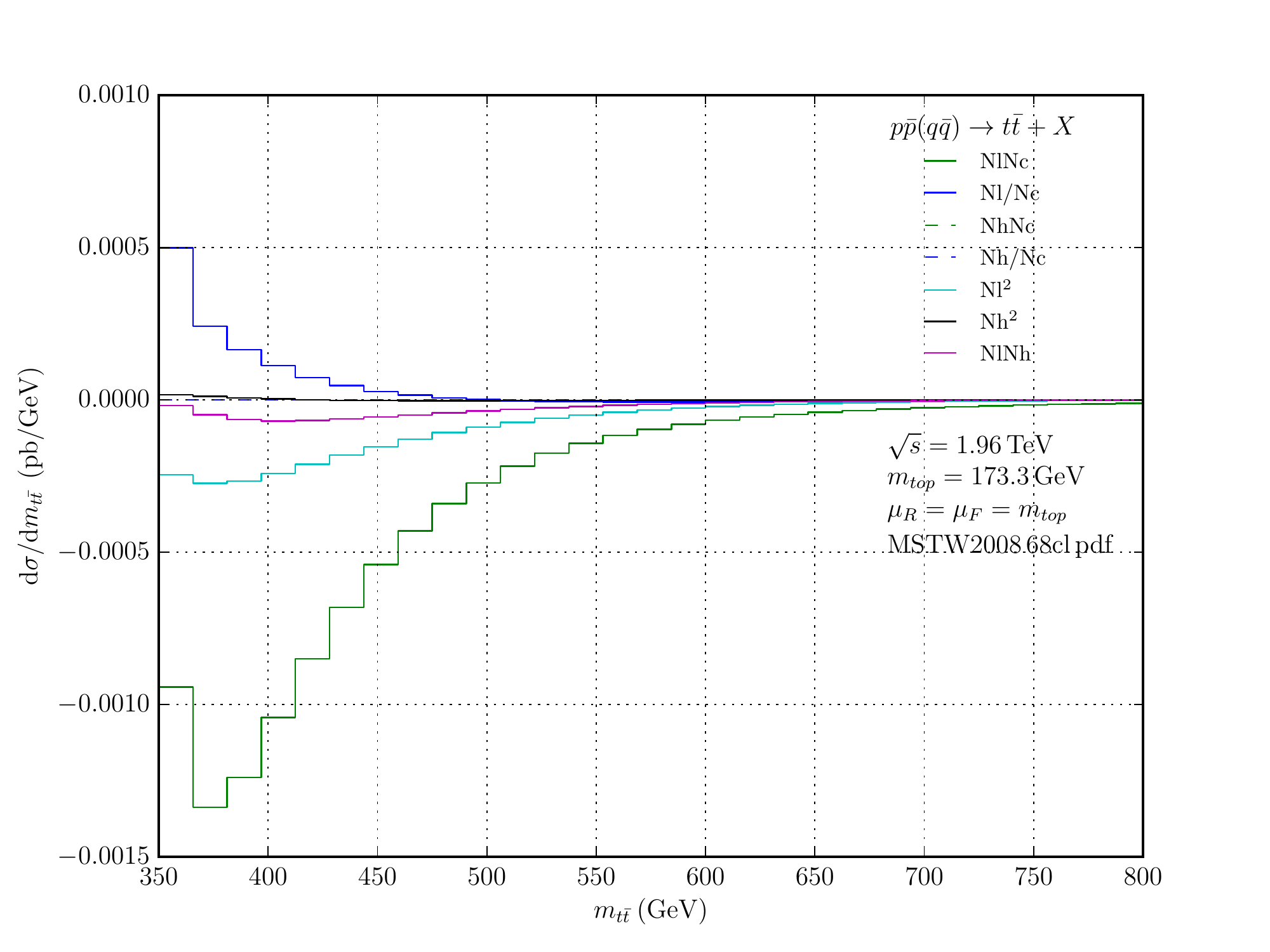}}
\end{tabular}
\caption{Contributions to the NNLO QCD corrections to $p\bar{p}(q\bar{q})\to t\bar{t}+X$ from fermionic colour factors.}
\label{fig.tevcolourl}
\end{figure}

In fig.\ref{fig.tevcolour} we show the breakdown of the NNLO corrections to the $y^t$ and $m_{t\bar{t}}$ 
distributions into colour factors. The same pattern found in fig.\ref{fig.LHC8colour} can be observed, with the 
$N_c^2$ colour factor contributing most significantly, followed by $N_lN_c$. As before, in fig.\ref{fig.tevcolourl}, 
we omit the leading color factor in order to further examine the size of the different fermionic colour factors.

\subsection{The forward-backward asymmetry at Tevatron}
The forward backward asymmetry in hadronic top quark pair production is an observable that measures the 
difference between forwardly and backwardly produced top quarks. It is defined as
\beq
\label{eq.defafb}
A_{FB}=\frac{\sigma(\Delta y^{t\bar{t}}_+)-\sigma(\Delta y^{t\bar{t}}_-)}
{\sigma(\Delta y^{t\bar{t}}_+)+\sigma(\Delta y^{t\bar{t}}_-)}
\eeq 
with $\Delta y^{t\bar{t}}=y^t-y^{\bar{t}}$ and $\Delta y^{t\bar{t}}_{\pm}=\theta(\pm\Delta y^{t\bar{t}})$. 

At leading-order $A_{FB}$ vanishes given the forward-backward symmetry of the Born matrix elements. At 
higher orders, however, loops and real emissions in the quark-antiquark and quark-gluon channels cause the
(anti)top to be more likely produced in the hemisphere defined by the direction of the incoming (anti)quark. 
Due to this fact, which was first observed in \cite{Kuhn:1998jr}, a non-vanishing $A_{FB}$ in $p\bar{p}$ 
collisions is predicted starting at next-to-leading order. 

Much attention was dedicated to the forward-backward asymmetry after measurements by the CDF 
collaboration showed a pronounced discrepancy with the Standard Model NLO prediction 
\cite{Aaltonen:2011kc,Aaltonen:2012it}. A recent publication by the D0 collaboration \cite{Abazov:2014cca}
showed measurements that differ from those by the CDF collaboration, and are in agreement with all SM 
predictions. In \cite{Czakon:2014xsa}, the NNLO QCD corrections to $A_{FB}$ were calculated for the first 
time, yielding a significant reduction in the scale uncertainty and confirming the agreement between the 
SM prediction and the latest result from D0.

Here we present our results for the forward-backward asymmetry at NNLO, computed in the 
so-called unexpanded form:
\beq
\label{eq.defafbunexp}
A_{FB}=\frac{\alpha_s^3\Delta\sigma_{NLO}+\alpha_s^4\Delta\sigma_{NNLO}+\order{\alpha_s^5}}{\alpha_s^2\sigma_{LO}+\alpha_s^3\sigma_{NLO}+\alpha_s^4\sigma_{NNLO}+\order{\alpha_s^5}}.
\eeq 

\begin{figure}[t]
\begin{tabular}{cc}
\subfloat[]{\includegraphics[width=0.5\textwidth]{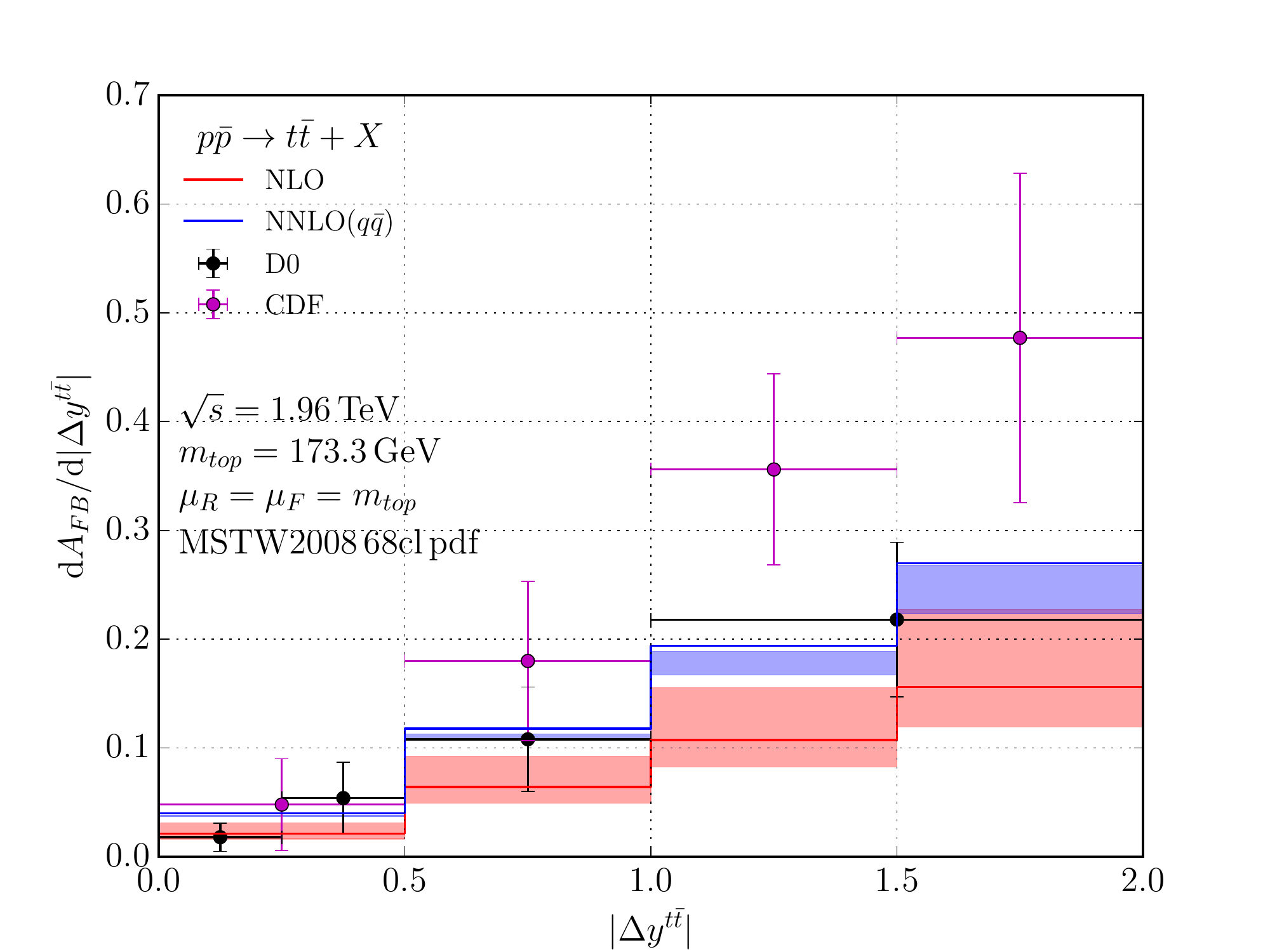}} & 
\subfloat[]{\includegraphics[width=0.5\textwidth]{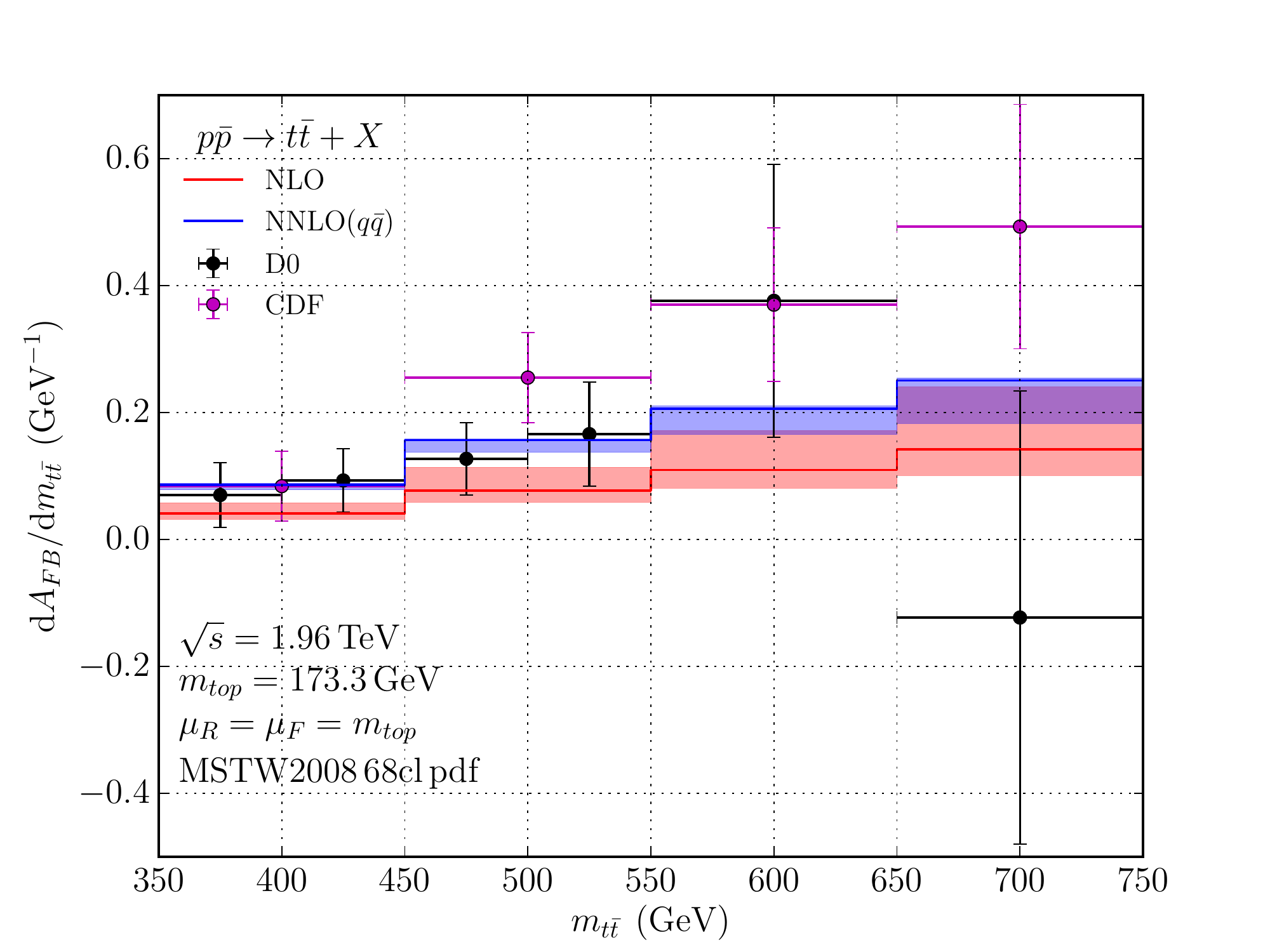}}\\
\multicolumn{2}{c}{
 \subfloat[]{\includegraphics[width=0.5\textwidth]{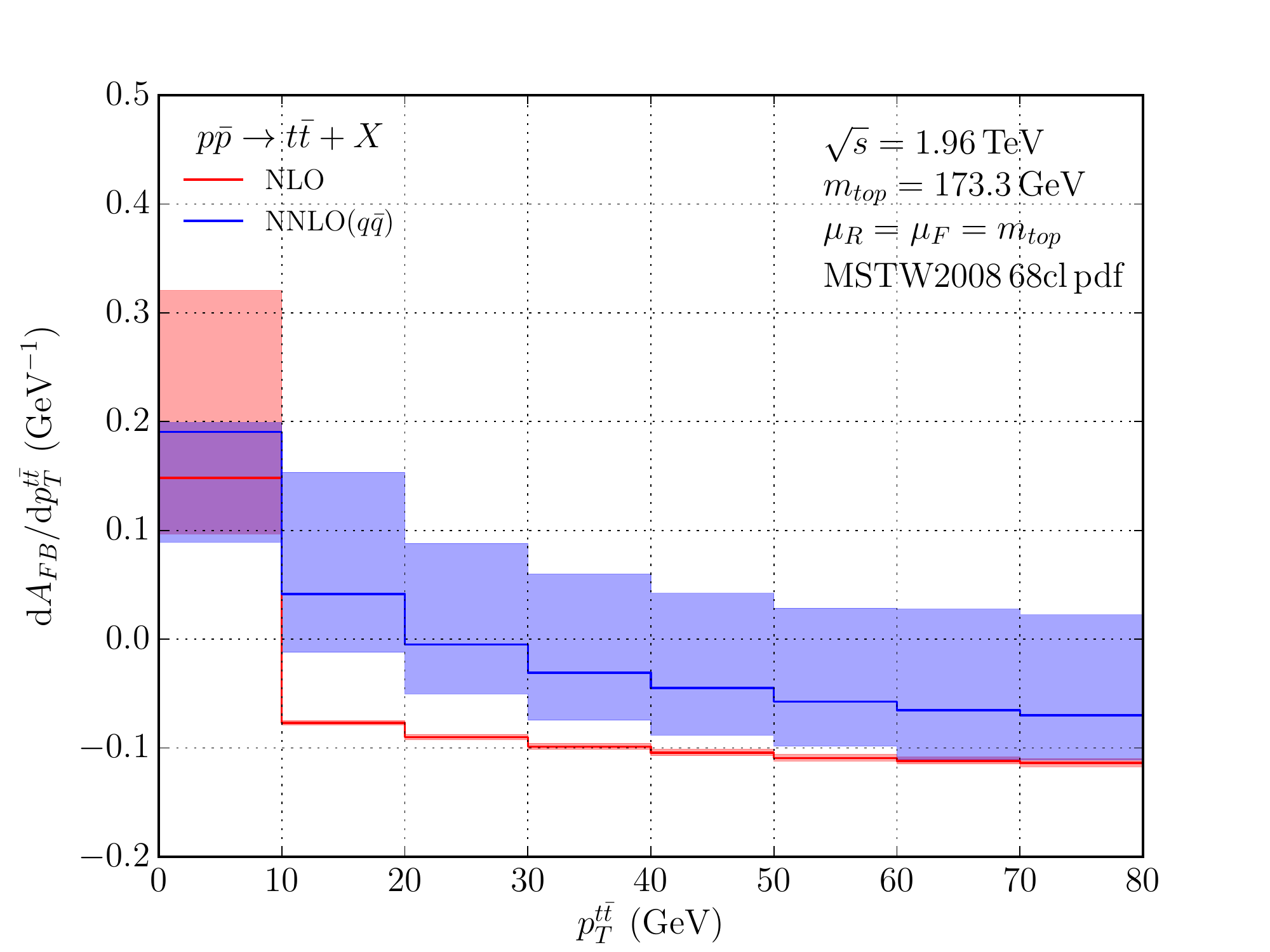}} }
\end{tabular}
\caption{Differential $A_{FB}$ in: (a) absolute rapidity difference $|\Delta y^{t\bar{t}}|=|y^t-y^{\bar{t}}|$, (b) invariant mass of the $t\bar{t}$ system $m_{t\bar{t}}$, and transverse momentm of the $t\bar{t}$ system $p_T^{t\bar{t}}$. The NNLO contributions included are in the quark-antiquark channel only. Renormalisation and factorisation scales are set equal $\mu_R=\mu_F=\mu$ and varied as $m_{top}/2\leq\mu\leq2m_{top}$. Experimental data points from CDF and D0 are taken from the results in the $\ell$ + jets channel presented in \cite{Aaltonen:2012it} and \cite{Abazov:2014cca} respectively.}
\label{fig.afb}
\end{figure}

In fig.\ref{fig.afb} we show $A_{FB}$ as a function of $|\Delta y^{t\bar{t}}|=|y^t-y^{\bar{t}}|$ as well as in 
$m_{t\bar{t}}$ and $p_T^{t\bar{t}}$. In the first two cases we find agreement with the D0 measurements, as 
well as a substantial reduction in the renormalisation and factorisation scale dependence at NNLO. The 
change from NLO to NNLO in the scale variation bands of the $p_T^{t\bar{t}}$ distribution is rather unusual, 
with a reduction in the scale dependence in the first bin, and an increase in all others.

We compared the results in fig.\ref{fig.afb} for the differential asymmetries with \cite{Czakon:2014xsa}.
Although at NNLO our computation only includes the quark-antiquark channel at NNLO (and it does not 
include all colour factors), when this channel is dominant, the two results are in agreement. Our results
appear to differ at the edges of the distributions. In those regions the top quark pair are produced mostly 
asymmetrically where we expect the quark-gluon channel, not included in our computation (but included in 
\cite{Czakon:2014xsa}), to play an important role. 

%
\section{Conclusions}
\label{sec:conclusions}
We presented the computation of the virtual-virtual two-parton final state contributions to top pair production 
in the quark-antiquark channel proportional to the leading colour factor $N_c^2$. Together with the 
three and four-parton contributions derived in a previous publication \cite{Abelof:2014fza}, this enabled us to 
complete the NNLO corrections to top pair hadro-production in the quark-antiquark channel for the 
phenomenologically most important colour factor. 

We derived the subtraction term needed at the two-parton level using the antenna subtraction method 
extended to deal with massive coloured final states. This required the integration of new massive
tree-level four-parton and three-parton one-loop antennae, which we presented here for the first time. We 
showed that the virtual-virtual counter-term has a similar structure to that found in the massless case 
\cite{Currie:2013vh}, and can be expressed in terms of integrated massive dipoles whose pole part is related 
to the massive infrared singularity operators of \cite{Catani:2000ef}. We furthermore demonstrated that the 
explicit poles of the double virtual matrix elements can be analytically cancelled against those of the 
corresponding counter-term, providing a crucial check on the correctness of our calculation.
In addition, we presented the computation of the heavy quark contributions related to the colour 
factors $N_hN_c$ and $N_h/N_c$. In particular, we presented our results using analytic 
expressions for the loop amplitudes in the so-called decoupling scheme, which were not 
available as such in the literature.

Together with the NNLO corrections proportional to the number of light quark flavours $N_l$ obtained in 
\cite{Abelof:2014jna}, as well as other easily added fermionic colour factors, we implemented the 
leading-colour and heavy quark NNLO contributions to $q\bar{q}\to t\bar{t}+X$ completed in this paper, 
in a fully differential Monte Carlo event generator. 
This program allowed us to produce several differential distributions for top pair production with NNLO 
corrections in the quark-antiquark channel for Tevatron and LHC energies. In particular, we obtained 
distributions in the top quark transverse momentum $p_T^t$ and rapidity $y^t$, as well as in the invariant mass 
and rapidity of the top-antitop system $m_{t\bar{t}}$ and $y^{t\bar{t}}$. As expected, we found that at the LHC, 
the NNLO corrections to the $q\bar{q}$ initiated process are not phenomenologically significant, except in the 
very forward and backward regions of the rapidity spectrum. For Tevatron, on the other hand, we found that 
these corrections are important and drastically reduce the scale dependence in all distributions considered.

In addition, we evaluated the NNLO corrections to the differential forward-backward asymmetries at Tevatron
and found good agreement with the results from D0 \cite{Abazov:2014cca}. In the regions where the 
contributions from the quark-gluon channel are expected to be small, we also found agreement with 
\cite{Czakon:2014xsa}.

%
\acknowledgments
We would like to thank Philipp Maierh\"ofer for assisting us with the installation and usage of {\tt OpenLoops} 
\cite{Cascioli:2011va}, Roberto Bonciani and Andreas von Manteuffel for their help with the 
implementation of the double virtual matrix elements and Matthias Steinhauser for detailed discussions on the ultraviolet 
renormalisation of massive loop amplitudes. 
G.A. is very grateful to the Institute for Theoretical Physics at ETH Z\"urich for its 
hospitality. He acknowledges support from the Swiss National Science Foundation (SNF) under contract 
PBEZP2-145917 and from the United States Department of Energy under grant DE-FG02-91ER40684 and 
contract DE-AC02-06CH11357. A.G acknowledges the support of the European 
Commission through the ERC Advanced Grant `MC@NNLO' (340983) and from the SNF under contract 
CRSII2-141847. I.M acknowledges the National Institute for International Education (NIIED) for 
supporting him with a Korean Government Scholarship (KGSP) under number KGSP-GRA-2014-244.

%
\appendix
%
%
\section{Master integrals for ${\cal A}^0_{q,Qgg}$}
\label{sec:misa04}
In this appendix we collect the six master integrals found in the IBP reduction of the integrated antenna 
${\cal A}^0_{q,Qgg}$. The Laurent expansion of these integrals can expressed in terms of  HPLs with arguments 
$x_i$ or $x_0$ and GPLs of argument $x_0$ and weights involving $1/x_i$. We keep the overall factor
\beq
\label{eq.NA04}
N_{X04}=C(\epsilon)^2\,(Q^2 +m_Q^2)^{-2\e} (1-x)^{-4\e}\frac{\pi}{2}
\eeq
unexpanded, in order to facilitate the introduction of the master integrals into the integrated antennae and the
subsequent expansion in distributions.

For the integrals $I_{[0]}$ and $I_{[-8]}$ we have the following all order expressions \cite{Abelof:2011jv}:
\beqa
&&\hspace{-0.2in}I_{[0]}=N_{X04}\,\frac{\Gamma^2(1-\e)}{\Gamma(4-4\e)}\, \exp^{2 \e \gamma_{E}}
\left(Q^2+m_Q^2\right)x_i^{-1+2\e}(1-x_i)^{3}(1-x_ix_0)^{-1+2\e}\nonumber\\
&&\hspace{0.5in}\times \gaussf{1-\e}{2-2\e}{4-4\e}{\frac{1-x_i}{1-x_ix_0}} \\ \nonumber\\ \nonumber\\
&&\hspace{-0.2in}I_{[-8]}=N_{X04} \,\frac{\Gamma^2(1-\e)}{2\Gamma(4-4\e)}\, \exp^{2 \e \gamma_{E}}
\left(Q^2+m^2\right)^{2}\,x_i^{-2+2\e}\,(1-x_i)^{4}(1-x_ix_0)^{-2+2\e}\nonumber\\
&&\hspace{0.5in}\times \gaussf{1-\e}{2-2\e}{5-4\e}{\frac{1-x_i}{1-x_ix_0}},
\eeqa
These all order expressions can be expanded up to 
the required order in $\e$ using the {\tt Mathematica} package {\tt HypExp} \cite{Huber:2005yg}. The expanded
forms written in terms of HPLs and GPLs are included in the ancillary file attached to the arXiv submission of 
this paper. 

The remaining master integrals are computed using differential equation techniques. We had already 
calculated $I_{[4]}$ in \cite{Abelof:2012he}. Written in terms of $x_i$ and $x_0$ it reads:
\beqa
&&\hspace{-0.1in}I_{[4]}=\frac{N_{X04}}{x_0}\,\bigg\{x_0G(0;x_i)-(1-x_0)G(1;x_0)+(1-x_0)G\left(\frac{1}{x_i};x_0\right)+\e\,\bigg[-\frac{2\pi^2x_0}{3}\nonumber\\
&&\hspace{0.1in}+5x_0G\left(0;x_i\right)-5(1-x_0)G\left(1;x_0\right)-
(1-x_0)G\left(0;x_i\right)G\left(1;x_0\right)+5(1-x_0)G\left(\frac{1}{x_i};x_0\right)\nonumber\\
&&\hspace{0.1in}+3(1-x_0)G\left(0;x_i\right)G\left(\frac{1}{x_i};x_0\right)+3x_0G\left(0,0;x_i\right)-2G\left(0,1;x_0\right)+
2G\left(0,\frac{1}{x_i};x_0\right)\nonumber\\
&&\hspace{0.1in}+4x_0G\left(1,0;x_i\right)+(1-x_0)G\left(1,1;x_0\right)-
3(1-x_0)G\left(1,\frac{1}{x_i};x_0\right)\nonumber\\
&&\hspace{0.1in}+(1-x_0)G\left(\frac{1}{x_i},1;x_0\right)+
(1-x_0)G\left(\frac{1}{x_i},\frac{1}{x_i};x_0\right)\bigg]+\order{\e^2}\bigg\}.
\eeqa 

The masters $I_{[4,8]}$, $I_{[4,5,8]}$, and $I_{[4,7,8]}$ are new. The first terms in their Laurent expansion 
are given by
\beqa
&&\hspace{-0.1in}I_{[4,8]}=N_{X04}\,\left(Q^2+m^2_Q\right)^{-1}\,x_i\,\bigg\{
G(0,1;x_0) G(0;x_i )-G(0;x_i ) G\left(\frac{1}{x_i
   },1;x_0\right)\nonumber\\
&&\hspace{0.1in}-\frac{ \pi ^2}{3} G\left(\frac{1}{x_i
   };x_0\right)+G(1;x_0) G(0,0;x_i )-G(0,0;x_i )
   G\left(\frac{1}{x_i };x_0\right)-G(1;x_0) G(1,0;x_i)\nonumber\\
&&\hspace{0.1in}+G(1,0;x_i ) G\left(\frac{1}{x_i
   };x_0\right)-G\left(0,0,\frac{1}{x_i
   };x_0\right)-G\left(\frac{1}{x_i
   },0,1;x_0\right)+G\left(\frac{1}{x_i },0,\frac{1}{x_i
   };x_0\right)\nonumber\\
&&\hspace{0.1in}+\frac{ \pi ^2}{3}
   G(1;x_0)+G(0,0,1;x_0)+\frac{\pi^2}{3} G(0;x_i
   )+G(0,0,0;x_i )-G(0,1,0;x_i )+2 \zeta_3+\order{\e}\bigg\}\nonumber\\ 
\\
&&\hspace{-0.1in}I_{[4,5,8]}=N_{X04}\,\left(Q^2+m^2_Q\right)^{-2}\,\frac{x_i^2}{1-x_i}\,\bigg\{\frac{1}{3\e^3}+\frac{1}{\e^2}\bigg[G(0;x_i)+\frac{1}{6}G(1;x_0)+\frac{1}{2}G\bigg(\frac{1}{x_i};x_0\bigg)\bigg]\nonumber\\
&&\hspace{0.1in}+\frac{1}{\e}\bigg[\frac{1}{2} G(1;x_0) G(0;x_i )+\frac{3}{2} G(0;x_i )
   G\left(\frac{1}{x_i };x_0\right)+\frac{1}{2}
   G\left(1,\frac{1}{x_i };x_0\right)+\frac{1}{2}
   G\left(\frac{1}{x_i },1;x_0\right)\nonumber\\
&&\hspace{0.1in}   +\frac{1}{2}
   G\left(\frac{1}{x_i },\frac{1}{x_i };x_0\right)-\frac{1}{6}
   G(1,1;x_0)+3 G(0,0;x_i )-\frac{5 \pi ^2}{18}\bigg]+\order{\e^0}\bigg\} \\ \nonumber\\
&&\hspace{-0.1in}I_{[4,7,8]}=N_{X04}\,\left(Q^2+m^2_Q\right)^{-2}\,x_i\,\bigg\{-\frac{1}{6\e^3}-\frac{1}{3\e^2}G(1;x_0)+\frac{1}{\e}\bigg[-\frac{7\pi^2}{36}-G\left(0,1;x_0\right)\nonumber\\
&&\hspace{0.1in}+G\left(0,\frac{1}{x_i};x_0\right)+2G\left(1,0;x_i\right)+\frac{1}{3}G\left(1,1;x_0\right)-G\left(1,\frac{1}{x_i};x_0\right)\bigg]+\order{\e^0}\bigg\}.
\eeqa

%
\section{Massive phase space for ${\cal A}^0_{q,Qgg}$}
\label{sec:psa04}
In order to provide a boundary condition for the system of differential equations involving the integrals 
$I_{[4,8]}$, $I_{[4,5,8]}$, and $I_{[4,7,8]}$ given in the previous section, we computed the soft limit
of $I_{[4,7,8]}$ via direct evaluation. This required a parametrization of the phase space associated to the 
DIS-like kinematics
\beq
p_2+q\to p_1+p_3+p_4
\eeq
with $p_2^2=p_3^2=p_4^2=0$, $p_1^2=m_Q^2$, $q^2=-Q^2<0$. As far as we know, this phase space has not
been derived previously in the literature. We present it below.

The starting point is
\beqa
&&\hspace{-0.2in}\int\dphi_3=\frac{1}{(2\pi)^{2d-3}}\int{\rm d}^dp_1{\rm d}^dp_3{\rm d}^dp_4\delta^+(p_1^2-m_Q^2)\delta^+(p_3^2)\delta^+(p_4^2)\delta^{(d)}(p_1+p_3+p_4-p_2-q)\nonumber\\
&&\hspace{-0.2in}\phantom{\int\dphi_3}=\frac{1}{4(2\pi)^{2d-3}}\int\frac{{\rm d}^{d-1}p_3}{E_3}\frac{{\rm d}^{d-1}p_4}{E_4}\delta^+((p_3+p_4-p_2-q)^2-m_Q^2)
\eeqa
where we have used the momentum-conserving delta function to integrate out $p_1$. Now, we parametrise  
the momenta as
\beqa
&&\hspace{-0.2in}p_2=E_2(1,\vec{0}_{(d-2)},1)\nonumber\\
&&\hspace{-0.2in}q=(E_{cm}-E_2,\vec{0}_{(d-2)},-E_2)\nonumber\\
&&\hspace{-0.2in}p_3=E_3(1,\vec{0}_{(d-3)},\sin\theta_1,\cos\theta_1)\nonumber\\
&&\hspace{-0.2in}p_4=E_4(1,\vec{0}_{(d-4)},\sin\theta_2\sin\theta_3,\sin\theta_2\cos\theta_3,\cos\theta_2)\\
&&\hspace{-0.2in}p_1=p_2+q-p_3-p_4,
\eeqa
such that,
\beqa
&&\hspace{-0.3in}\int{\rm d}^{d-1}p_3=\int_0^\infty{\rm d}E_3\,E_3^{d-2}\int_{-1}^1{\rm d}(\cos\theta_1)\sin^{d-4}\theta_1\int{\rm d}\Omega_{d-2}\nonumber\\
&&\hspace{-0.3in}\int{\rm d}^{d-1}p_4=\int_0^\infty{\rm d}E_4\,E_4^{d-2}\int_{-1}^1{\rm d}(\cos\theta_2)\sin^{d-4}\theta_2\int_{-1}^1{\rm d}(\cos\theta_3)\sin^{d-5}\theta_3\int{\rm d}\Omega_{d-3}.
\eeqa
The phase space measure takes the form
\beqa
&&\hspace{-0.2in}\int\dphi_3=\frac{\Omega_{d-2}\Omega_{d-3}}{4(2\pi)^{2d-3}}\int{\rm d}E_3{\rm d}E_4{\rm d}(\cos\theta_1){\rm d}(\cos\theta_2){\rm d}(\cos\theta_3)\nonumber\\
&&\hspace{0.4in}\times\Big[E_3E_4\Big]^{d-3}\Big[\sin\theta_1\sin\theta_2\Big]^{d-4}\sin^{d-5}\theta_3,
\eeqa
with the solid angle given by
\beq
\Omega_d=\frac{2\pi^{d/2}}{\Gamma(d/2)}.
\eeq
We can now change variables to the following invariants:
\beqa
&&\hspace{-0.2in} s_{23}=2p_2\cdot p_3=2E_2E_3(1-\cos\theta_1)\nonumber\\
&&\hspace{-0.2in} s_{24}=2p_2\cdot p_3=2E_2E_4(1-\cos\theta_2)\nonumber\\
&&\hspace{-0.2in} s_{34}=2p_3\cdot p_4=2E_3E_4(1-\sin\theta_1\sin\theta_2\cos\theta_3-\cos\theta_1\cos\theta_2)\nonumber\\
&&\hspace{-0.2in} s_{13}=2p_1\cdot p_3=2p_2\cdot p_3+2q\cdot p_3-2p_3\cdot p_4=2E_3E_{cm}-s_{34}\nonumber\\
&&\hspace{-0.2in} s_{14}=2p_1\cdot p_4=2p_2\cdot p_4+2q\cdot p_4-2p_3\cdot p_4=2E_4E_{cm}-s_{34}.
\eeqa
The corresponding Jacobian is
\beq
J=\bigg[\det\bigg(\frac{\partial(s_{13},s_{14},s_{23},s_{24},s_{34})}{\partial(E_3,E_4,\cos\theta_1,\cos\theta_2,\cos\theta_3)} \bigg) \bigg]^{-1}=\frac{1}{32E_{cm}^2E_2^2E_3^2E_4^2\sin\theta_1\sin\theta_2}.
\eeq
Note that the energy $E_2$ is fixed such that $2 p_2 \cdot q= E_{cm}^2 +Q^2$ and $s_{12}$ is not an 
independent integration variable. It is given by
\beq
s_{12}=2 p_1\cdot p_2= (E_{cm}^2+Q^2)-s_{23}-s_{24}.
\eeq

Using the above expressions, the phase space measure in $d=4 -2\e$ dimensions reads:
\beqa
\label{eq.phase1}
\int\dphi_3&=&\frac{2 (4 \pi)^{-4 +2 \e}}{\Gamma(1-2 \e)} \, (E_{cm}^2 +Q^2)^{-1+2\e}
\int  {\rm d}s_{12}\,{\rm d}s_{13} \, {\rm d}s_{14} \, {\rm d}s_{23} \, {\rm d}s_{24} \, {\rm d}s_{34}
(-\Delta_4)^{-1/2 -\e} 
\nonumber\\
& & \times \Theta(-\Delta_4)\, \delta(E_{cm}^2 -m_Q^2 - s_{13}-s_{14}-s_{34})
\,\delta(E_{cm}^2+Q^2 -s_{12}-s_{23}-s_{24})
\nonumber\\
\eeqa
where the dependence on $s_{12}$ has been introduced through a delta function, and the Gram determinant
is given by
\beqa
\Delta_4=\lambda(s_{12}s_{34},s_{13}s_{24},s_{14}s_{23})+4m_Q^2s_{23}s_{24}s_{34}
\eeqa
with
\beq
\lambda(x,y,z)=x^2+y^2+z^2-2xy-2xz-2yz.
\eeq

We now define the following dimensionless variables:
\beq
u_0=\frac{m_Q^2}{E_{cm}^2 -m_Q^2}, \hspace{2cm} u=\frac{E_{cm}^2 +Q^2}{E_{cm}^2 -m_Q^2}
\eeq
and 
\beqa
&&\hspace{-0.2in}z_{12}=\frac{1}{u (E_{cm}^2 -m_Q^2)} \, s_{12}  \hspace{1cm}
z_{13}=\frac{1}{(E_{cm}^2 -m_Q^2)}\, s_{13} \hspace{1cm}
z_{14}=\frac{1}{(E_{cm}^2 -m_Q^2)}\, s_{14} \nonumber\\
&&\hspace{-0.2in}z_{23}=\frac{1}{u (E_{cm}^2 -m_Q^2)}\, s_{23}  \hspace{1cm}
z_{24}=\frac{1}{u(E_{cm}^2 -m_Q^2)}\, s_{24} \hspace{1cm}
z_{34}=\frac{1}{(E_{cm}^2 -m_Q^2)}\, s_{34} \nonumber\\
\eeqa
with
\beq
(E_{cm}^2 -m_Q^2)=\frac{(1-x_i)}{x_i}(Q^2 +m_Q^2)
\eeq
where 
\beq
x_i=\frac{Q^2+m_Q^2}{2 p_2 \cdot q}=\frac{Q^2+m_Q^2}{(E_{cm}^2 +m_Q^2)} \hspace{1cm}\text{and} 
\hspace{1cm}x_0=\frac{Q^2}{Q^2+m_Q^2}.
\eeq 

In terms of the dimensionless variables $z_{ij}$ the phase space reads, 
\beqa
\label{eq.phase1}
\int\dphi_3&=&\frac{2 (4 \pi)^{-4 +2 \e}}{\Gamma(1-2 \e)} \, (E_{cm}^2 +Q^2)^{1-2\e}
\int  {\rm d}z_{12}\,{\rm d}z_{13} \, {\rm d}z_{14} \, {\rm d}z_{23} \, {\rm d}z_{24} \, {\rm d}z_{34}
\nonumber\\
& \times& (-\bar{\Delta}_4)^{-1/2 -\e} \, \Theta(-\bar{\Delta}_4)\,\delta(1- z_{13}-z_{14}-z_{34})
\,\delta(1-z_{12}-z_{23}-z_{24})
\nonumber\\
\eeqa
with
\beq
\bar{\Delta}_4=\lambda(z_{12}z_{34},z_{13}z_{24},z_{14}z_{23})+4u_0z_{23}z_{24}z_{34}.
\eeq
We use the delta functions to integrate out $z_{12}$ and $z_{34}$, and reparametrise the remaining 
integration variables as
\beqa
&&z_{13}=\chi_1\nonumber\\
&&z_{24}=\frac{\chi_2 (1-\chi_1)}{(1+u_0)}\nonumber\\
&&z_{23}=\frac{\chi_1 \chi_3(1+u_0\,-\chi_2(1-\chi_1))}{(1+u_0) (u_0+ \chi_1)}\nonumber\\
&&z_{14}=z_{14}^{-}+\chi_4(z_{14}^{+}-z_{14}^{-}),
\eeqa
where $z_{14}^{\pm}$ are the roots of $\bar{\Delta}_4$ viewed as a function of $z_{14}$. The Jacobian for 
this reparametrisation is
\beq
{\rm d}z_{13} {\rm d}z_{23} {\rm d}z_{24}=\left \{ \frac{\chi_1(1-\chi_1)(1+u_0-(1-\chi_1)\chi_2)}
{(1+u_0)^2 (u_0+\chi_1)} \right \} \,{\rm d}\chi_{1} {\rm d}\chi_{2} {\rm d} \chi_3,
\eeq
and the integration region gets mapped to the hypercube $0\leq \chi_i\leq 1$. We finally arrive at
\beqa
\label{eq.phase2}
\int \dphi_3&=&\frac{\pi^{-4 +2 \e}}{128\, \Gamma(1-2 \e)}
(Q^2+m_Q^2)^{1-2\e}(1-x_i)^{1-2\e}x_i^{-1+2\e}
\int {\rm d} \chi_1 {\rm d} \chi_2 {\rm d} \chi_3 {\rm d} \chi_4 \nonumber \\
& \times & \Big[ \chi_1(1-\chi_1)\Big]^{1- 2\e} 
\Big[ \chi_2(1-\chi_2)\Big]^{-\e} \Big[\chi_3(1-\chi_2)\Big]^{-\e}\Big [\chi_4(1-\chi_4)\Big]^{1/2- \e} 
\nonumber \\
&& \Big[(1-x_0) +\chi_{1}\,(1-x_i) \Big]^{-1+\e}. 
\eeqa

In the soft limit $x_i\to 1$, after the trivial integration over $\chi_4$ is performed, our phase space parametrisation completely factorises as, 
\beqa
\label{eq.phasesoft}
&&\hspace{-0.2in}\int\dphi_{3,soft}=
2^{-7+4 \e}\,\pi^{-3+2 \e} \frac{1}{\Gamma^2(1-\e)}
(Q^2+m_Q^2)^{1-2\e}(1-x_i)^{3-4\e}(1-x_0)^{-2 +2 \e}\nonumber \\
&&\hspace{0.2in}\times \int {\rm d} \chi_1 {\rm d} \chi_2 {\rm d} \chi_3 
\times  \Big[ \chi_1(1-\chi_1) \Big]^{1- 2\e} 
\,\Big[\chi_2(1-\chi_2)\Big]^{-\e} \Big[\chi_3(1-\chi_3)\Big]^{-\e}, 
\eeqa
and the invariants take the following simple form
\beqa 
\label{eq.invsoft}
s_{12}& \to&(Q^2+m_Q^2) \nonumber\\
s_{13}& \to&(Q^2+m_Q^2) \,(1-x_i) \chi_1\nonumber\\
s_{23}& \to&(Q^2+m_Q^2) \chi_1 \chi_3 \frac{(1-x_i)}{(1-x_0)} \nonumber\\
s_{24}&\to &(Q^2+m_Q^2) \chi_2 (1-\chi_1) \frac{(1-x_i)}{(1-x_0)}\nonumber\\
s_{14}& \to &(Q^2+m_Q^2)\frac{1-x_i}{x_i}z_{14}\nonumber\\
s_{34}& \to &(Q^2+m_Q^2)\frac{(1-x_i)}{x_i}z_{34}\nonumber\\
s_{234}&\to &-(Q^2+m_Q^2)\frac{(1-x)}{(1-x_0)}\Big[\chi_1\chi_3 +(1- \chi_1)\chi_2 \Big].
\eeqa

Using eqs.(\ref{eq.phasesoft}) and (\ref{eq.invsoft}) we can find explicit all order expression for all master
integrals that do not involve the invariants $s_{14}$ or $s_{34}$. As a boundary condition for the integrals 
found in the reduction of ${\cal A}^0_{q,Qgg}$ we only needed to compute the soft limit of $I_{[4,7,8]}$,
which reads
\beqa
I^{(soft)}_{[4,7,8]}&=&N_{X04}\frac{1}{(Q^2+m_Q^2)^2}(1-x_0)^{+2 \e}\,\frac{\Gamma^2(1-\e)}{\Gamma(1-2\e)} \exp^{2 \e \gamma_E} \nonumber\\
&&\hspace{0.2in}  \times \left[-\frac{1}{2\e^3} +\frac{1}{6 \e^3}\Fthreetwo{1}{-2\e}{1-\e}{1-3\e}{1-2\e}{1} \right].
\eeqa

%
\section{Master integrals for ${\cal A}^{1,lc}_{q,Qg}$}
\label{sec:misa13}
In this appendix we collect the seven master integrals found in the IBP reduction of the integrated antenna 
${\cal A}^{1,lc}_{q,Qg}$. The Laurent expansion of these integrals can expressed in terms of  HPLs with arguments $x_i$ or $x_0$ and GPLs of argument $x_0$ and weights involving $1/x_i$.

The phase space associated to the DIS-like kinematics of this antenna, namely
\beq
p_2+q\to p_1+p_3
\eeq
with $p_2^2=p_3^2=0$, $p_1^2=m_Q^2$, $q^2=-Q^2<0$, was derived in \cite{Abelof:2011jv}. It is given by
\beq\label{eq.antnlopsif}
\int\dphi_{X_{i,jk}}=(2 \pi) \frac{(4\pi)^{-2+\e}}{\Gamma(1-\e)}\,(Q^2 +m_Q^2)^{-\e}\,x_i^{\e}
\,(1-x_i)^{1-2\e}(1-x_i\,x_{0})^{-1+\e}\int_0^1 {\rm d}y \,y^{-\e}(1-y)^{-\e},
\eeq
with
\beq
x_0=\frac{Q^2}{Q^2+m_Q^2}\hspace{2cm}x_i=\frac{Q^2+m_Q^2}{2p_2\cdot q}.
\eeq

The inclusive phase space integral, denoted as $I_{[0]}$, is obtained by performing the $y$ integration 
in eq.(\ref{eq.antnlopsif}): 
\beq
I_{[0]}=(Q^2 +m_Q^2)^{-\e}\,
(4\pi)^{-1+\e}\frac{\Gamma(1-\e)}{2\Gamma(2-2 \e)}\,x_i^{\e}
(1-x_i)^{1-2\e}(1-x_i\,x_{0})^{-1+\e}.
\eeq

With this parametrization of the phase space, the invariants can be expressed as
\beq\label{eq.antnlopssij}
s_{12}=(Q^2+m_Q^2)\bigg(\frac{1-y-x_ix_0+x_iy}{x(1-x_ix_0)}\bigg)\hspace{0.7in}s_{23}=(Q^2+m_Q^2)\bigg(\frac{(1-x_i)y}{x_i(1-x_ix_0)}\bigg).
\eeq
The third invariant $s_{13}$ can be obtained using momentum conservation. It reads
\beq
s_{13}=s_{12}+s_{23}-Q^2-m_Q^2=(Q^2+m_Q^2)\bigg(\frac{1-x_i}{x_i}\bigg).
\eeq 

The evaluation of the master integrals $I_{[3]}$, $I_{[3,5]}$, $I_{[3,6]}$ and $I_{[4,6]}$ is straightforward. It can  
be done to all orders in $\e$ by employing an all order representation of the tadpole and bubble integrals 
involved, and integrating over the phase space in eq.(\ref{eq.antnlopsif}). The loop integrals that we need are
\beqa
&&\hspace{-0.5in}{\rm Tad}(m_Q^2)=\loopint\frac{1}{l^2-m^2}=-i\,C_{\Gamma}\frac{\Gamma(1-2\e)\Gamma(-1+\e)}{\Gamma^2(1-\e)\Gamma(1+\e)}\,\left(m_Q^2\right)^{1-\e}\\ \nonumber\\
&&\hspace{-0.5in}{\rm Bub}(k^2;0,0)=\loopint\frac{1}{l^2(l-k)^2}=i\frac{C_{\Gamma}}{\e(1-2\e)}\left(-k^2\right)^{-\e}\\ \nonumber\\
&&\hspace{-0.5in}{\rm Bub}(k^2;0,m_Q^2)=\loopint\frac{1}{(l^2-m^2)(l-k)^2}\nonumber\\
&&\hspace{-0.5in}\phantom{{\rm Bub}(k^2;0,m_Q^2)}=i\frac{C_{\Gamma}}{\e(1-\e)}\frac{\Gamma(1-2\e)}{\Gamma^2(1-\e)}\,\left(m^2\right)^{-\e}\gaussf{\e}{1}{2-\e}{\frac{k^2}{m_Q^2}}
\eeqa
with
\beq
C_{\Gamma}=\frac{1}{\left(4\pi\right)^{2-\e}}\frac{\Gamma^2(1-\e)\Gamma(1+\e)}{\Gamma(1-2\e)}.
\eeq

Integrating these loop integrals over the antenna phase space in eq.(\ref{eq.antnlopsif}) we find
\beqa
&&\hspace{-0.3in}I_{[3]}=\int\dphi_{X_{i,jk}}\,{\rm Tad}(m_Q^2)\nonumber\\
&&\hspace{-0.3in}\phantom{I_{[3]}}=
N_{X13}\,S_{\e}\,\left( Q^2+m_Q^2\right) \frac{x_i^\e(1-x_i)^{1-2\e}(1-x_0)^{1-\e}(1-x_ix_0)^{-1+\e}}{\e(1-\e)(1-2\e)}\\ \nonumber\\
&&\hspace{-0.3in}I_{[3,5]}=\int\dphi_{X_{i,jk}}\,{\rm Bub}(s_{13}+m^2_Q,0,m_Q^2)\nonumber\\
&&\hspace{-0.3in}\phantom{I_{[3,5]}}=\frac{N_{X13}\,S_{\e}}{\e(1-2\e)^2}\bigg[
\frac{\cos(2\pi\e)\Gamma(1-\e)\Gamma(1+2\e)}{2\Gamma(1+\e)}x^{2\e}(1-x_i)^{2-4\e}(1-x_ix_0)^{-2+2\e}\nonumber\\
&&\hspace{0.2in}+x_i^{1+\e}(1-x_i)^{1-2\e}(1-x_0)^{1-\e}(1-x_ix_0)^{-2+\e}\gaussf{1}{\e}{2\e}{\frac{1-x}{1-xx_0}}\bigg]\\ \nonumber\\
&&\hspace{-0.3in}I_{[3,6]}=\int\dphi_{X_{i,jk}}\,{\rm Bub}(-Q^2,0,m^2)\nonumber\\ 
&&\hspace{-0.3in}\phantom{I_{[3,6]}}=\frac{N_{X13}\,S_\e}{\e(1-\e)(1-2\e)}x_i^\e(1-x_i)^{1-2\e}(1-x_ix_0)^{-1+\e}\gaussf{1-\e}{\e}{2-\e}{x_0}\\ \nonumber\\
&&\hspace{-0.3in}I_{[4,6]}=\int\dphi_{X_{i,jk}}\,{\rm Bub}(-s_{23},0,0)\nonumber\\
&&\hspace{-0.3in}\phantom{I_{[4,6]}}=N_{X13}\,S_{\e}\,\frac{\Gamma(1-\e)\Gamma(1-2\e)}{\e(1-2\e)\Gamma(2-3\e)}x_i^{2\e}(1-x_i)^{1-3\e}(1-x_ix_0)^{-1+2\e},\\ \nonumber
\eeqa
with the overall factors $N_{X13}$ and $S_\e$ given by
\beq
N_{X13}=i\,\cep^2\,(Q^2+m_Q^2)^{-2\e}
\eeq
and
\beq
S_\e=C_\Gamma\frac{e^{\e\gamma_E}}{2\Gamma(1-\e)}.
\eeq

\begin{figure}[t]
\centering
\includegraphics[width=0.9\textwidth]{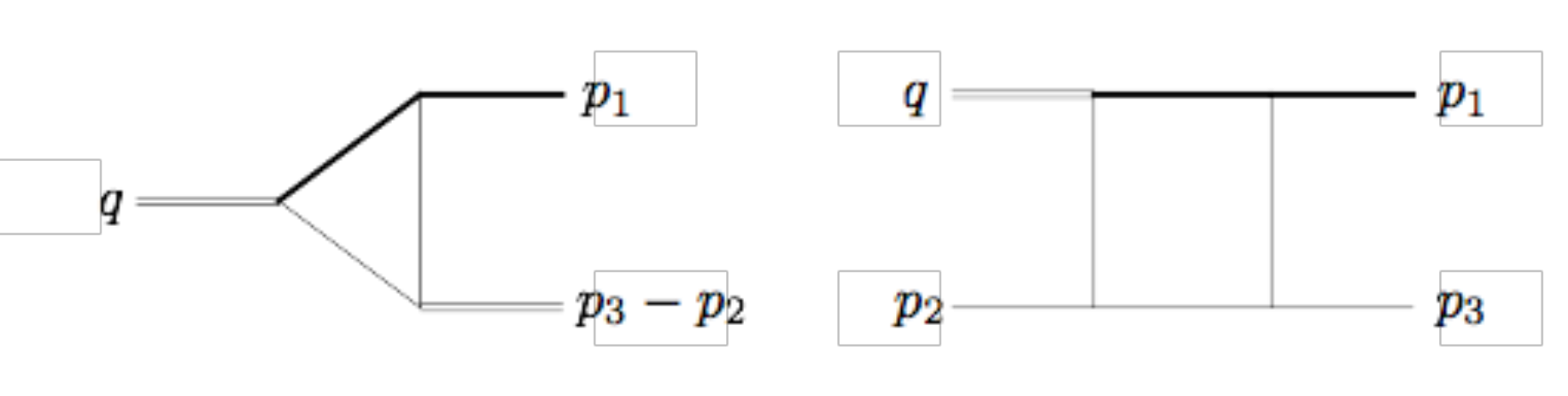}
\caption{Triangle and box one-loop integrals}
\label{fig.boxtri}
\end{figure}

The remaining three master integrals, namely $I_{[3,4,6]}$, $I_{[3,4,5,6]}$ and $I_{[3,4,6,7]}$, cannot be 
computed in this way. This would require all order expressions for the triangle and box integrals depicted in 
fig.\ref{fig.boxtri} and given by 
\beqa
&&\hspace{-0.5in}{\rm Tri}(m_Q^2,-s_{23},-Q^2;m_Q^2,0,0)=\loopint\frac{1}{(l^2-m_Q^2)(l-p_1)^2(l-p_1+p_2-p_3)^2}\\ \nonumber\\
&&\hspace{-0.5in}{\rm Box}(m_Q^2,0,0,-Q^2;s_{13}+m_Q^2,-s_{23};m_Q^2,0,0,0)=\nonumber\\
&&\hspace{0.5in}\loopint\frac{1}{(l^2-m_Q^2)(l-p_1)^2(l-p_1-p_3)^2(l-p_1+p_2-p_3)^2},
\eeqa
which are not known.

We computed these mixed one-loop phase space integrals with differential equations, using their soft limits,
independently evaluated to all orders in $\e$, as boundary conditions. For $I_{[3,4,5,6]}$ we used the soft 
limit of the underlying one-loop box given in \cite{Brucherseifer:2013iv}, whereas for $I_{[3,4,6]}$ and 
$I_{[3,4,6,7]}$ we calculated the soft limit of the one-loop triangle to all orders in $\e$ using a Mellin-Barnes 
representation. 

We express the integrals in terms of the regular functions $R_{\alpha}^{(n)}$ as in eq.(\ref{eq.ialpha}) in
order to facilitate the expansion of ${\cal A}^{1,lc}_{q,Qg}$ in plus distributions. The decomposition of 
$I_{[3,4,6]}$, $I_{[3,4,5,6]}$ and $I_{[3,4,6,7]}$ into regular functions $R_{\alpha}^{(n)}$ reads
\beqa
&&\hspace{-0.3in}I_{[3,4,6]}=N_{X13}\Big(
(1-x)^{-2 \e} R_{[3,4,6]}^{(2)} + (1-x)^{-3 \e} R_{[3,4,6]}^{(3)} + (1-x)^{-4 \e} R_{[3,4,6]}^{(4)}
\Big ) \\ \nonumber\\
&&\hspace{-0.3in}I_{[3,4,6,7]}=N_{X13}\Big(
(1-x)^{-2 \e} R_{[3,4,6,7]}^{(2)} + (1-x)^{-3 \e} R_{[3,4,6,7]}^{(3)} + (1-x)^{-4 \e} R_{[3,4,6,7]}^{(4)}
\Big ) \\ \nonumber\\
&&\hspace{-0.3in}I_{[3,4,5,6]}=N_{X13}(1-x)^{-4 \e}R_{[3,4,5,6]}^{(4)}.
\eeqa
The first few terms in the Laurent expansion the $R_{\alpha}^{(n)}$'s in the above equation are given by:
\beqa
&&\hspace{-0.05in}R_{[3,4,6]}^{(2)}=\frac{1}{Q^2+m_Q^2}x_i\bigg\{\frac{1}{8\e^2}\bigg[G(1;x_0)-G\left(\frac{1}{x_i};x_0\right)\bigg]+\frac{1}{8\e}\bigg[4G\left(0,\frac{1}{x_i};x_0\right)-4G\left(0,1;x_0\right)\nonumber\\
&&\hspace{0.2in}-G\left(1,1;x_0\right)+3G\left(1,\frac{1}{x_i};x_0\right)+
G\left(\frac{1}{x_i},1;x_0\right)-3G\left(\frac{1}{x_i},\frac{1}{x_i};x_0\right)\bigg]+\order{\e^0}\bigg\}\nonumber\\
\\
&&\hspace{-0.05in}R_{[3,4,6]}^{(3)}=\frac{1}{Q^2+m_Q^2}x_i\bigg\{-\frac{1}{4\e^2}\bigg[G(1;x_0)-G\left(\frac{1}{x_i};x_0\right)\bigg]+\frac{1}{4\e}\bigg[3G\left(0,\frac{1}{x_i};x_0\right)\nonumber\\
&&\hspace{0.2in}-3G\left(0,1;x_0\right)+G\left(1,1;x_0\right)-3G\left(1,\frac{1}{x_i};x_0\right)+2G\left(\frac{1}{x_i},\frac{1}{x_i};x_0\right)\bigg]+\order{\e^0}\bigg\}\nonumber\\
\\
&&\hspace{-0.05in}R_{[3,4,6]}^{(4)}=\frac{1}{Q^2+m_Q^2}x_i\bigg\{\frac{1}{8\e^2}\bigg[G(1;x_0)-G\left(\frac{1}{x_i};x_0\right)\bigg]+\frac{1}{8\e}\bigg[2G\left(0,\frac{1}{x_i};x_0\right)-2G\left(0,1;x_0\right)\nonumber\\
&&\hspace{0.2in}-G\left(1,1;x_0\right)+3G\left(1,\frac{1}{x_i};x_0\right)-
G\left(\frac{1}{x_i},1;x_0\right)-G\left(\frac{1}{x_i},\frac{1}{x_i};x_0\right)\bigg]+\order{\e^0}\bigg\}\nonumber\\
\\
&&\hspace{-0.05in}R_{[3,4,6,7]}^{(2)}=\frac{1}{(Q^2+m_Q^2)^2}x_i\bigg\{-\frac{1}{8\e^3}-\frac{1}{4\e^2}G(1;x_0)+\frac{1}{4\e}\bigg[ G(1,1;x_0)-3G(0,1;x_0)\nonumber\\
&&\hspace{0.2in}+2G\left(0,\frac{1}{x_i};x_0\right)-2G\left(1,\frac{1}{x_i};x_0\right)+\frac{\pi^2}{12}  \bigg]+\order{\e^0}\bigg\}
\\ \nonumber\\
&&\hspace{-0.05in}R_{[3,4,6,7]}^{(3)}=\frac{1}{(Q^2+m_Q^2)^2}x_i\bigg\{\frac{1}{8\e^3}+\frac{1}{4\e^2}G(1;x_0)+\frac{1}{4\e}\bigg[-G(1,1;x_0)+3G(0,1;x_0)\nonumber\\
&&\hspace{0.2in}-3G\left(0,\frac{1}{x_i};x_0\right)+3G\left(1,\frac{1}{x_i};x_0\right) -\frac{\pi^2}{4} \bigg]+\order{\e^0}\bigg\}
\\ \nonumber\\
&&\hspace{-0.2in}R_{[3,4,6,7]}^{(4)}=\frac{1}{(Q^2+m_Q^2)^2}x_i\bigg\{-\frac{1}{24\e^3}-\frac{1}{12\e^2}G(1;x_0)+\frac{1}{4\e}\bigg[\frac{1}{3}G(1,1;x_0)-G(0,1;x_0)\nonumber\\
&&\hspace{0.2in}+G\left(0,\frac{1}{x_i};x_0\right)-3G\left(1,\frac{1}{x_i};x_0\right) -\frac{\pi^2}{36} \bigg]+\order{\e^0}\bigg\}
\\ \nonumber\\
&&\hspace{-0.2in}R_{[3,4,5,6]}^{(4)}=\frac{1}{(Q^2+m_Q^2)^2}\frac{x_i^2}{1-x_i}\bigg\{\frac{5}{24\e^3}
+\frac{1}{24\e^2}\bigg[G(1;x_0)+15G(0;x_0)+9G\left(\frac{1}{x_i};x_0\right)\nonumber\\
&&\hspace{0.2in}+\frac{1}{24\e}\bigg[ 
3G\left(0;x_i\right)G\left(1;x_0\right)+27G\left(0;x_i\right)G\left(\frac{1}{x_i};x_0\right)+39G\left(0,0;x_i\right)+6G\left(0,1;x_i\right)\nonumber\\
&&\hspace{0.2in}-G\left(1,1;x_0\right)+3G\left(1,\frac{1}{x_i};x_0\right)+
3G\left(\frac{1}{x_i},1;x_0\right)+15G\left(\frac{1}{x_i},\frac{1}{x_i};x_0\right)-\frac{25}{6}\pi^2
\bigg]+\order{\e^0}\bigg\}.\nonumber\\
\eeqa


\bibliography{bibliography}
\end{document}